\documentclass[final,1p,times,numbers]{elsarticle}
\usepackage{amsmath}
\usepackage{amssymb}
\usepackage{amsthm}
\usepackage{algorithm}
\usepackage{algpseudocode}
\usepackage{tcolorbox}
\tcbuselibrary{breakable,skins}
\usepackage{graphicx}
\usepackage[export]{adjustbox}
\usepackage{booktabs}
\usepackage{multirow}
\usepackage{array}
\usepackage{float}
\usepackage{subcaption}
\usepackage{tabularx}
\usepackage{longtable}
\usepackage{geometry}
\usepackage{setspace}
\algrenewcommand\algorithmicrequire{\textbf{Require:}}

\AtBeginDocument{%
  \newgeometry{
    left=1.4cm,
    right=1.4cm,
    top=2cm,
    bottom=2cm
  }%
}

\usepackage[colorlinks=true,citecolor=blue,linkcolor=blue,urlcolor=blue]{hyperref}

\biboptions{sort&compress}
\usepackage{placeins}

\makeatletter

\expandafter\def\csname legbar@Figures/All_aluminum/legends/legend_theta_Err.pdf\endcsname{{1.}{0.55}}
\expandafter\def\csname legbar@Figures/All_aluminum/legends/legend_theta_True.pdf\endcsname{{1.1}{0.6}}
\expandafter\def\csname legbar@Figures/All_aluminum/legends/legend_u_Err.pdf\endcsname{{1.}{0.55}}
\expandafter\def\csname legbar@Figures/All_aluminum/legends/legend_u_True.pdf\endcsname{{1.1}{0.55}}
\expandafter\def\csname legbar@Figures/All_aluminum/legends/legend_w_Err.pdf\endcsname{{1.}{0.55}}
\expandafter\def\csname legbar@Figures/All_aluminum/legends/legend_w_True.pdf\endcsname{{1.}{0.5}}

\expandafter\def\csname legbar@Figures/All_blood/legends/legend_p_Err.pdf\endcsname{{1.}{0.4781}}
\expandafter\def\csname legbar@Figures/All_blood/legends/legend_p_True.pdf\endcsname{{1.}{0.4781}}
\expandafter\def\csname legbar@Figures/All_blood/legends/legend_theta_Err.pdf\endcsname{{1.05}{0.55}}
\expandafter\def\csname legbar@Figures/All_blood/legends/legend_theta_True.pdf\endcsname{{1.1}{0.60}}
\expandafter\def\csname legbar@Figures/All_blood/legends/legend_u_Err.pdf\endcsname{{1.05}{0.55}}
\expandafter\def\csname legbar@Figures/All_blood/legends/legend_u_True.pdf\endcsname{{0.9}{0.45}}
\expandafter\def\csname legbar@Figures/All_blood/legends/legend_w_Err.pdf\endcsname{{1.05}{0.5481}}
\expandafter\def\csname legbar@Figures/All_blood/legends/legend_w_True.pdf\endcsname{{1.}{0.52}}

\expandafter\def\csname legbar@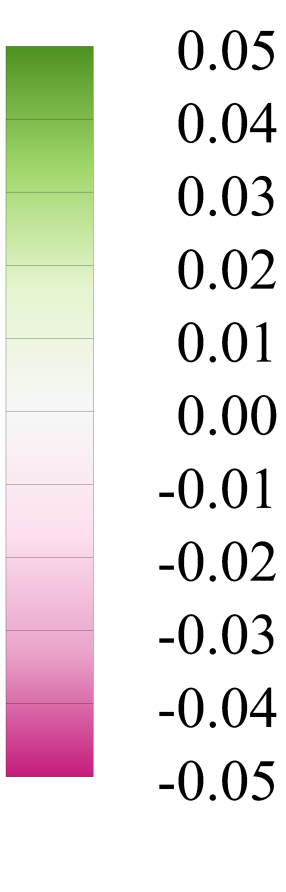\endcsname{{1.}{0.55}}
\expandafter\def\csname legbar@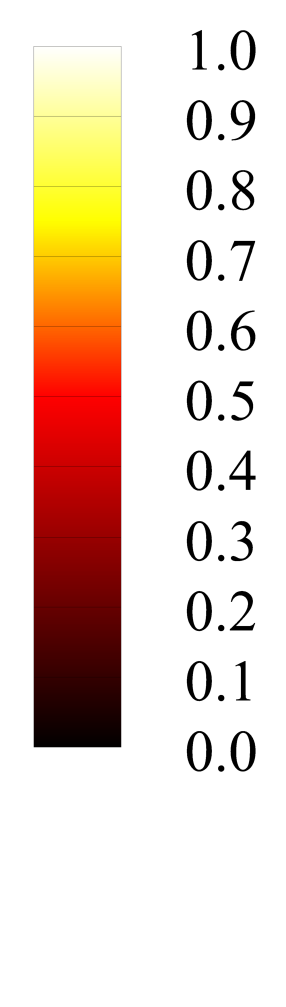\endcsname{{1.}{0.62}}
\expandafter\def\csname legbar@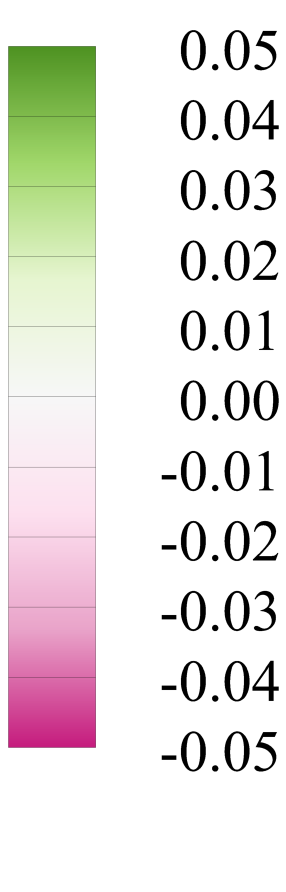\endcsname{{1.}{0.55}}
\expandafter\def\csname legbar@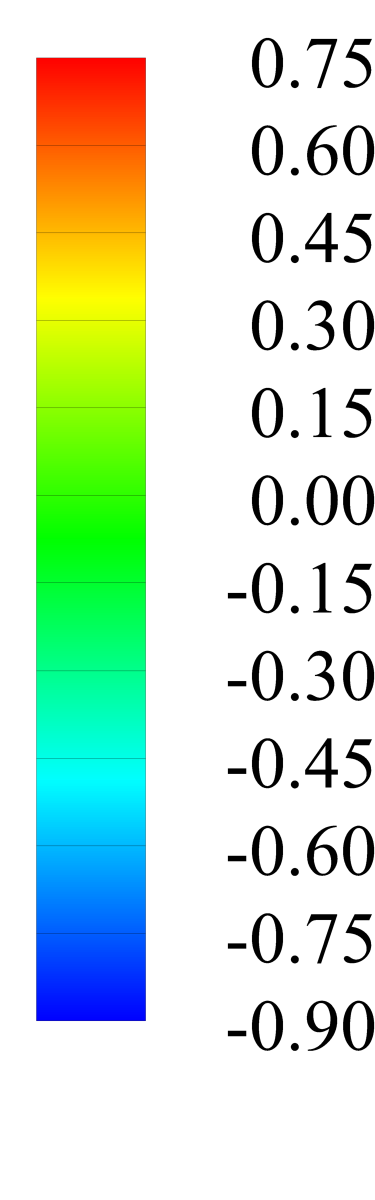\endcsname{{1.}{0.55}}
\expandafter\def\csname legbar@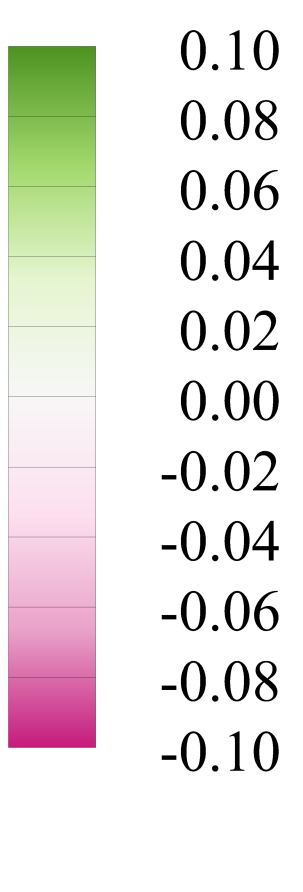\endcsname{{1.}{0.55}}
\expandafter\def\csname legbar@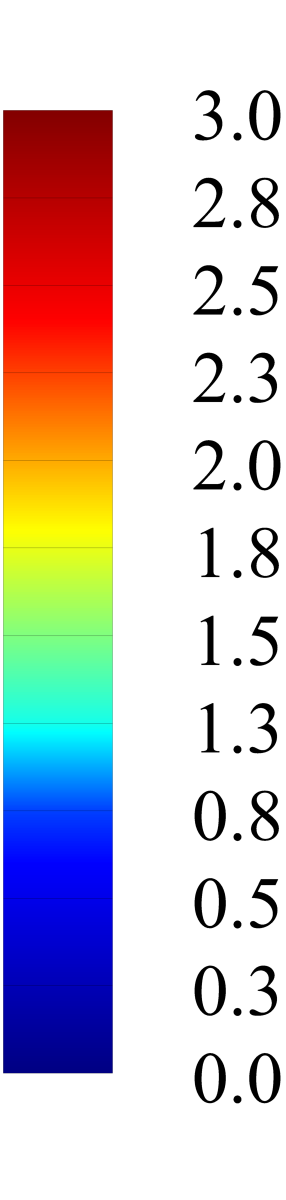\endcsname{{1.}{0.5}}

\expandafter\def\csname legbar@Figures/All_gold/legends/legend_theta_Err.pdf\endcsname{{1.}{0.55}}
\expandafter\def\csname legbar@Figures/All_gold/legends/legend_theta_True.pdf\endcsname{{1.}{0.6}}
\expandafter\def\csname legbar@Figures/All_gold/legends/legend_u_Err.pdf\endcsname{{1.}{0.55}}
\expandafter\def\csname legbar@Figures/All_gold/legends/legend_u_True.pdf\endcsname{{1.}{0.55}}
\expandafter\def\csname legbar@Figures/All_gold/legends/legend_w_Err.pdf\endcsname{{1.}{0.55}}
\expandafter\def\csname legbar@Figures/All_gold/legends/legend_w_True.pdf\endcsname{{1.}{0.5}}

\expandafter\def\csname legbar@Figures/All_silver/legends/legend_theta_Err.pdf\endcsname{{1.}{0.55}}
\expandafter\def\csname legbar@Figures/All_silver/legends/legend_theta_True.pdf\endcsname{{1.}{0.65}}
\expandafter\def\csname legbar@Figures/All_silver/legends/legend_u_Err.pdf\endcsname{{1.}{0.55}}
\expandafter\def\csname legbar@Figures/All_silver/legends/legend_u_True.pdf\endcsname{{1.}{0.47}}
\expandafter\def\csname legbar@Figures/All_silver/legends/legend_w_Err.pdf\endcsname{{1.}{0.55}}
\expandafter\def\csname legbar@Figures/All_silver/legends/legend_w_True.pdf\endcsname{{1.}{0.5}}

\expandafter\def\csname legbar@Figures/All_aluminum/u0001/legend_p_Err.pdf\endcsname{{1.0}{0.55}}
\expandafter\def\csname legbar@Figures/All_aluminum/u0001/legend_p_True.pdf\endcsname{{0.9}{0.4781}}
\expandafter\def\csname legbar@Figures/All_aluminum/u0005/legend_p_Err.pdf\endcsname{{1.}{0.55}}
\expandafter\def\csname legbar@Figures/All_aluminum/u0005/legend_p_True.pdf\endcsname{{1.1}{0.55}}
\expandafter\def\csname legbar@Figures/All_aluminum/u005/legend_p_Err.pdf\endcsname{{1.}{0.55}}
\expandafter\def\csname legbar@Figures/All_aluminum/u005/legend_p_True.pdf\endcsname{{1.05}{0.55}}
\expandafter\def\csname legbar@Figures/All_aluminum/u02/legend_p_Err.pdf\endcsname{{1.}{0.55}}
\expandafter\def\csname legbar@Figures/All_aluminum/u02/legend_p_True.pdf\endcsname{{1.05}{0.55}}

\expandafter\def\csname legbar@Figures/All_blood/u0001/legend_p_Err.pdf\endcsname{{1.}{0.55}}
\expandafter\def\csname legbar@Figures/All_blood/u0001/legend_p_True.pdf\endcsname{{1.}{0.5}}
\expandafter\def\csname legbar@Figures/All_blood/u0005/legend_p_Err.pdf\endcsname{{1.0}{0.55}}
\expandafter\def\csname legbar@Figures/All_blood/u0005/legend_p_True.pdf\endcsname{{1.0}{0.55}}
\expandafter\def\csname legbar@Figures/All_blood/u005/legend_p_Err.pdf\endcsname{{1.0}{0.55}}
\expandafter\def\csname legbar@Figures/All_blood/u005/legend_p_True.pdf\endcsname{{1.15}{0.55}}
\expandafter\def\csname legbar@Figures/All_blood/u02/legend_p_Err.pdf\endcsname{{1.}{0.55}}
\expandafter\def\csname legbar@Figures/All_blood/u02/legend_p_True.pdf\endcsname{{1.15}{0.55}}

\expandafter\def\csname legbar@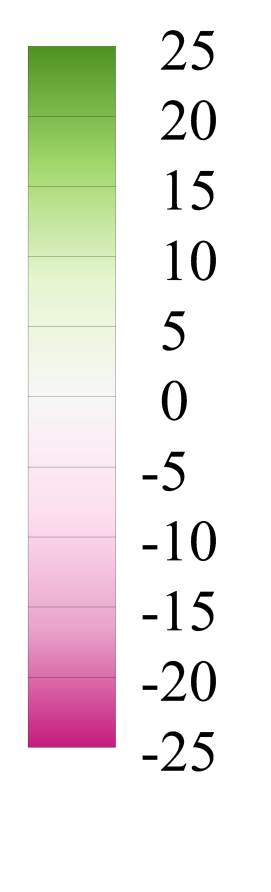\endcsname{{1.}{0.55}}
\expandafter\def\csname legbar@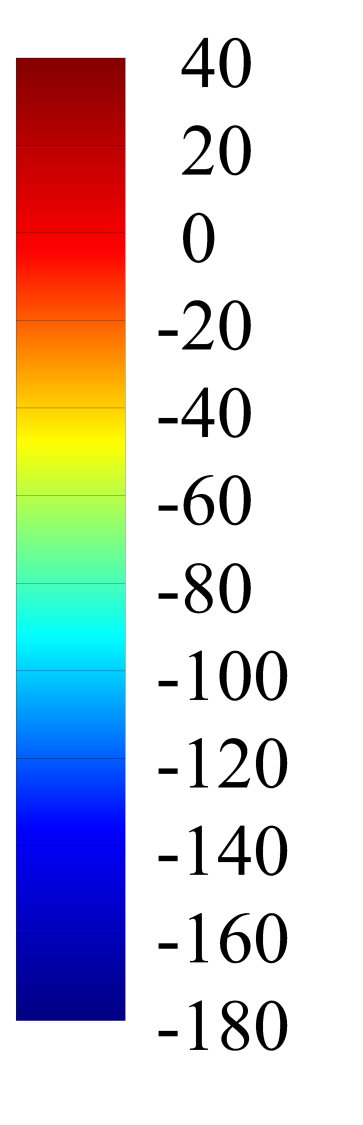\endcsname{{.95}{0.55}}
\expandafter\def\csname legbar@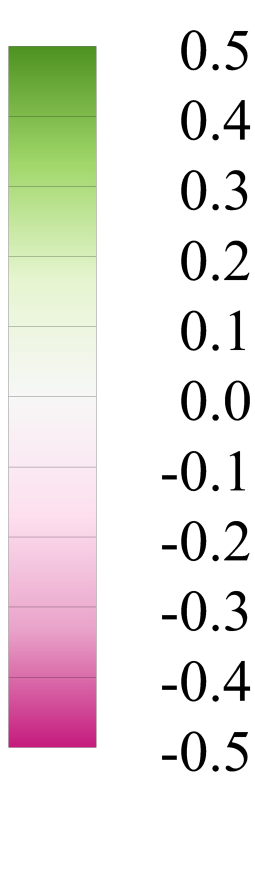\endcsname{{1.}{0.55}}
\expandafter\def\csname legbar@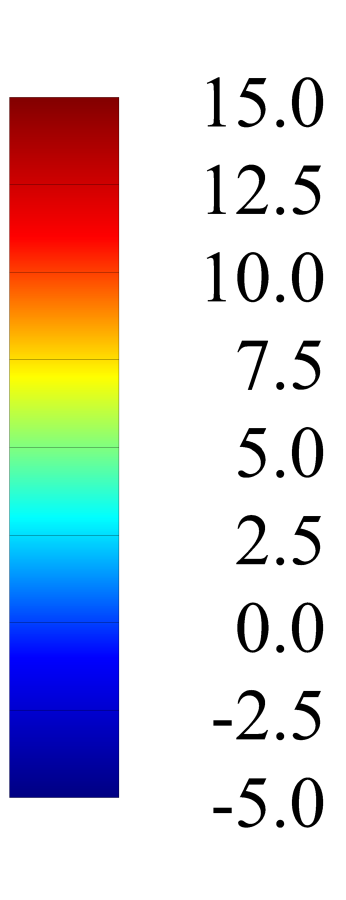\endcsname{{1.}{0.5}}
\expandafter\def\csname legbar@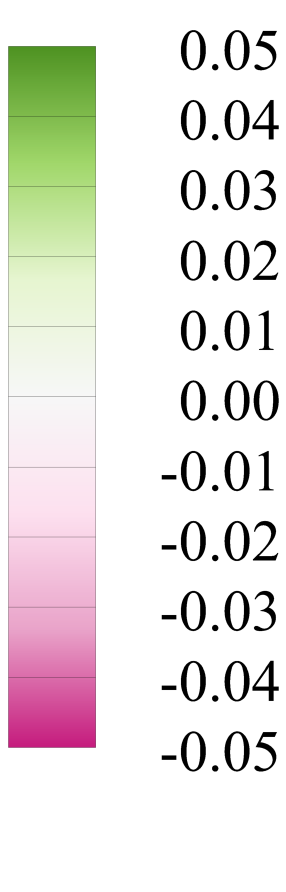\endcsname{{1.}{0.55}}
\expandafter\def\csname legbar@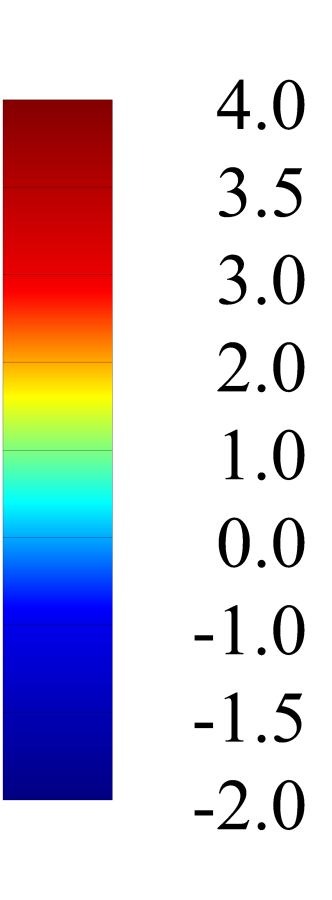\endcsname{{1.}{0.5}}
\expandafter\def\csname legbar@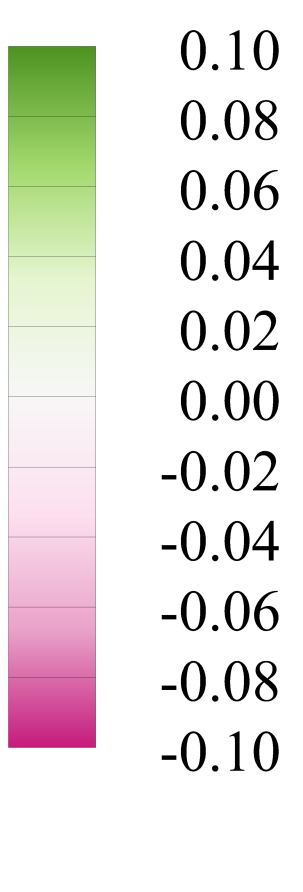\endcsname{{1.}{0.55}}
\expandafter\def\csname legbar@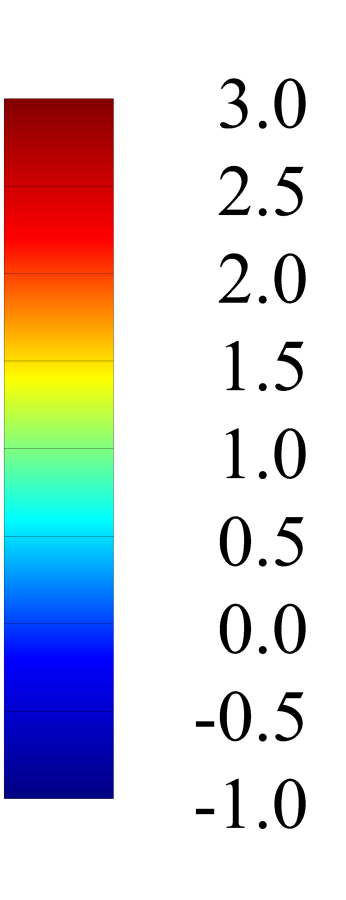\endcsname{{1.}{0.5}}

\expandafter\def\csname legbar@Figures/All_gold/u0001/legend_p_Err.pdf\endcsname{{0.9}{0.55}}
\expandafter\def\csname legbar@Figures/All_gold/u0001/legend_p_True.pdf\endcsname{{0.8}{0.4781}}
\expandafter\def\csname legbar@Figures/All_gold/u0005/legend_p_Err.pdf\endcsname{{1.}{0.55}}
\expandafter\def\csname legbar@Figures/All_gold/u0005/legend_p_True.pdf\endcsname{{0.85}{0.4781}}
\expandafter\def\csname legbar@Figures/All_gold/u005/legend_p_Err.pdf\endcsname{{1.}{0.55}}
\expandafter\def\csname legbar@Figures/All_gold/u005/legend_p_True.pdf\endcsname{{0.9}{0.4781}}
\expandafter\def\csname legbar@Figures/All_gold/u02/legend_p_Err.pdf\endcsname{{1.}{0.55}}
\expandafter\def\csname legbar@Figures/All_gold/u02/legend_p_True.pdf\endcsname{{0.85}{0.4781}}

\expandafter\def\csname legbar@Figures/All_silver/u0001/legend_p_Err.pdf\endcsname{{1.0}{0.55}}
\expandafter\def\csname legbar@Figures/All_silver/u0001/legend_p_True.pdf\endcsname{{1.05}{0.55}}
\expandafter\def\csname legbar@Figures/All_silver/u0005/legend_p_Err.pdf\endcsname{{1.}{0.55}}
\expandafter\def\csname legbar@Figures/All_silver/u0005/legend_p_True.pdf\endcsname{{1.1}{0.6}}
\expandafter\def\csname legbar@Figures/All_silver/u005/legend_p_Err.pdf\endcsname{{1.}{0.55}}
\expandafter\def\csname legbar@Figures/All_silver/u005/legend_p_True.pdf\endcsname{{1.}{0.50}}
\expandafter\def\csname legbar@Figures/All_silver/u02/legend_p_Err.pdf\endcsname{{1.}{0.55}}
\expandafter\def\csname legbar@Figures/All_silver/u02/legend_p_True.pdf\endcsname{{1.1}{0.55}}

\expandafter\def\csname legbar@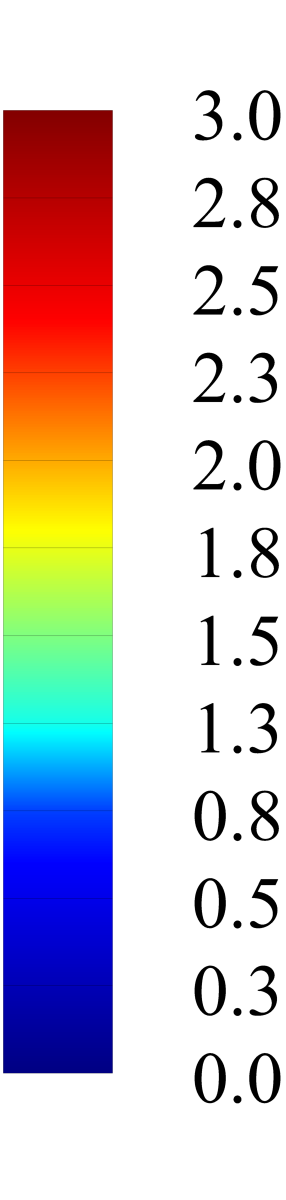\endcsname{{1.}{0.5069}}
\expandafter\def\csname legbar@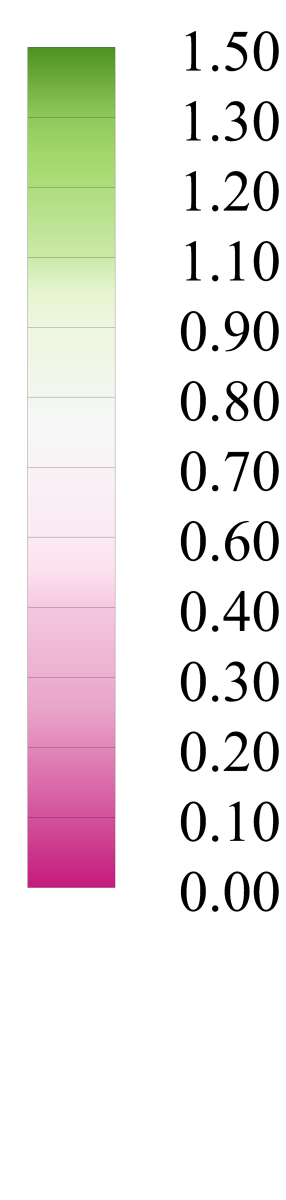\endcsname{{1.15}{0.6106}}
\expandafter\def\csname legbar@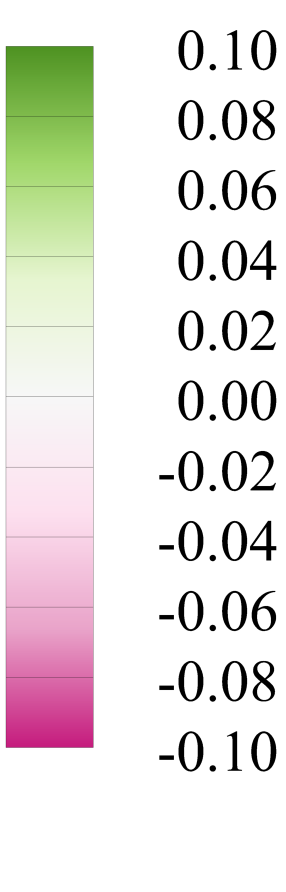\endcsname{{1.}{0.5546}}

\expandafter\def\csname legbar@Figures/All_aluminum/legends/legend_stream_Err.pdf\endcsname{{1}{0.5357}}
\expandafter\def\csname legbar@Figures/All_aluminum/legends/legend_stream_True.pdf\endcsname{{1}{0.5128}}
\expandafter\def\csname legbar@Figures/All_blood/legends/legend_stream_Err.pdf\endcsname{{1.}{0.5462}}
\expandafter\def\csname legbar@Figures/All_blood/legends/legend_stream_True.pdf\endcsname{{1.}{0.5128}}
\expandafter\def\csname legbar@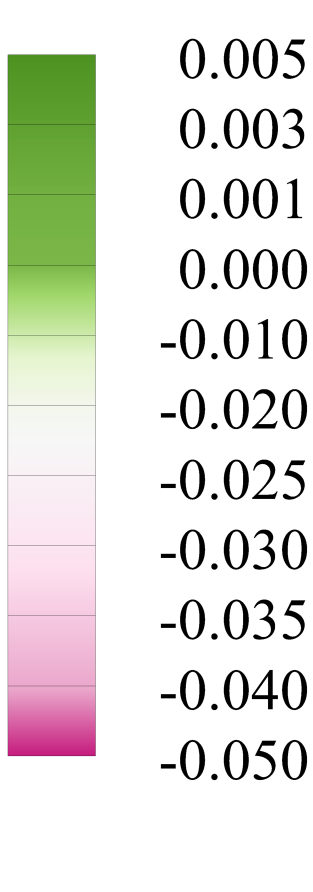\endcsname{{1}{0.5462}}
\expandafter\def\csname legbar@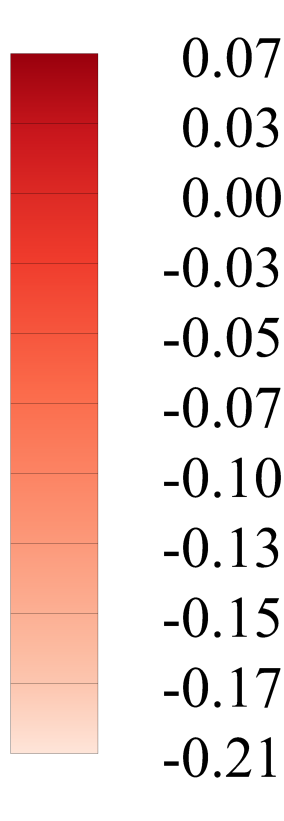\endcsname{{1}{0.5128}}
\expandafter\def\csname legbar@Figures/All_gold/legends/legend_stream_Err.pdf\endcsname{{1}{0.5462}}
\expandafter\def\csname legbar@Figures/All_gold/legends/legend_stream_True.pdf\endcsname{{1}{0.5128}}
\expandafter\def\csname legbar@Figures/All_silver/legends/legend_stream_Err.pdf\endcsname{{1}{0.5462}}
\expandafter\def\csname legbar@Figures/All_silver/legends/legend_stream_True.pdf\endcsname{{1}{0.5128}}

\newlength{\leg@H}\newlength{\leg@h}
\newcommand{\legscale}{1}
\def\leg@domain{0.5063}
\newcommand{\alignedleg}[3][\legscale]{%
  \IfFileExists{#2}{%
    \@ifundefined{legbar@#2}%
      {\includegraphics[height=#3]{#2}}%
      {\expandafter\expandafter\expandafter\alignedleg@\csname legbar@#2\endcsname{#1}{#2}{#3}}}%
    {\fbox{\parbox[b][#3][c]{0.2\fieldh}{\centering\tiny legend\\pending}}}}
\def\alignedleg@#1#2#3#4#5{
  \setlength{\leg@H}{#5}\setlength{\leg@h}{#1\leg@H}\setlength{\leg@h}{#3\leg@h}
  \raisebox{\dimexpr\leg@domain\leg@H-#2\leg@h\relax}{\includegraphics[height=\leg@h]{#4}}}
\makeatother
\newlength{\fieldh}
\newcommand{\fieldimg}[2][\fieldh]{\IfFileExists{#2}{\includegraphics[height=#1]{#2}}{\fbox{\parbox[b][#1][c]{0.42#1}{\centering\tiny missing}}}}
\newcommand{\fieldrow}[3]{\fieldimg{Figures/#1/u0001/#2_#3.pdf}\hspace{1pt}\fieldimg{Figures/#1/u0005/#2_#3.pdf}\hspace{1pt}\fieldimg{Figures/#1/u005/#2_#3.pdf}\hspace{1pt}\fieldimg{Figures/#1/u02/#2_#3.pdf}}
\newcommand{\rowlab}[1]{\rotatebox{90}{\makebox[\fieldh][c]{\scriptsize #1}}\hspace{1pt}}
\newsavebox{\labbox}
\newcommand{\labimg}[3]{\sbox{\labbox}{\fieldimg[#1]{#2}}\usebox{\labbox}\llap{\makebox[\wd\labbox][c]{\raisebox{-1.1em}[0pt][0pt]{\scriptsize #3}}}}
\makeatletter
\newcommand{\Rehead}[1]{\rlap{\raisebox{\dimexpr\fieldh+3pt}[0pt][0pt]{\expandafter\expandafter\expandafter\Rehead@\csname Re@#1\endcsname}}}
\newcommand{\Rehead@c}[1]{\hspace{0.1425\fieldh}\makebox[0.3136\fieldh][c]{\scriptsize $Re=#1$}\hspace{0.0127\fieldh}}
\newcommand{\Rehead@}[4]{\Rehead@c{#1}\hspace{1pt}\Rehead@c{#2}\hspace{1pt}\Rehead@c{#3}\hspace{1pt}\Rehead@c{#4}}
\expandafter\def\csname Re@All_blood\endcsname{{0.46}{2.28}{22.8}{91.1}}
\expandafter\def\csname Re@All_silver\endcsname{{0.53}{2.67}{26.7}{106.9}}
\expandafter\def\csname Re@All_copper\endcsname{{0.52}{2.58}{25.8}{103.2}}
\expandafter\def\csname Re@All_aluminum\endcsname{{0.46}{2.28}{22.8}{91.4}}
\expandafter\def\csname Re@All_gold\endcsname{{0.64}{3.20}{32.0}{127.9}}
\makeatother
\newcommand{\fieldblock}[4][\legscale]{%
  \ifnum#4=1
    \vspace*{0.9em}\rowlab{True}\Rehead{#2}\fieldrow{#2}{True}{#3}\hspace{3pt}\smash{\alignedleg[#1]{Figures/#2/legends/legend_#3_True.pdf}{\fieldh}}\\[2pt]
    \rowlab{Predicted}\fieldrow{#2}{Pred}{#3}\hspace{3pt}\smash{\alignedleg[#1]{Figures/#2/legends/legend_#3_True.pdf}{\fieldh}}\\[2pt]
    \rowlab{Error}\fieldrow{#2}{Err}{#3}\hspace{3pt}\smash{\alignedleg[#1]{Figures/#2/legends/legend_#3_Err.pdf}{\fieldh}}%
  \else
    \vspace*{0.9em}\rowlab{True}\Rehead{#2}\fieldrow{#2}{True}{#3}\\[2pt]
    \rowlab{Predicted}\fieldrow{#2}{Pred}{#3}%
    \ifnum#4=2 \\[2pt]\rowlab{Error}\fieldrow{#2}{Err}{#3}\fi
  \fi}
\newlength{\presh}
\newcommand{\presrow}[3][\legscale]{\makebox[\linewidth][c]{\labimg{\presh}{Figures/#2/#3/True_p.pdf}{True}\hspace{1pt}\labimg{\presh}{Figures/#2/#3/Pred_p.pdf}{Predicted}\hspace{2pt}\alignedleg[#1]{Figures/#2/#3/legend_p_True.pdf}{\presh}\hspace{5pt}\labimg{\presh}{Figures/#2/#3/Err_p.pdf}{Error}\hspace{2pt}\alignedleg[#1]{Figures/#2/#3/legend_p_Err.pdf}{\presh}}\par\vspace{1.1em}}

\usepackage{orcidlink}

\journal{Journal of Computational Physics}
\begin{document}
\begin{frontmatter}
\title{SAGE-PINN: A Singularity-free Axisymmetric Geometry-Encoded Physics-Informed Neural Network for Axisymmetric Multiphysics Flow in Cylindrical Coordinates}

\author[iitm]{Raj Maurya\orcidlink{0009-0005-8225-5852}}
\author[iitm]{Joyprakash Akhuli\orcidlink{0009-0004-0905-5398}}
\author[iitm]{Doyel Pandey\corref{cor1}\orcidlink{0000-0002-8331-5424}}
\ead{doyel@iitmandi.ac.in}
\cortext[cor1]{Corresponding author}

\affiliation[iitm]{organization={Indian Institute of Technology Mandi},
            addressline={Mandi},
            postcode={175001},
            state={Himachal Pradesh},
            country={India}}


\begin{abstract}
Physics-informed neural networks (PINNs) provide a mesh-free framework for multiphysics modeling, yet fluid flow in axisymmetric cylindrical coordinates remains largely unexplored because the governing operators are singular at the axis and momentum and energy are strongly coupled through advection and buoyancy. We introduce \textbf{SAGE-PINN}, a \textbf{Singularity-free Axisymmetric Geometry-Encoded Physics-Informed Neural Network} for this class of problems. SAGE-PINN removes the coordinate singularity by construction: the parity-preserving variable $s=r^2$ and a modified stream-function representation eliminate explicit $1/r$ terms from the trained residuals and satisfy mass conservation identically. Boundary-fitted analytical ans\"atze enforce the boundary conditions exactly, while staged momentum-thermal-joint training, residual normalization, P\'eclet- and Richardson-number continuation, and a physics-scaled drift guard stabilize the coupled optimization. The framework is trained without labelled data, using only physics and boundary constraints, and its predictions are validated against finite-element reference solutions for steady nanofluid flow and heat transfer through a stenosed artery over a wide range of flow and thermal regimes, without case-specific tuning. Mass conservation remains at single-precision round-off, and over the central operating range the axial velocity and temperature are reproduced within $1.6\%$ and $9.1\%$ relative $L_2$ error, respectively. The main deterioration occurs only in the limiting regimes of strongly buoyancy-influenced slow flow and highly advection-dominated heat transfer. Embedding cylindrical geometry, conservation structure, and boundary physics directly into the network provides a robust route to PINN simulation of axisymmetric multiphysics flows.
\end{abstract}

\end{frontmatter}

\section{Introduction}
\label{sec:introduction}

Axisymmetric flows through constricted channels arise in a wide range of biomedical and industrial systems, including stenosed arteries, nozzles, orifices, and pipe fittings. In the cardiovascular context, arterial stenosis is a major manifestation of atherosclerotic cardiovascular disease \cite{channon2006endothelium,bentzon2014plaque}, and the associated alterations in wall shear stress and pressure drop depend strongly on the shape and severity of the constriction \cite{mandal2011stenoses,young1973stenoses}. For idealized stenosed vessels, the constriction can be represented as an axisymmetric perturbation of a cylindrical geometry, making the axisymmetric $(r,z)$ formulation a natural and computationally efficient setting: it reduces the full three-dimensional problem to a two-dimensional meridional plane while retaining the essential flow physics. Nanoparticle-based drug delivery and nanofluid transport have also attracted considerable interest in stenosed arterial configurations \cite{ahmed2016antimicrobials,akbar2016nanoparticles,nadeem2016drug}; however, their numerical investigation has so far relied predominantly on mesh-based finite-element and finite-volume solvers \cite{abdelsalam2025synergistic,zuberi2025copper,turabi2026inflamed}, for which changes in geometry or configuration generally require repeated mesh generation or adaptation \cite{sun2020surrogate,jin2021nsfnets}.

Physics-informed neural networks (PINNs) have emerged as a mesh-free alternative for solving partial differential equations by embedding the governing equations and boundary conditions directly into the training loss and evaluating the associated residuals at arbitrary collocation points through automatic differentiation, thereby avoiding the need for labelled simulation data \cite{raissi2019physics,cai2021pinnreview}. Boundary constraints can also be enforced exactly by construction through problem-specific ans\"atze \cite{berg2018unified,sun2020surrogate}. PINNs have consequently been applied to a broad range of fluid and thermal problems, including nanofluid and two-phase heat transfer \cite{kim2026casson,jalili2024twophase}, natural convection in cavities \cite{hashemi2024cavity,fowler2024cavity}, heat transfer with partially unknown boundary data \cite{cai2021heattransfer}, coupled flow and heat transfer in wavy channels \cite{zhang2022pinnffht,roy2025wavy}, and reconstruction of flow and temperature fields \cite{zhang2025ces,yang2025reconstruction}. In haemodynamics, PINNs have been used to reconstruct near-wall flow and wall shear stress from sparse data \cite{arzani2021nearwall,wang2025gan}, predict pressure drops across stenoses \cite{gamilov2026pressure}, and model coronary fluid--structure interaction \cite{alzhanov2024fsi,aghaee2025aortic,akinwamide2026review}. However, none of these studies addresses the fully coupled Navier-Stokes and energy equations in axisymmetric cylindrical coordinates for a constricted domain, for which FEM/FVM solvers still remain the prevailing computational approach \cite{nadeem2016drug,zuberi2025copper,turabi2026inflamed}.

A standard PINN faces five major difficulties for the coupled axisymmetric problem considered here. \emph{Axis singularity:} in cylindrical coordinates, the governing equations contain $1/r$ and $1/r^{2}$ terms that become singular at the axis $r=0$, making residual evaluation near the axis ill-conditioned. Moreover, a standard multilayer perceptron defined directly on $(r,z)$ does not inherently respect the required radial parity: the radial velocity is odd in $r$, whereas the axial velocity, pressure, and temperature are even in $r$, so the network must effectively learn cancellations among singular terms. This singular behaviour is intrinsic to axisymmetric stream-function formulations \cite{nowakowski2023weighted} and also persists in polar-coordinate PINNs \cite{zhou2025polar}; existing workaround formulations avoid the difficulty only by leaving cylindrical coordinates \cite{fowler2024cavity}. \emph{Unbalanced gradients:} deep PINNs are susceptible to spectral bias and to strongly unbalanced, vanishing, or exploding residual gradients, which can slow or stall convergence; self-adaptive loss balancing can alleviate this imbalance but does not remove singular or intrinsically stiff terms \cite{xiang2021selfadaptive}. \emph{Multiphysics coupling:} the energy equation advects temperature through the unknown velocity field, while buoyancy feeds the temperature field back into the momentum equations. The resulting residuals can differ substantially in magnitude and stiffness, so fully joint training may allow one field to destabilize the learning of the other, as reported in coupled flow and heat-transfer problems \cite{hashemi2024cavity,jalili2024twophase,roy2025wavy}. \emph{Physical admissibility:} unconstrained neural networks do not automatically guarantee physical properties such as mass conservation or bounded thermal fields; conservation-embedded architectures can therefore improve the physical reliability of the solution \cite{jnini2025riemann}. \emph{Cost in advection-dominated regimes:} PINN training may require a large number of iterations, particularly when the residual contains weak diffusive information; convergence can depend strongly on learning-rate scheduling and second-order refinement, such as L-BFGS, while residual formulations involving lower-order derivatives can reduce the associated training cost \cite{su_fvpin}.

Recent studies have introduced several architectural and training strategies to address some of the difficulties outlined above. Hard constraints imposed through problem-specific ans\"atze allow boundary and interface conditions to be satisfied by construction rather than through penalty terms \cite{berg2018unified,sun2020surrogate}; in particular, physics-constrained, boundary-encoded networks have been shown to outperform soft-constrained formulations for internal vascular flows \cite{sun2020surrogate}. Complementary soft- and hard-constrained strategies have also been developed for Navier--Stokes problems with complex boundary conditions \cite{zhou2024softhard}, while architectures based on divergence-free symmetric tensors can enforce conservation of mass and momentum to machine precision \cite{jnini2025riemann}. Alternative PDE formulations have likewise been exploited to embed incompressibility structurally, as in the velocity-pressure and velocity-vorticity NSFnets \cite{jin2021nsfnets}. Beyond architectural constraints, self-adaptive loss weighting has been used to improve the balance among competing residuals \cite{xiang2021selfadaptive}, domain decomposition has been employed to distribute training across subdomains of complex flows \cite{gu2024domain}, parametric PINNs treating flow parameters as additional inputs have enabled rapid prediction at unseen parameter values \cite{naderibeni2024parametric}, and finite-volume-type residual formulations with reduced derivative order have been proposed to improve accuracy and training efficiency \cite{su_fvpin}. Nevertheless, these developments address different subsets of the difficulties encountered in the present problem. Table~\ref{tab:pinn_mapping} summarizes representative PINN studies in terms of the coordinate system, governing physics, and the limitation most relevant to the present work.
\begin{table}[htbp]
\centering
\setlength{\tabcolsep}{10pt} 
\renewcommand{\arraystretch}{1.5}
\begin{tabularx}{\textwidth}{@{}>{\raggedright\arraybackslash}p{3.4cm}>{\raggedright\arraybackslash}p{2.9cm}>{\raggedright\arraybackslash}p{4.2cm}X@{}}
\toprule
\textbf{Method} & \textbf{Coordinate system} & \textbf{Physics} & \textbf{Limitation} (relevant to present study) \\
\midrule
NSFnets \cite{jin2021nsfnets} & Cartesian 2-D/3-D & Incompressible NS (velocity-pressure-vorticity) & No cylindrical axis; no thermal coupling. \\
Domain-decomposition PINN \cite{gu2024domain} & Cartesian 2-D benchmarks & Incompressible NS & Eases training on large domains only; no energy equation. \\
Complementary soft/hard PINN \cite{zhou2024softhard} & Cartesian 2-D (blocked cavity) & NS with complex boundary conditions & 2-D Cartesian; no cylindrical axis or thermal coupling. \\
FV-PINN \cite{su_fvpin} & Cartesian 2-D & Steady incompressible NS (integral residual form) & Lowers derivative order but no cylindrical axis or energy equation. \\
Polar-coordinate SWE PINN \cite{zhou2025polar} & Polar 2-D & Shallow-water equations & Origin/boundary discontinuity handled approximately; not NS+energy. \\
Coupled NS+energy PINNs \cite{roy2025wavy,hashemi2024cavity,fowler2024cavity} & Cartesian 2-D & NS + energy (wavy channels, cavities) & No constricted axisymmetric geometry or axis singularity. \\
\bottomrule
\end{tabularx}
\caption{Representative PINN studies for fluid flow and heat transfer, classified according to the coordinate system, governing physics, and the limitation relevant to the present axisymmetric multiphysics problem.}
\label{tab:pinn_mapping}
\end{table}

The studies summarized in Table~\ref{tab:pinn_mapping} reveal a clear gap. To the best of our knowledge, no PINN framework has yet addressed the coupled steady Navier--Stokes and energy equations in axisymmetric cylindrical coordinates $(r,z)$ for a constricted or stenosed domain. Existing PINN studies of arterial flows are largely Cartesian and often data-driven \cite{arzani2021nearwall,wang2025gan,gamilov2026pressure,alzhanov2024fsi,aghaee2025aortic}, whereas axisymmetric stenosis studies involving nanofluids and heat transfer still rely on conventional FEM/FVM solvers \cite{nadeem2016drug,akbar2016nanoparticles,zuberi2025copper,turabi2026inflamed,abdelsalam2025synergistic}; recent reviews likewise report no cylindrical-axisymmetric PINN for coupled Navier--Stokes and energy transport \cite{cai2021pinnreview,akinwamide2026review}. This motivates three questions: (Q1) can the axis singularity be removed exactly through a suitable coordinate transformation and constrained network representation? (Q2) can the essential symmetry, wall, inlet, and thermal conditions be enforced by construction rather than learned through penalty terms? and (Q3) can the coupled flow-thermal system be trained robustly across strongly varying transport regimes? The present work is developed around these three questions.

\subsection*{Main Contributions}
To address these challenges, we develop SAGE-PINN, a singularity-free, axisymmetric, geometry-encoded physics-informed neural network for coupled flow and heat transfer in constricted cylindrical domains. The framework combines a coordinate transformation tailored to axisymmetric geometry, physics-informed hard constraints, and a staged training strategy for coupled momentum and thermal transport. Its performance is assessed systematically for stenosed-artery nanofluid flow over a wide range of flow and thermal regimes using a single network configuration. The main contributions of the present study are summarized as follows.
\begin{itemize}
    \item \textbf{A PINN for the coupled Navier--Stokes and energy equations in axisymmetric cylindrical coordinates.} SAGE-PINN, a singularity-free, axisymmetric, geometry-encoded physics-informed neural network, solves the coupled continuity, momentum, and energy equations for nanofluid flow through a stenosed artery; to the best of our knowledge, no PINN has previously been developed for this class of problems.
    \item \textbf{Parity-preserving coordinate and modified stream function.} Replacing the radial coordinate by $s=r^{2}$ and writing the Stokes stream function as $\psi=r^{2}\varphi(s,z)$, with $\varphi$ predicted by the network, preserves the required radial parity, enforces continuity identically for any network output, and removes the $1/r$ and $1/r^{2}$ terms from the trained residuals.
    \item \textbf{Hard flow and thermal constraints on a curved boundary.} A boundary-fitted coordinate and gated analytical ans\"atze satisfy the axis-symmetry, no-slip, parabolic-inlet, flux, wall-heating, inlet-temperature, and inlet zero-flux conditions exactly, thereby eliminating the corresponding penalty terms from the training objective.
    \item \textbf{Two-way continuation.} P\'eclet continuation stabilizes training of the energy equation in the advection-dominated regime, while Richardson continuation stabilizes the momentum stage in the buoyancy-dominated regime, where the body force can exceed the viscous force by up to $294$ times; checkpoints are selected only at the target $Ri$ and $Pe$.
    \item \textbf{Staged training with physics-scaled stabilization.} Two independent network trunks are trained through a momentum--thermal--joint schedule with per-term residual normalization, detached coupling gradients, an objective-gated physics-scaled drift guard, and final L-BFGS refinement; all case-dependent scales are determined from $(Re,Ri,Pe)$ alone.
    \item \textbf{Systematic validation across flow regimes.} Forty cases: pure blood and $3\%$ Ag, Cu, Au, and Al$_2$O$_3$ nanofluids at eight inlet velocities, spanning $0.46 \le Re \le 128$, $10.5 \le Pe \le 2330$, and $0.2 \le Ri \le 2.8\times10^{4}$, are solved using a single configuration and compared with COMSOL 6.4 finite-element solutions, both field-wise and through wall- and throat-based quantities, together with an analysis of where and why the method loses accuracy.
\end{itemize}

The remainder of this article is organized as follows. Section~\ref{sec:governing_equations} presents the mathematical modeling, governing equations, nanofluid mixture rules, non-dimensional formulation, and the singularity-free axisymmetric representation. Section~\ref{sec:methodology} describes the SAGE-PINN framework, including the network architecture, hard constraints, residual normalization, continuation strategies, and staged training procedure. Section~\ref{sec:numerical_experiment} outlines the geometry, collocation strategy, test cases, error metrics, and implementation details. Section~\ref{sec:results_discussion} presents the comparison with the reference finite-element solutions, together with the predicted fields, wall and throat quantities, training behaviour, mass conservation, and ablation studies. Section~\ref{sec:limitations_future} discusses the limitations and future directions, and Section~\ref{sec:conclusion} summarizes the main conclusions.

\section{Governing Equations}
\label{sec:governing_equations}

This section presents the nanofluid constitutive relations, governing flow and energy equations, associated boundary conditions, and their non-dimensional and singularity-free axisymmetric formulations.

The working fluids are pure blood (base fluid, subscript $_{bf}$) and four nanofluids (subscript $_{nf}$) obtained by dispersing copper (Cu), silver (Ag), gold (Au), or aluminum oxide (Al$_2$O$_3$) nanoparticles (subscript $_s$) in blood at a fixed volume fraction $\Phi = 0.03$; pure blood corresponds to $\Phi = 0$. The addition of nanoparticles modifies the effective density $\rho_{nf}$, dynamic viscosity $\mu_{nf}$, thermal conductivity $K_{nf}$, volumetric heat capacity $(\rho c_p)_{nf}$, and thermal-expansion coefficient $(\rho\alpha)_{nf}$ of the fluid. Following the standard nanofluid mixture rules \cite{nadeem2016drug,ahmed2016antimicrobials,akbar2016nanoparticles}, with the Brinkman viscosity model and Maxwell-Garnett thermal-conductivity model, the effective properties are given by
\begin{align}
\mu_{nf} &= \frac{\mu_{bf}}{(1-\Phi)^{2.5}}, \label{eq:mu_nf}\\[2pt]
\rho_{nf} &= (1-\Phi)\rho_{bf} + \Phi\rho_s, \label{eq:rho_nf}\\[2pt]
(\rho\alpha)_{nf} &= (1-\Phi)(\rho\alpha)_{bf} + \Phi(\rho\alpha)_s, \label{eq:rhoalpha_nf}\\[2pt]
(\rho c_p)_{nf} &= (1-\Phi)(\rho c_p)_{bf} + \Phi(\rho c_p)_s, \label{eq:rhocp_nf}\\[2pt]
\frac{K_{nf}}{K_{bf}} &= \frac{K_s + 2K_{bf} - 2\Phi(K_{bf} - K_s)}{K_s + 2K_{bf} + \Phi(K_{bf} - K_s)}. \label{eq:K_nf}
\end{align}
All nanofluid properties are treated as constants evaluated at the reference temperature $T_0$. Supplementary Table~S3 lists the base-fluid and particle properties used here (blood $\mu_{bf} = 3.5\times10^{-3}$~Pa$\cdot$s).

\begin{figure}[htbp]
\centering
\includegraphics[height=6cm]{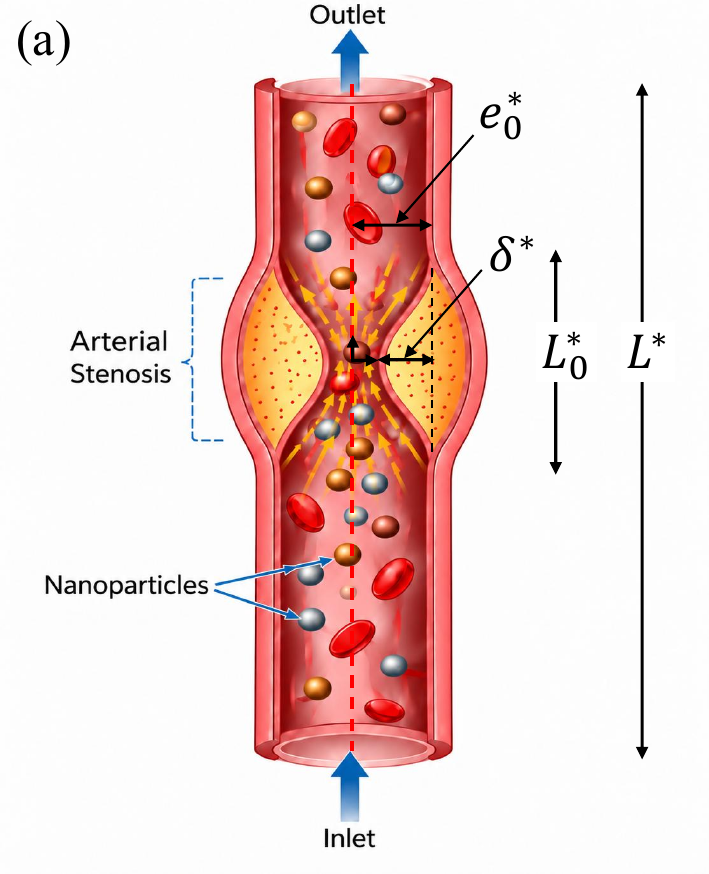}\hspace{0.5cm}
\includegraphics[height=6cm]{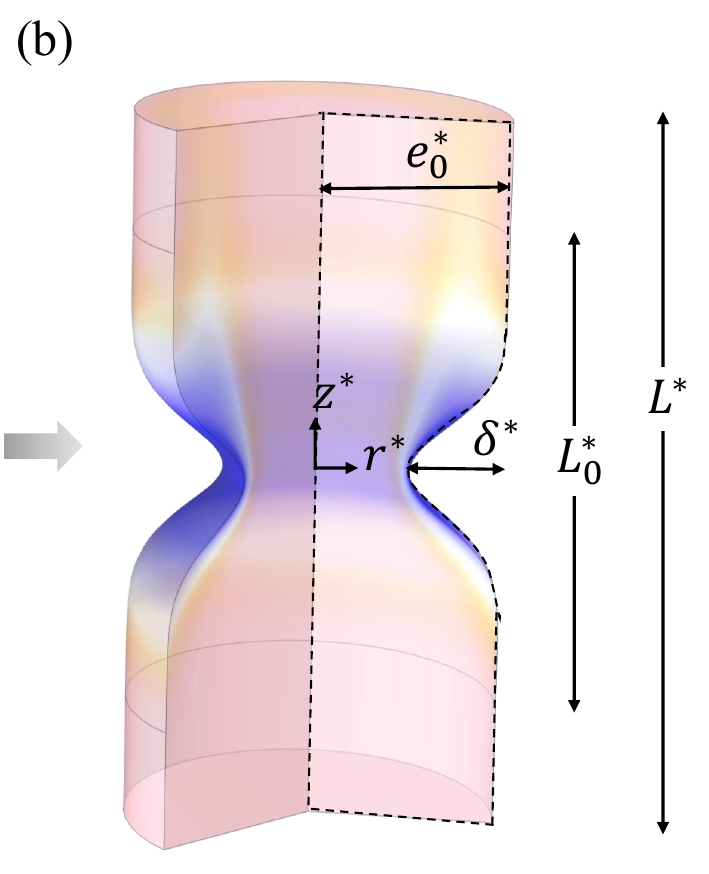}\hspace{0.5cm}
\includegraphics[height=6cm]{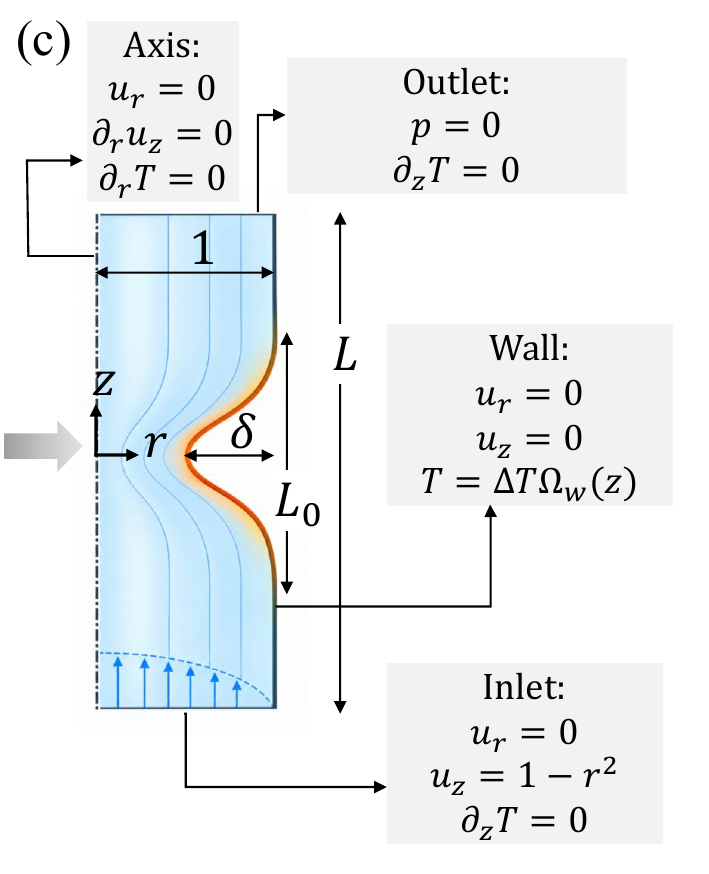}
\caption{Schematic representation of the stenosed-artery problem: (a) physical configuration of a stenosed artery carrying a blood-based nanofluid, showing the inlet, outlet, stenosed region, and characteristic geometric parameters; (b) corresponding axisymmetric mathematical representation of the physical domain in cylindrical coordinates $(r^*,z^*)$, where the computation domain is highlighted by the black dashed boundary; and (c) non-dimensional computational domain, $0\le r\le h(z)$ and $|z|\le L/2$, together with the prescribed non-dimensional boundary conditions at the axis, inlet, outlet, and stenosed wall.}
\label{fig:geometry}
\end{figure}

We consider steady, laminar, incompressible, axisymmetric flow of a homogeneous Newtonian nanofluid through a cylindrical artery of healthy radius $e_0^*$ containing a smooth bell-shaped stenosis, as shown schematically in Fig.~\ref{fig:geometry}(a). The problem is formulated in cylindrical coordinates $(r^*,z^*)$, where the $z^*$-axis coincides with the vessel axis and $r^*$ denotes the radial distance from it. The stenosed wall is described by the smooth profile \cite{nadeem2016drug}
\begin{equation}
h(z^*) =
\begin{cases}
e_0^* - \delta^*\exp\!\left(-\omega^2 4 z^{*2}/(L_{0}^{*})^2\right), & |z^*| \le L_{0}^{*}/2, \\[2pt]
e_0^*, & \text{otherwise},
\end{cases}
\label{eq:wall_profile}
\end{equation}
where $\delta^*=0.0075$~m is the maximum stenosis height, $e_0^*=0.015$~m is the healthy artery radius, $L_{0}^{*}=0.03$~m is the axial length of the stenosis, $L^{*}=0.045$~m is the total computational length of the artery, and $\omega=3.5$ is the dimensionless shape parameter. The resulting $50\%$ reduction of the healthy radius at the stenosis throat lies within the range commonly considered for arterial stenosis \cite{young1973stenoses,mandal2011stenoses}. Owing to rotational symmetry, the three-dimensional configuration is reduced to the two-dimensional meridional domain $(r^*,z^*)$, highlighted in Fig.~\ref{fig:geometry}(b). The origin of the coordinate system is placed at the midpoint of the axis of symmetry, so that the inlet and outlet are located at $z^*=-L^*/2$ and $z^*=L^*/2$, respectively. The geometric parameters are summarized in Supplementary Table~S4.

The dimensional formulation comprises the continuity equation together with the radial and axial momentum equations and the energy equation governing heat transport in the nanofluid, given by \cite{nadeem2016drug,ahmed2016antimicrobials}
\begin{align}
\frac{\partial u_r^*}{\partial r^*} + \frac{u_r^*}{r^*} + \frac{\partial u_z^*}{\partial z^*} &= 0, \label{eq:cont_dim}\\[2pt]
\rho_{nf}\!\left(u_r^*\frac{\partial u_r^*}{\partial r^*} + u_z^*\frac{\partial u_r^*}{\partial z^*}\right)
&= -\frac{\partial p^*}{\partial r^*}
+ \frac{1}{r^*}\frac{\partial}{\partial r^*}\!\left(2r^*\mu_{nf}\frac{\partial u_r^*}{\partial r^*}\right)
+ \frac{\partial}{\partial z^*}\!\left[\mu_{nf}\!\left(\frac{\partial u_r^*}{\partial z^*} + \frac{\partial u_z^*}{\partial r^*}\right)\right]
- 2\mu_{nf}\frac{u_r^*}{r^{*2}}, \label{eq:rmom_dim}\\[2pt]
\rho_{nf}\!\left(u_r^*\frac{\partial u_z^*}{\partial r^*} + u_z^*\frac{\partial u_z^*}{\partial z^*}\right)
&= -\frac{\partial p^*}{\partial z^*}
+ \frac{1}{r^*}\frac{\partial}{\partial r^*}\!\left[r^*\mu_{nf}\!\left(\frac{\partial u_r^*}{\partial z^*} + \frac{\partial u_z^*}{\partial r^*}\right)\right]
+ \frac{\partial}{\partial z^*}\!\left(2\mu_{nf}\frac{\partial u_z^*}{\partial z^*}\right)
+ g(\rho\alpha)_{nf}(T^* - T_0), \label{eq:zmom_dim}\\[2pt]
(\rho c_p)_{nf}\!\left(u_r^*\frac{\partial T^*}{\partial r^*} + u_z^*\frac{\partial T^*}{\partial z^*}\right)
&= K_{nf}\!\left(\frac{\partial^2 T^*}{\partial r^{*2}} + \frac{1}{r^*}\frac{\partial T^*}{\partial r^*} + \frac{\partial^2 T^*}{\partial z^{*2}}\right). \label{eq:energy_dim}
\end{align}
Here, $u_r^*$ and $u_z^*$ denote the radial and axial velocity components, respectively, $p^*$ is the pressure, $T^*$ is the temperature, and $g$ is the gravitational acceleration. The momentum equations are written in stress-divergence form, with buoyancy incorporated through the Boussinesq term $g(\rho\alpha)_{nf}(T^*-T_0)$ in the axial momentum equation. With the positive $z$-direction taken upward, this term represents buoyancy-driven acceleration of the heated fluid along $+z$. In the energy equation \eqref{eq:energy_dim}, viscous dissipation, thermal radiation, and nanoparticle transport associated with Brownian motion and thermophoresis are neglected for the present dilute suspension with $\Phi=3\%$.

The governing system is completed by prescribing the corresponding boundary conditions on the wall, axis, inlet, and outlet. At the stenosed wall, $r^*=h(z^*)$, the no-slip condition $u_r^*=u_z^*=0$ is imposed together with a localized Gaussian heating profile centered at the stenosis throat,
\begin{equation}
T^* =
\begin{cases}
T_0 + (T_S - T_0)\exp\!\left(-\dfrac{z^{*2}}{(\sigma^{*})^2}\right)
\left(1-\dfrac{4z^{*2}}{(L_{0}^{*})^2}\right)^{2},
& |z^*| \le L_{0}^{*}/2,\\[6pt]
T_0, & \text{otherwise},
\end{cases}
\label{eq:wall_T_dim}
\end{equation}
where $T_0\,(=309$~K) and $T_S\,(=314$~K) denote the reference and peak wall temperatures, respectively, and $\sigma^{*}(=0.01$~m) controls the width of the heated region. The polynomial factor has a double root at $|z^*|=L_{0}^{*}/2$, ensuring that the heated and unheated portions of the wall join with continuous temperature and slope. Along the axis of symmetry, $r^*=0$, the conditions $u_r^*=0$ and $\partial u_z^*/\partial r^*=\partial T^*/\partial r^*=0$ are imposed. At the inlet, $z^*=-L^*/2$, a fully developed parabolic profile $u_z^*=u_0^*[1-(r^*/e_0^*)^2]$ is prescribed with $u_r^*=0$ and zero axial heat flux, $\partial T^*/\partial z^*=0$, where $u_0^*$ is the inlet centreline velocity. At the outlet, $z^*=L^*/2$, the pressure is fixed at $p^*=0$ together with the zero axial heat-flux condition $\partial T^*/\partial z^*=0$.

The governing equations are non-dimensionalised for computational convenience and to express the problem in terms of the dominant flow and thermal transport parameters. Using the healthy artery radius $e_0^*$, inlet centreline velocity $u_0^*$, inertial pressure scale $\rho_{nf} (u_0^*)^2$, and reference temperature $T_0$, we define the dimensionless variables as follows
\begin{equation}
r = \frac{r^*}{e_0^*}, \quad z = \frac{z^*}{e_0^*}, \quad u_r = \frac{u_r^*}{u_0^*}, \quad u_z = \frac{u_z^*}{u_0^*}, \quad p = \frac{p^*}{\rho_{nf}(u_0^*)^2}, \quad T = \frac{T^* - T_0}{T_0}. \label{eq:scaling}
\end{equation}
The corresponding dimensionless continuity, momentum conservation, and enery equations are obtained as
\begin{align}
\frac{\partial u_r}{\partial r} + \frac{u_r}{r} + \frac{\partial u_z}{\partial z} &= 0, \label{eq:cont_nd}\\[2pt]
u_r\frac{\partial u_r}{\partial r} + u_z\frac{\partial u_r}{\partial z}
&= -\frac{\partial p}{\partial r}
+ \frac{1}{Re}\!\left[\frac{1}{r}\frac{\partial}{\partial r}\!\left(2r\frac{\partial u_r}{\partial r}\right)
+ \frac{\partial}{\partial z}\!\left(\frac{\partial u_r}{\partial z} + \frac{\partial u_z}{\partial r}\right)
- 2\frac{u_r}{r^2}\right], \label{eq:rmom_nd}\\[2pt]
u_r\frac{\partial u_z}{\partial r} + u_z\frac{\partial u_z}{\partial z}
&= -\frac{\partial p}{\partial z}
+ \frac{1}{Re}\!\left[\frac{1}{r}\frac{\partial}{\partial r}\!\left(r\left(\frac{\partial u_r}{\partial z} + \frac{\partial u_z}{\partial r}\right)\right)
+ \frac{\partial}{\partial z}\!\left(2\frac{\partial u_z}{\partial z}\right)\right]
+ Ri\,T, \label{eq:zmom_nd}\\[2pt]
u_r\frac{\partial T}{\partial r} + u_z\frac{\partial T}{\partial z}
&= \frac{1}{Pe}\!\left(\frac{\partial^2 T}{\partial r^2} + \frac{1}{r}\frac{\partial T}{\partial r} + \frac{\partial^2 T}{\partial z^2}\right), \label{eq:energy_nd}
\end{align}
The dimensionless temperature $T$ varies from $0$ in the unheated fluid to a maximum value
$\Delta T=(T_S-T_0)/T_0=5/309\approx0.0162$ at the peak of the heated wall. The dimensionless momentum equations retain the stress-divergence form of their dimensional counterparts, while the flow and thermal behaviour is governed by the Reynolds, Richardson, and P\'eclet numbers,
\begin{equation}
Re = \frac{\rho_{nf}\,e_0^* u_0^*}{\mu_{nf}}, \qquad
Ri = g\,\frac{(\rho\alpha)_{nf}\,e_0^* T_0}{\rho_{nf}\,(u_0^*)^{2}}, \qquad
Pe = \frac{(\rho c_p)_{nf}\,e_0^* u_0^*}{K_{nf}},
\label{eq:groups}
\end{equation}
respectively. Here, $Re$ represents the ratio of inertial to viscous effects, $Ri$ characterizes the relative importance of buoyancy and inertia, and $Pe$ measures the relative strength of thermal advection and diffusion. The subscript $_{nf}$ denotes the effective nanofluid properties defined earlier; for pure blood ($\Phi=0$), these reduce to the corresponding base-fluid properties. The associated Grashof number, $Gr=Ri\,Re^2$, is independent of the inlet velocity $u_0^*$ and is reported for reference in Supplementary Section~S6. For a fixed geometry and fluid composition, $Re$ and $Pe$ increase linearly with $u_0^*$, whereas $Ri\propto (u_0^*)^{-2}$. Hence, varying the inlet velocity systematically spans the operating range from buoyancy, viscous, and diffusion-dominated conditions at low velocities to inertia- and advection-dominated conditions at higher velocities. The non-dimensionalisation and thermal formulation follow conventions used in nanofluid stenosis and PINN heat-transfer studies \cite{nadeem2016drug,cai2021heattransfer,roy2025wavy}.

In dimensionless form, the computational domain is defined by
$0\le r\le h(z)$ and $z\in[-L/2,L/2]$, with $L/2=1.5$ and
$h(z)=h(z^*)/e_0^*$. The corresponding non-dimensional boundary conditions for the axisymmetric multiphysics problem are prescribed as follows (cf. Fig.~\ref{fig:geometry}(c)). At the stenosed wall, $r=h(z)$, the no-slip condition $u_r=u_z=0$ is imposed together with the prescribed wall-temperature distribution
\begin{equation}
T = \Delta T\,\Omega_w(z), \qquad
\Omega_w(z) =
\begin{cases}
\exp\!\left(-\dfrac{z^2}{\sigma^{2}}\right)
\left(1-\dfrac{z^2}{z_{\mathrm{heat}}^2}\right)^{2},
& |z| \le z_{\mathrm{heat}}, \\[6pt]
0, & \text{otherwise},
\end{cases}
\label{eq:theta_wall_nd}
\end{equation}
where $\sigma=\sigma^*/e_0^*=2/3$ is the dimensionless heating width and
$z_{\mathrm{heat}}=L_0^*/(2e_0^*)=1$ denotes the dimensionless half-length of the heated region. Along the axis of symmetry, $r=0$, regularity requires
\begin{equation}
u_r=0, \qquad
\frac{\partial u_z}{\partial r}=0, \qquad
\frac{\partial T}{\partial r}=0.
\label{eq:axis_nd}
\end{equation}
At the inlet, $z=-L/2$, a dimensionless parabolic velocity profile is prescribed as
$u_z=1-r^2$ with $u_r=0$, together with the zero axial heat-flux condition
$\partial T/\partial z=0$. At the outlet, $z=L/2$, the reference pressure is set to $p=0$, while the thermal field again satisfies the zero axial heat-flux condition $\partial T/\partial z=0$. The non-dimensional governing equations (\ref{eq:cont_nd})-(\ref{eq:energy_nd}), together with the corresponding scaled boundary conditions, are implemented in COMSOL to obtain the finite-element solutions, which are taken as the ``true'' reference solutions throughout the present study.

\subsection{Modified Non-Dimensionalized Equation}
\label{sec:modified_equation}
Non-dimensionalization places the governing variables and residuals on comparable scales, which is beneficial for stable PINN training. However, the dimensionless equations (\ref{eq:cont_nd})--(\ref{eq:energy_nd}) remain challenging for a standard PINN because the cylindrical operators contain $1/r$ and $1/r^{2}$ terms that become singular at the axis $r=0$. To eliminate these coordinate singularities, we introduce the parity-preserving radial coordinate $s=r^{2}$ together with the modified stream-function representation
\begin{equation}
s = r^{2}, \qquad
\psi(r,z) = r^{2}\,\varphi(s,z),
\label{eq:psi_ansatz}
\end{equation}
where $\varphi(s,z)$ is a smooth scalar field. The corresponding velocity components are then obtained from the axisymmetric stream function as
\begin{equation}
u_z = \frac{1}{r}\frac{\partial\psi}{\partial r}
= 2\varphi + 2s\frac{\partial\varphi}{\partial s},
\qquad
u_r = -\frac{1}{r}\frac{\partial\psi}{\partial z}
= -r\frac{\partial\varphi}{\partial z}.
\label{eq:vel_modified}
\end{equation}
Defining
$A=\partial\varphi/\partial z$ and $B=u_z$, and using the chain rule
$\partial/\partial r = 2r\,\partial/\partial s$, Eqs.~(\ref{eq:rmom_nd})--(\ref{eq:energy_nd}) can be rewritten in the $(s,z)$ coordinates in the following singularity-free form:
\begin{align}
A^{2} + 2sAA_{s} - BA_{z}
&= -2p_{s}
+ \frac{1}{Re}\Big[4\big(C_{s} - A_{s}\big) + 2B_{sz} - A_{zz}\Big],
\label{eq:res_r}\\[3pt]
BB_{z} - 2sAB_{s}
&= -p_{z}
+ \frac{1}{Re}\Big[2\big(E + sE_{s}\big) + 2B_{zz}\Big]
+ Ri\, T,
\label{eq:res_z}\\[3pt]
-2sA\,T_{s} + B\,T_{z}
&= \frac{1}{Pe}\Big[4T_{s} + 4s T_{ss} + T_{zz}\Big],
\label{eq:res_theta}
\end{align}
where $C_s=-(3A_{s}+2sA_{ss})$ and $E=2B_{s}-A_{z}$ are intermediate quantities, and subscripts denote the corresponding partial derivatives evaluated through automatic differentiation. Because $u_r$ and $u_z$ are derived from the stream-function representation (\ref{eq:vel_modified}), the continuity equation (\ref{eq:cont_nd}) is satisfied identically for any sufficiently smooth $\varphi$ (proof in~\ref{sec:appendix_stream_transform}) and therefore does not need to be included in the training loss. It is retained only as a diagnostic, for which the converged solutions exhibit residuals at the level of single-precision round-off, $O(10^{-7})$. The transformation $s=r^{2}$ also regularizes the axis naturally. Since $\partial/\partial r = 2r\,\partial/\partial s$, any smooth function of $s$ satisfies $\partial(\cdot)/\partial r=0$ at $r=0$; consequently, the symmetry conditions on $u_z$, $p$, and $T$ are automatically satisfied at the axis. Similarly, the cylindrical operator $f_{rr}+f_r/r$ is transformed into the regular polynomial form $4f_s+4sf_{ss}$, eliminating any explicit division by $r$. The physical boundary conditions and dimensionless parameters remain unchanged under this transformation. Equations~(\ref{eq:res_r})-(\ref{eq:res_theta}) therefore constitute the singularity-free governing equations used for PINN training throughout the remainder of this work. The detailed transformation from Eqs.~(\ref{eq:rmom_nd})-(\ref{eq:energy_nd}) to Eqs.~(\ref{eq:res_r})-(\ref{eq:res_theta}) is provided in ~\ref{sec:modified_eqns_derivations}.

\section{Methodology}
\label{sec:methodology}
A standard PINN approximates the solution fields $\mathcal{F}(r,z)=(u_r,u_z,p,T)$ using a fully connected neural network and determines its parameters by minimizing a weighted sum of the mean-squared residuals of the non-dimensional governing equations (\ref{eq:cont_nd})-(\ref{eq:energy_nd}) and the boundary-condition residuals, without requiring labelled solution data \cite{raissi2019physics,cai2021pinnreview}. For the present axisymmetric multiphysics problem, however, this formulation presents two fundamental difficulties. First, the cylindrical residuals contain $1/r$ and $1/r^2$ terms that become singular at the axis $r=0$, making residual evaluation near the axis ill-conditioned. Second, the momentum and energy residuals can differ substantially in scale and stiffness, leading to an unbalanced optimization problem. SAGE-PINN retains the residual-minimization principle of a conventional PINN while addressing these difficulties through the coordinate transformation, constrained network representations, and training strategy described below. All derivatives required to evaluate the governing residuals are computed using automatic differentiation, as detailed in Section~\ref{sec:autodiff}.

\subsection{Proposed Method: SAGE-PINN}
\label{sec:proposed_method}
\begin{figure}[htbp]
\centering
\includegraphics[width=\textwidth]{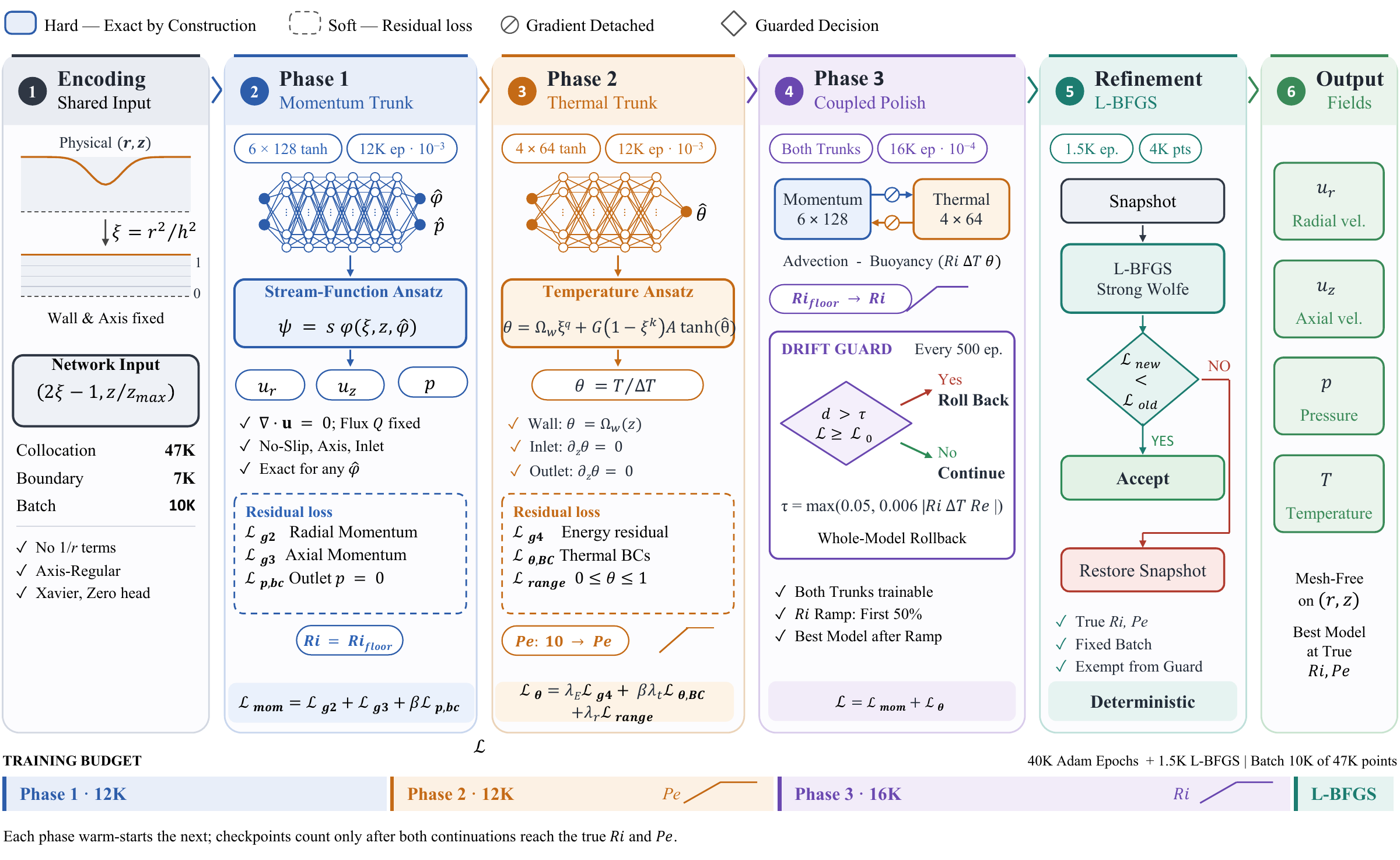}
\caption{Overview of SAGE-PINN: boundary-fitted input encoding, the momentum-pressure and thermal trunks, the hard-constraint output ans\"atze, and the staged training with continuation.}
\label{fig:architecture}
\end{figure}

SAGE-PINN is a singularity-free, axisymmetric, geometry-encoded PINN developed for the coupled flow and thermal problem considered here. Its formulation is built around two principal ideas: (i) a parity-preserving coordinate transformation combined with a modified stream-function representation, which removes the explicit $1/r$ and $1/r^2$ singular terms and satisfies continuity identically; and (ii) boundary-fitted analytic ans\"atze that embed the essential axis, wall, inlet, and thermal constraints directly into the network representation. These constructions substantially reduce the number of conditions that must be imposed through the loss function and allow the network optimization to focus primarily on the transformed momentum and energy residuals. The complete SAGE-PINN architecture and training procedure are described in detail below and summarized schematically in Fig.~\ref{fig:architecture}.

\subsubsection{Model Architecture}
\label{sec:model_architecture}

Instead of a single monolithic network, SAGE-PINN employs two independent multilayer perceptron (MLP) trunks with a common input representation but separate outputs for the flow and thermal fields. The momentum--pressure trunk predicts the raw stream-function correction $\hat{\varphi}$ and raw pressure $\hat{p}$, whereas the smaller thermal trunk predicts the raw thermal variable $\hat{\theta}$. This separation enables the staged training strategy described in Section~\ref{sec:training_strategy}: either trunk can be frozen while the other is optimized, thereby isolating the momentum and thermal updates during the corresponding training stages. Both trunks are fully connected networks of the form
\begin{equation}
\hbar^{(0)} = x, \qquad
\hbar^{(\ell)} =
f\!\left(W^{(\ell)}\hbar^{(\ell-1)} + b^{(\ell)}\right),
\quad \ell=1,\ldots,N,
\qquad
y = W^{(N+1)}\hbar^{(N)} + b^{(N+1)},
\label{eq:mlp}
\end{equation}
where $f\equiv\tanh$ is used in all hidden layers and the output layer is linear. Here, $x$ denotes the normalized network input defined in Eq.~(\ref{eq:input}), $y$ is the corresponding raw trunk output, and $W^{(\ell)}$ and $b^{(\ell)}$ are the trainable weights and biases. The smoothness of the $\tanh$ activation permits the higher-order derivatives required in the governing residuals to be evaluated by automatic differentiation.

\begin{figure}[htbp]
\centering
\includegraphics[width=\textwidth]{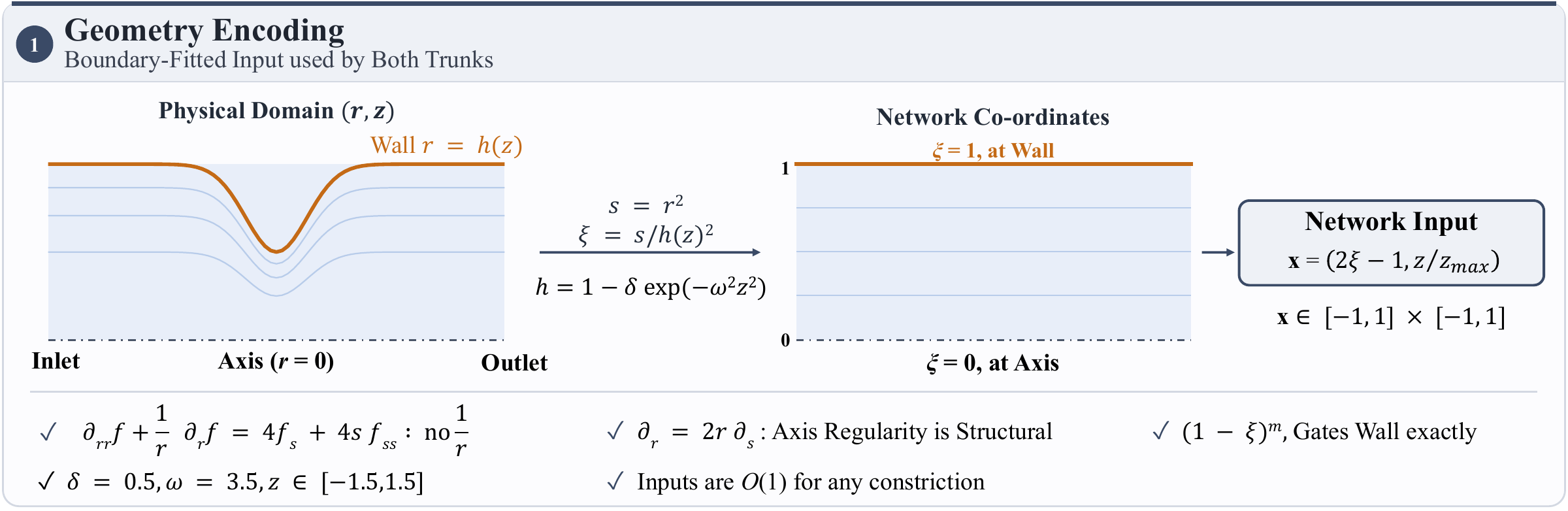}
\caption{Boundary-fitted input encoding. The squared radial coordinate $s=r^2$ is normalized by the local squared wall radius $S(z)=h(z)^2$, such that $\xi=s/S(z)$ maps the axis to $\xi=0$ and the curved wall to $\xi=1$ for every $z$. The resulting network input is $(2\xi-1,\,z/z_{\max})\in[-1,1]^2$.}
\label{fig:geom_encoding}
\end{figure}

\textbf{Input layer:}
The raw dimensionless coordinates are first mapped to a boundary-fitted representation. With $s=r^2$ and the local squared wall radius
\begin{equation}
S(z) = h(z)^2
=
\left[1-\delta\exp\!\left(-(\omega z)^2\right)\right]^2,
\qquad
\xi=\frac{s}{S(z)}\in[0,1],
\label{eq:wall_coord}
\end{equation}
the curved physical domain is transformed so that $\xi=0$ corresponds to the axis and $\xi=1$ to the wall for every axial location. Here, $\delta=0.5$ and $\omega=3.5$ are the dimensionless stenosis height and shape parameter, respectively. The input to both trunks is then normalized as
\begin{equation}
x=\big(2\xi-1,\;z/z_{\max}\big)\in[-1,1]^2,
\label{eq:input}
\end{equation}
with $z_{\max}=L/2=1.5$ in the present study. This mapping keeps both input coordinates on comparable scales and, more importantly, provides a geometry-aware coordinate that is subsequently used to impose the curved-wall constraints analytically through factors of $(1-\xi)^m$, which vanishes at the wall. The boundary-fitted encoding is illustrated in Fig.~\ref{fig:geom_encoding}.

\textbf{Output layer:}
The momentum-pressure trunk returns the raw pair $(\hat{\varphi},\hat{p})$, while the thermal trunk returns the raw scalar $\hat{\theta}$. These quantities are not used directly as the solution fields; instead, they enter the hard-constrained constructions introduced in Sections~\ref{sec:mom_ansatz}--\ref{sec:theta_ansatz}. All hidden-layer weights are initialized using Xavier-normal initialization with zero biases, while the final linear layer of each trunk is initialized to zero. Consequently, the initial network outputs vanish and the reconstructed fields reduce exactly to their prescribed analytic base states. Training therefore begins from the boundary-consistent base flow and thermal field, with the networks learning only the required corrections. The architecture of the two trunks is summarized in Table~\ref{tab:architecture}.

\begin{table}[htbp]
\centering
\setlength{\tabcolsep}{15pt}
\renewcommand{\arraystretch}{1.3}
\begin{tabular}{lccccc}
\toprule
Trunk & Input & Hidden layers & Raw output & Constrained output & Parameters \\
\midrule
Momentum--pressure & 2 & $6 \times 128$ (tanh) & $(\hat{\varphi},\hat{p})$ & $(\varphi,p)$ & 83{,}202 \\
Thermal & 2 & $4 \times 64$ (tanh) & $\hat{\theta}$ & $\theta$ & 12{,}737 \\
\bottomrule
\end{tabular}
\caption{Network architecture used in SAGE-PINN. The momentum--pressure and thermal trunks contain $83{,}202$ and $12{,}737$ trainable parameters, respectively, for a total of $95{,}939$ parameters.}
\label{tab:architecture}
\end{table}

\subsubsection{Parity-Preserving Stream-Function Ansatz}
\label{sec:mom_ansatz}
\begin{figure}[htbp]
\centering
\includegraphics[width=\textwidth]{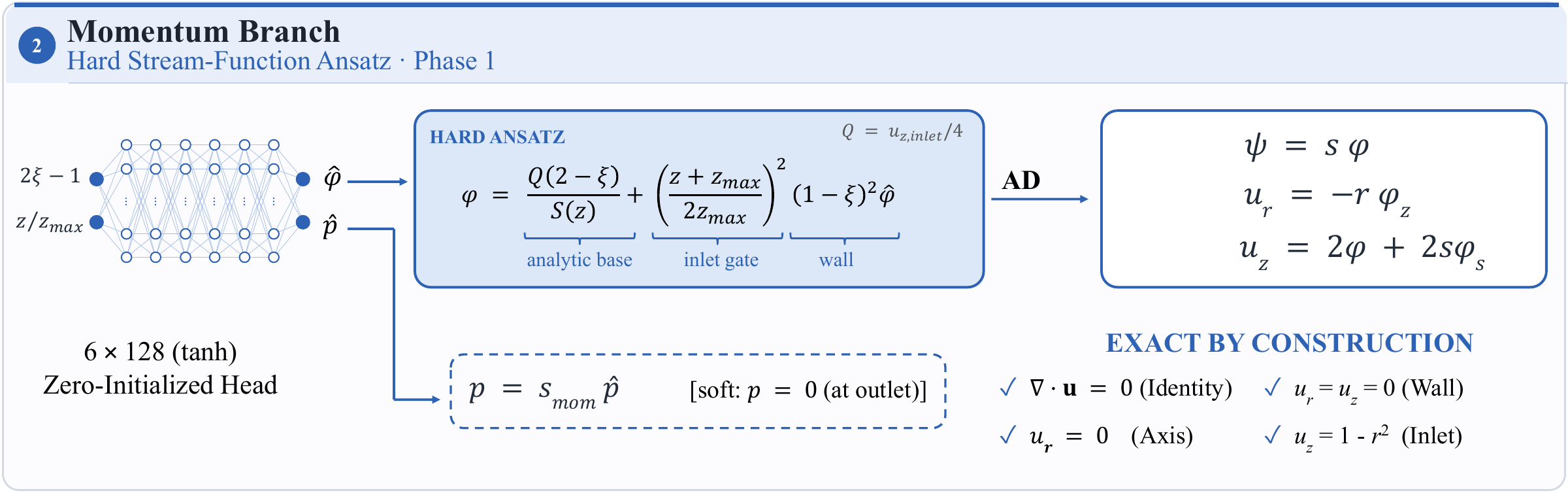}
\caption{Momentum-pressure trunk and flow-boundary ansatz: the network output $n$ enters the modified stream function through an analytic base and an inlet- and wall-gated correction, from which $u_r$ and $u_z$ follow by automatic differentiation; continuity, flux, no-slip, axis, and inlet conditions hold by construction.}
\label{fig:trunks}
\label{fig:trunk_momentum}
\end{figure}

The momentum field is represented through the modified stream-function variable $\varphi(s,z)$, with the Stokes stream function given by $\psi=s\, \varphi$, as introduced in Eq.~\eqref{eq:psi_ansatz}. The velocity components are then recovered analytically from Eq.~\eqref{eq:vel_modified}, as illustrated in Fig.~\ref{fig:trunk_momentum}. This representation provides two important structural properties. First, any sufficiently smooth $\varphi$ generates an axisymmetric divergence-free velocity field, so the continuity equation, Eq.~\eqref{eq:cont_nd}, is satisfied identically and need not be included in the training loss; it is retained only as a diagnostic, with the corresponding proof given in \ref{sec:appendix_stream_transform}. Second, since $u_r$ contains an explicit factor of $r$, the axis condition $u_r|_{r=0}=0$ is satisfied identically. Moreover, all terms in the transformed residuals, Eqs.~\eqref{eq:res_r}-\eqref{eq:res_theta}, remain regular at $s=0$, because the explicit $1/r$ and $1/r^2$ coefficients of the cylindrical formulation are removed by the transformation described in Section~\ref{sec:modified_equation}. Consequently, the network is not required to learn cancellations among individually singular terms, and collocation points can be placed arbitrarily close to the symmetry axis without introducing coordinate singularities.

\subsubsection{Hard-Constraint Momentum (Flow-Boundary) Ansatz}
\label{sec:flow_ansatz}
The raw network output $\hat{\varphi}(\xi,z)$ is incorporated into the modified stream-function variable through
\begin{equation}
\varphi(\xi,z)
=
\underbrace{\frac{Q(2-\xi)}{S(z)}}_{\varphi_{\mathrm{base}}}
+
\underbrace{
\left(\frac{z+z_{\max}}{2z_{\max}}\right)^{2}
(1-\xi)^2
\hat{\varphi}(\xi,z)
}_{\varphi_{\mathrm{NN}}},
\qquad
Q=\frac{u_{z,\mathrm{inlet}}}{4},
\label{eq:shi_ansatz}
\end{equation}
where $\varphi_{\mathrm{base}}$ is an analytic base field and $\varphi_{\mathrm{NN}}$ is the network correction. The correction is modulated by two geometric gates. The axial factor
$G_z(z)=[(z+z_{\max})/(2z_{\max})]^2$ vanishes at the inlet $z=-z_{\max}$ and therefore preserves the prescribed parabolic inflow, whereas the wall factor $G_\xi(\xi)=(1-\xi)^2$ vanishes at $\xi=1$ and enforces the corresponding wall constraint independently of the network output. At the wall, $\xi=1$ and $s=S(z)$, so Eq.~\eqref{eq:shi_ansatz} gives
\begin{equation}
\psi\big|_{\xi=1}
=
s\, \varphi
=
S(z)\frac{Q}{S(z)}
=
Q .
\end{equation}
Hence, the wall is a constant-stream-function surface. Since
\begin{equation}
2\pi\int_{0}^{h(z)} u_z\,r\,dr
=
2\pi\left[\psi(h(z),z)-\psi(0,z)\right]
=
2\pi Q,
\end{equation}
the volumetric flow rate is conserved exactly and is independent of the network correction. For the prescribed dimensionless inlet profile $u_z=1-r^2$, the corresponding reduced flux is $Q=1/4$.

Together with the stream-function representation of Section~\ref{sec:mom_ansatz}, this ansatz satisfies the axis condition, wall no-slip condition, prescribed parabolic inlet profile, and axial flux constraint exactly for any $\hat{\varphi}$. The algebraic verification is provided in \ref{sec:appendix_hard_bc}. The associated velocity boundary residuals are therefore used only as diagnostic quantities and are not included in the training objective. In the converged solutions, the wall stream-function drift remains $O(10^{-8})$, while direct radial quadrature recovers the prescribed flux to within $O(10^{-5})$ at every axial station. Pressure is not constrained by the stream-function ansatz and remains determined only up to an additive constant. This null degree of freedom is removed through the softly imposed outlet reference condition $p=0$, as described in Section~\ref{sec:loss_balancing}.

\subsubsection{Hard-Constraint Thermal Ansatz}
\label{sec:theta_ansatz}
The thermal variable in the governing equations is the dimensionless temperature
$T=(T^*-T_0)/T_0$ defined in Eq.~\eqref{eq:scaling}. Its maximum prescribed wall value is $\Delta T=(T_S-T_0)/T_0\approx 0.0162.$ To keep the thermal network output and the associated loss terms at $O(1)$, we introduce the normalized temperature
\begin{equation}
\theta=\frac{T}{\Delta T},
\label{eq:theta_hat}
\end{equation}
for which the physically admissible temperature range is $0\leq\theta\leq1$. This is a fixed rescaling of the thermal field rather than an additional physical variable. Since the energy equation is linear in $T$, Eq.~\eqref{eq:res_theta} retains the same form in terms of $\theta$, while the buoyancy contribution in Eq.~\eqref{eq:res_z} becomes $Ri\,\Delta T\,\theta$.

\begin{figure}[htbp]
\centering
\includegraphics[width=\textwidth]{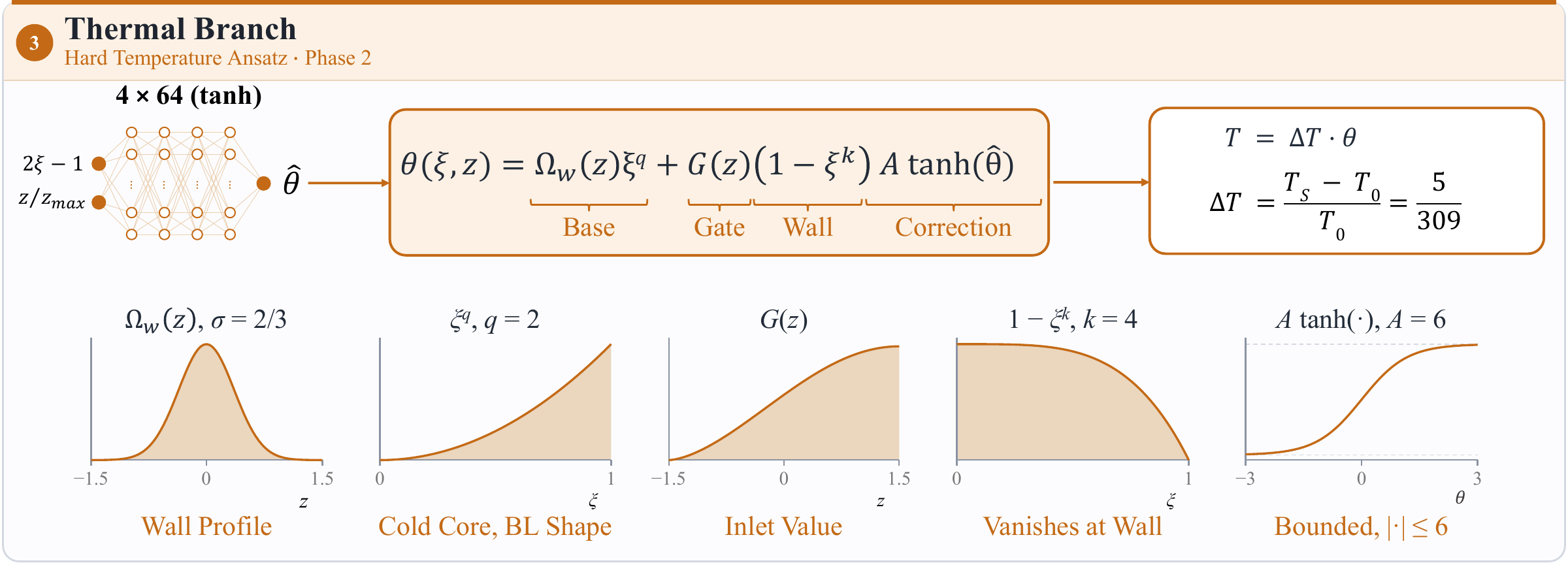}
\caption{Thermal trunk and hard-constraint temperature ansatz. The analytic wall-heating profile $\Omega_w(z)$ is carried by the radial base $\xi^q$, while the bounded network correction is modulated by the inlet gate $G(z)$ and the wall factor $1-\xi^k$. The constituent functions $\Omega_w(z)$, $\xi^q$, $G(z)$, $1-\xi^k$, and $A\tanh(\cdot)$ are shown separately.}
\label{fig:trunk_thermal}
\end{figure}

The raw thermal-network output $\hat{\theta}(\xi,z)$, shown schematically in Fig.~\ref{fig:trunk_thermal}, is mapped to the constrained temperature field through
\begin{equation}
\theta(\xi,z)
=
\underbrace{\Omega_w(z)\,\xi^{q}}_{\text{base}}
+
\underbrace{G(z)}_{\text{inlet gate}}
\underbrace{\big(1-\xi^{k}\big)}_{\text{wall factor}}
\underbrace{A\,\tanh\!\big(\hat{\theta}(\xi,z)\big)}_{\text{network correction}},
\label{eq:theta_ansatz}
\end{equation}
where $\Omega_w(z)$ is the normalized wall-heating profile defined in Eq.~\eqref{eq:theta_wall_nd}, $q=2$ is the radial-base exponent, $k=4$ is the wall-factor exponent, and $A=6$ controls the maximum magnitude of the network correction. The inlet gate is defined as
\begin{equation}
G(z)
=
1-\exp\!\left[
-\left(\frac{z+z_{\max}}{2z_{\max}}\right)^2
\right].
\label{eq:theta_gate}
\end{equation}
The individual factors in Eq.~\eqref{eq:theta_ansatz} serve the following purposes:
\begin{itemize}
\item \textbf{Wall constraint:}
At $\xi=1$, the correction factor $(1-\xi^k)$ vanishes while $\xi^q=1$, giving $\theta\big|_{\xi=1}=\Omega_w(z)$ for any value of the network output. The prescribed wall-temperature condition is therefore satisfied exactly. Because $\Omega_w(z)$ has a double root at $|z|=z_{\mathrm{heat}}$, the heated and unheated portions of the wall join with continuous temperature and slope.
\item \textbf{Inlet gate:}
At $z=-z_{\max}$, both $G(z)$ and $G'(z)$ vanish. Since $\Omega_w(-z_{\max})=0$, the ansatz yields $\theta=0,~~\partial\theta/\partial z=0$ at the inlet. Thus, the prescribed zero axial heat-flux condition is satisfied exactly, while the inlet temperature level is additionally anchored to the reference state. This anchoring removes the constant-temperature freedom associated with a Neumann condition alone; without the inlet gate, the unconstrained inlet level was observed to reach
$\langle |\theta| \rangle \approx 0.8$, compared with the reference value of $O(10^{-13})$.
\item \textbf{Radial base:}
The factor $\xi^q$ provides the analytic base field with a radially varying thermal profile, with $\xi^q=0$ on the axis and $\xi^q=1$ at the wall. This avoids initializing the entire cross-section at the prescribed wall temperature. Near the axis, $\xi\propto s$, so a contribution proportional to $\xi^q$ behaves as $s^q$. Since the transformed diffusion operator contains
$4\theta_s+4s\theta_{ss}$, both contributions scale as $s^{q-1}$ near $s=0$. Hence, $q\geq1$ ensures regularity, and $q=2$ is adopted here.
\item \textbf{Correction amplitude:}
Because the hyperbolic tangent is bounded, $A$ limits the magnitude of the network correction. The best attainable area-weighted relative $L_2$ errors of $\theta$, evaluated in the worst case over the eight inlet velocities, are
$2.8\times10^{-1}$,
$1.2\times10^{-1}$,
$5.1\times10^{-2}$,
$1.8\times10^{-2}$, and
$1.3\times10^{-4}$
for $A=1,2,3,4,$ and $6$, respectively. The required correction increases with $Pe$ as the thermal boundary layer becomes thinner. We therefore use $A=6$, for which the representable correction is sufficient throughout the parameter sweep while the bounded $\tanh$ activation suppresses the unphysical temperature excursions, reaching up to approximately $200$ times the physical temperature scale in unconstrained variants.
\end{itemize}
The admissible range $0\leq\theta\leq1$ is further promoted through the range penalty introduced in Section~\ref{sec:loss_balancing}.

\subsubsection{Residual Normalisation}
\label{sec:residual_norm}
The momentum and energy residuals can differ substantially in magnitude, leading to strongly unbalanced gradient contributions during training. In the absence of normalization, approximately $99.99\%$ of the total governing-equation residual loss is contributed by the axial-momentum equation, while the radial-momentum and energy residuals contribute negligibly. To reduce this imbalance, the momentum and energy residuals are scaled using
\begin{equation}
s_{\mathrm{mom}}
=
\max\!\left(
1,\,
\frac{1}{Re},\,
|Ri\,\Delta T|
\right),
\qquad
s_{\mathrm{energy}}
=
\max\!\left(
1,\,
\frac{1}{Pe}
\right),
\label{eq:res_scales}
\end{equation}
where the three terms entering $s_{\mathrm{mom}}$ represent the characteristic coefficients associated with inertia, viscous diffusion, and buoyancy in the axial-momentum balance. The same momentum scale is applied to both radial- and axial-momentum residuals to maintain a consistent normalization of the coupled momentum system.

The pressure field is scaled consistently as
\begin{equation}
p=p_{\mathrm{scale}}\hat{p},
\qquad
p_{\mathrm{scale}}=s_{\mathrm{mom}},
\end{equation}
so that the raw network output $\hat{p}$ remains of $O(1)$ over the full parameter range. Using $1/Re$ alone as the pressure scale would require $\hat{p}$ to reach approximately $230$ times its natural scale at $u_0^*=10^{-4}$~m/s, where buoyancy is dominant, and approximately $100$ times at $u_0^*=2\times10^{-2}$~m/s, where inertial effects dominate. The outlet pressure residual is normalized by the same $p_{\mathrm{scale}}$, so this rescaling does not alter its relative contribution to the loss.

During the P\'eclet and Richardson-number continuation procedures described in Sections~\ref{sec:pe_continuation} and \ref{sec:ri_continuation}, the normalization factors are evaluated using the instantaneous effective values of $Pe$ and $Ri$ rather than their final target values. The residual scaling therefore evolves consistently with the continuation path and avoids artificial changes in the loss magnitude arising solely from the parameter ramp.

\subsubsection{Gradient-Path Control}
\label{sec:grad_path}
The momentum and thermal fields remain physically coupled through buoyancy and advection: the temperature enters the axial-momentum residual through the buoyancy term, while the velocity field enters the energy residual through advection. In SAGE-PINN, these couplings are retained in the forward evaluation of the governing residuals, but the corresponding cross-trunk gradient paths are detached during back-propagation,
\begin{equation}
\frac{\partial \mathcal{R}_z}
{\partial(\text{thermal parameters})}
=0,
\qquad
\frac{\partial \mathcal{R}_\theta}
{\partial(\text{momentum parameters})}
=0.
\label{eq:detach}
\end{equation}
Thus, the axial-momentum residual updates only the momentum--pressure trunk, whereas the energy residual updates only the thermal trunk. The physical coupling is therefore preserved in the residual values, while gradient transfer between the two parameter blocks is suppressed. In the joint training stage, this yields a block-decoupled optimization in which each trunk responds to its own governing residual evaluated with the current state of the other field. This separation reduces cross-contamination of the optimization gradients when the momentum and thermal residuals have markedly different stiffness or convergence rates.

\subsubsection{P\'eclet Continuation}
\label{sec:pe_continuation}
\begin{figure}[htbp]
\centering
\includegraphics[width=0.9\textwidth]{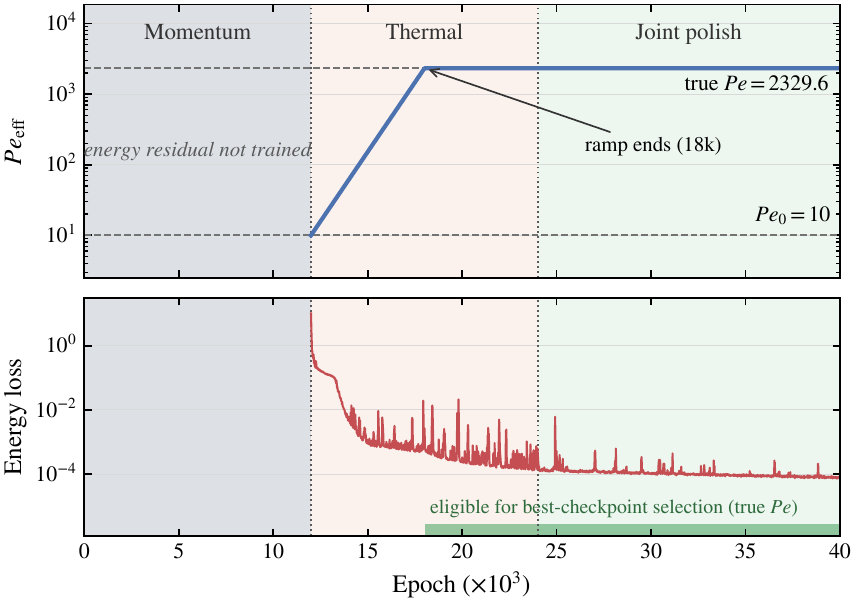}
\caption{P\'eclet continuation for pure blood at $u_0^*=2\times10^{-2}$~m/s ($Pe=2329.6$). Top: $Pe_{\mathrm{eff}}$ is ramped geometrically from $Pe_0=10$ to the true value over the first half of the Thermal Stage. Bottom: Energy-Residual Loss $\mathcal{L}_{g,\theta}$ wrt Epochs.}
\label{fig:pe_ramp}
\end{figure}

At the upper end of the parameter range, the energy equation becomes strongly advection dominated, with $Pe$ reaching approximately $2.3\times10^{3}$. As $Pe$ increases, the diffusive contribution in Eq.~\eqref{eq:res_theta} becomes progressively weaker relative to advection, resulting in a more difficult optimization problem and slower convergence of the thermal field. To improve robustness in this regime, SAGE-PINN employs a continuation strategy in which the thermal problem is first trained at a lower effective P\'eclet number and is then gradually transitioned to the target value. During the thermal stage, the effective P\'eclet number is varied geometrically according to
\begin{equation}
Pe_{\mathrm{eff}}(t)
=
Pe_0
\left(
\frac{Pe}{Pe_0}
\right)^{\min(1,\,t/f_{\mathrm{ramp}})},
\qquad
t=
\frac{\text{local epoch}-1}
{\text{stage length}-1}
\in[0,1],
\label{eq:pe_ramp}
\end{equation}
where $Pe_0=10$ and $f_{\mathrm{ramp}}=0.5$. The stronger diffusive contribution at the beginning of the stage provides a smoother auxiliary thermal problem. As $Pe_{\mathrm{eff}}$ increases, the network is progressively adapted to the target advection-dominated operator, after which the remaining half of the stage is performed at the true $Pe$, as illustrated in Fig.~\ref{fig:pe_ramp}. The continuation is disabled when $Pe\leq1.5Pe_0$, for which the target problem is already sufficiently close to the starting value. Under the present parameter sweep, the P\'eclet continuation is therefore active in $35$ of the $40$ cases.

\subsubsection{Richardson Continuation}
\label{sec:ri_continuation}

\begin{figure}[htbp]
\centering
\includegraphics[width=0.9\textwidth]{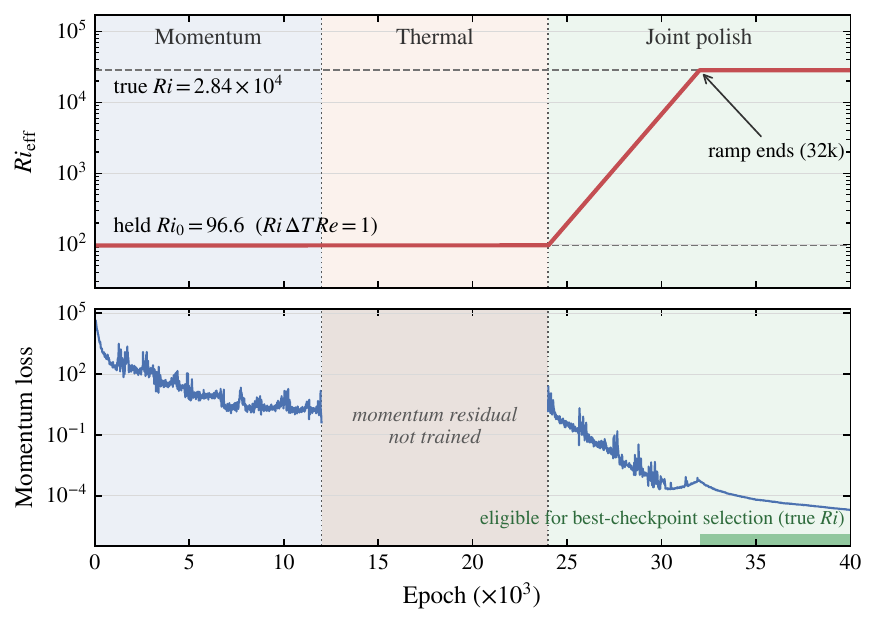}
\caption{Richardson continuation for the 3\% Au nanofluid at $u_0^*=10^{-4}$~m/s. $Ri_{\mathrm{eff}}$ is held at $Ri_{\mathrm{floor}}$ (Eq.~\ref{eq:ri_floor}) during the Momentum and Thermal Stages, then ramped geometrically to the true $Ri$ over the first half of the Joint-Polish Stage (top). Momentum loss $\mathcal{L}_{g,r}+\mathcal{L}_{g,z}$ (bottom) rises briefly as the Buoyancy term grows during the ramp, and decreases once $Ri$ reaches its true value.}
\label{fig:ri_ramp}
\end{figure}

At low inlet velocities, the buoyancy contribution can become large relative to viscous diffusion in the axial-momentum balance. For $u_0^*=10^{-4}$~m/s, the characteristic ratio $Ri\,\Delta T\,Re$ ranges from approximately $60$ for blood to $294$ for the Au nanofluid, indicating a strongly buoyancy-dominated regime. During the initial momentum-training stage, however, the thermal field has not yet been resolved in the interior, so applying the full buoyancy coupling at this stage can substantially hinder convergence of the momentum solution. SAGE-PINN therefore begins from a reduced Richardson number,
\begin{equation}
Ri_{\mathrm{floor}}
=
\min\!\left(
Ri,\,
\frac{1}{\Delta T\,Re}
\right),
\label{eq:ri_floor}
\end{equation}
such that, whenever the reduction is active, $Ri_{\mathrm{floor}}\,\Delta T\,Re=1$. This choice limits the characteristic buoyancy contribution to the scale of the viscous term during the decoupled momentum and thermal stages, thereby providing a better-conditioned auxiliary problem. The target Richardson number is restored geometrically during the joint-training stage according to
\begin{equation}
Ri_{\mathrm{eff}}(t)
=
Ri_{\mathrm{floor}}
\left(
\frac{Ri}{Ri_{\mathrm{floor}}}
\right)^{\min(1,\,t/f_{\mathrm{ramp}})},
\qquad
t=
\frac{\text{local epoch}-1}
{\text{stage length}-1}
\in[0,1],
\label{eq:ri_ramp}
\end{equation}
with $f_{\mathrm{ramp}}=0.5$. As shown in Fig.~\ref{fig:ri_ramp}, the effective Richardson number is held at $Ri_{\mathrm{floor}}$ during the momentum and thermal stages, and then increased smoothly to the target value over the first half of the joint stage. The continuation is introduced in the joint stage because both the momentum--pressure and thermal trunks are trainable there, allowing the two fields to adjust simultaneously as the physical buoyancy coupling is progressively restored. The second half of this stage is performed at the true $Ri$, and the subsequent L-BFGS refinement always uses the target parameter value. The continuation is omitted when the target Richardson number lies within a factor of $1.5$ of $Ri_{\mathrm{floor}}$, for which the auxiliary and target problems are already sufficiently close. Under the present parameter sweep, the Richardson continuation is active in $32$ of the $40$ cases.

\textbf{Checkpoint gating:} Models obtained before completion of either continuation correspond to auxiliary parameter values rather than to the target physical problem. Best-checkpoint selection is therefore activated only after both $Pe_{\mathrm{eff}}$ and $Ri_{\mathrm{eff}}$ have reached their prescribed target values.

\subsubsection{Phased and Joint Training}
\label{sec:phased_joint}
\begin{figure}[htbp]
\centering
\includegraphics[width=\textwidth]{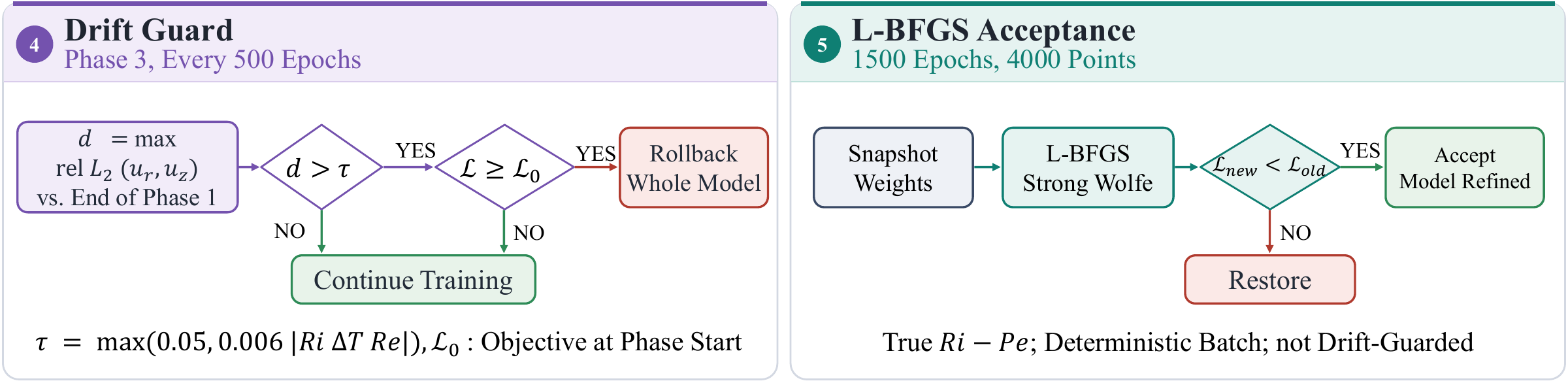}
\caption{Training safeguards of SAGE-PINN: (left) the objective-gated, physics-scaled drift guard of the joint polish stage, which rolls back the whole model; (right) the acceptance test of the L-BFGS refinement, which keeps the refined model only if the loss decreases.}
\label{fig:safeguards}
\end{figure}

Rather than optimizing both trunks jointly from the outset, SAGE-PINN uses three sequential training stages:
\begin{enumerate}
\item \textbf{Momentum stage:}
Only the momentum--pressure trunk is trainable, while the thermal trunk remains fixed at its initialized state. The momentum equations are solved at $Ri_{\mathrm{floor}}$, and the active objective contains the radial- and axial-momentum residuals together with the outlet pressure condition. This stage establishes a stable velocity and pressure field before the thermal field is trained.
\item \textbf{Thermal stage:}
The momentum--pressure trunk is frozen and only the thermal trunk is optimized. The velocity field obtained from the preceding stage therefore acts as a fixed advection field in the energy equation, while $Pe_{\mathrm{eff}}$ is increased according to the continuation strategy of Section~\ref{sec:pe_continuation}. This decoupled stage allows the thermal field to converge without simultaneously perturbing the momentum solution.
\item \textbf{Joint polish stage:}
Both trunks are made trainable and optimized together at a reduced learning rate. During the first half of this stage, $Ri_{\mathrm{eff}}$ is increased from $Ri_{\mathrm{floor}}$ to the target $Ri$ according to Section~\ref{sec:ri_continuation}, after which training proceeds at the full physical coupling. This stage reconciles the momentum and thermal fields before the final L-BFGS refinement.
\end{enumerate}
\textbf{Drift guard:}
When the momentum trunk is reactivated in the joint stage, the velocity field is expected to change as the buoyancy coupling is restored. Large changes, however, may also indicate optimization instability. To distinguish these cases, the relative $L_2$ drift of $(u_r,u_z)$ from the velocity field obtained at the end of the momentum stage is evaluated every $500$ Adam steps on $8{,}000$ sampled points and compared with the physics-scaled tolerance
\begin{equation}
\tau
=
\max\!\left(
0.05,\,
0.006\,|Ri\,\Delta T\,Re|
\right),
\label{eq:drift_budget}
\end{equation}
which increases with the characteristic strength of the buoyancy coupling. Exceeding this tolerance alone does not trigger a rollback, since a substantial velocity adjustment may be physically required. The guard is activated only when the excessive drift is accompanied by a failure of the objective to improve relative to its value at the beginning of the joint stage,
\begin{equation}
\text{trip}
\iff
\big(\text{drift}>\tau\big)
\wedge
\big(\mathcal{L}\geq\mathcal{L}_{\mathrm{stage\,start}}\big).
\label{eq:drift_trip}
\end{equation}
When this condition is met, the complete coupled model is rolled back rather than the momentum trunk alone, thereby preserving consistency between the velocity and temperature fields. The subsequent L-BFGS refinement is not subject to the drift guard because it is protected by the separate acceptance criterion described in Section~\ref{sec:training_strategy}. The two training safeguards are summarized in Fig.~\ref{fig:safeguards}.

\subsubsection{Loss Balancing Strategy}
\label{sec:loss_balancing}

\begin{figure}[htbp]
\centering
\includegraphics[width=\textwidth]{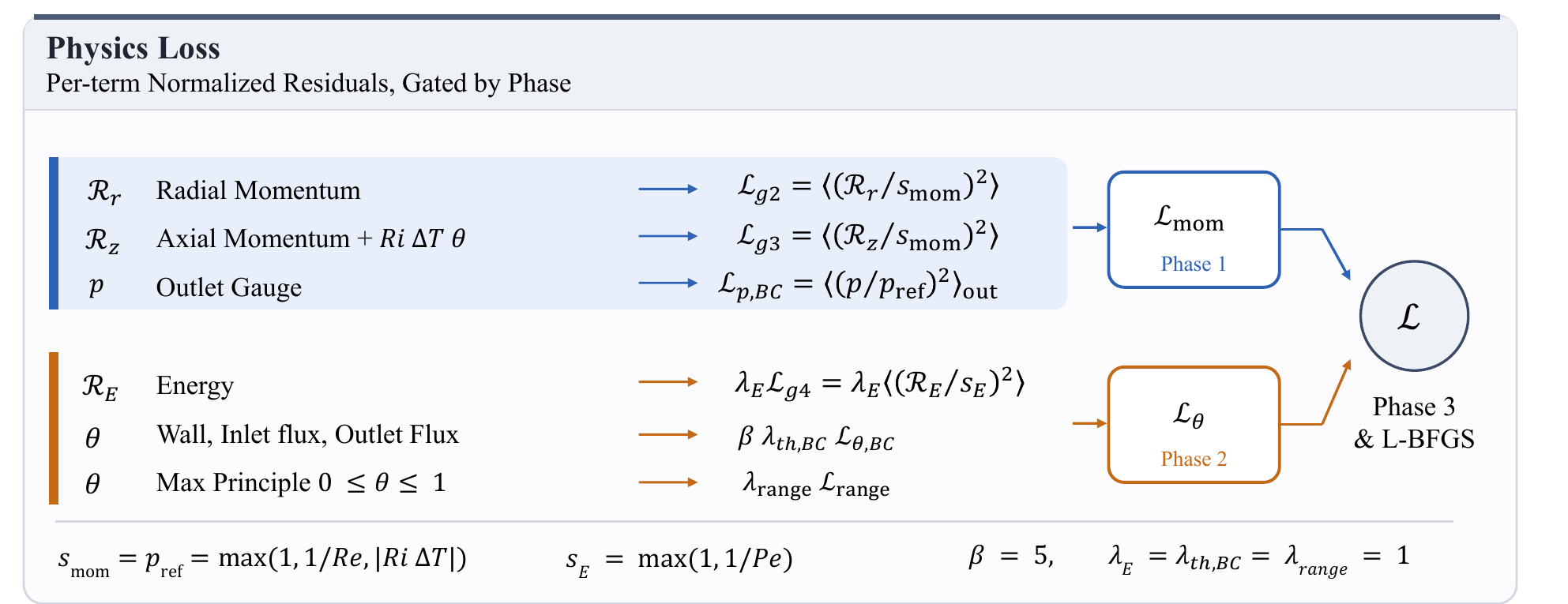}
\caption{Composition of the normalised objective in each training stage: momentum residuals and outlet pressure (momentum stage); energy residual, thermal boundary terms with the P\'eclet-weighted outlet flux, and range penalty (thermal stage); and the full objective (joint polish and L-BFGS).}
\label{fig:loss_phases}
\end{figure}

The complete objective is assembled from the normalized governing-equation residuals, the remaining soft boundary conditions, and the thermal-range penalty:
\begin{equation}
\mathcal{L}_{\mathrm{total}}
=
\underbrace{\mathcal{L}_{g,r}+\mathcal{L}_{g,z}}_{\text{momentum residuals}}
+
\lambda_{\mathrm{energy}}\mathcal{L}_{g,\theta}
+
\beta
\left(
\mathcal{L}_{p,\mathrm{BC}}
+
\lambda_{\mathrm{th,BC}}\mathcal{L}_{\theta,\mathrm{BC}}
\right)
+
\lambda_{\mathrm{range}}\mathcal{L}_{\mathrm{range}} .
\label{eq:total_loss}
\end{equation}
The normalized governing-equation losses are
\begin{equation}
\mathcal{L}_{g,r}
=
\left\langle
\left(\frac{\mathcal{R}_r}{s_{\mathrm{mom}}}\right)^2
\right\rangle,
\qquad
\mathcal{L}_{g,z}
=
\left\langle
\left(\frac{\mathcal{R}_z}{s_{\mathrm{mom}}}\right)^2
\right\rangle,
\qquad
\mathcal{L}_{g,\theta}
=
\left\langle
\left(\frac{\mathcal{R}_\theta}{s_{\mathrm{energy}}}\right)^2
\right\rangle,
\label{eq:governing_losses}
\end{equation}
Here, $\langle f\rangle_{\mathcal{S}}$ denotes the arithmetic mean of $f$ over the corresponding point set $\mathcal{S}$,
\begin{equation}
\langle f\rangle_{\mathcal{S}}
=
\frac{1}{N_{\mathcal{S}}}
\sum_{i=1}^{N_{\mathcal{S}}}
f(\mathbf{x}_i),
\label{eq:mean_operator}
\end{equation}
where $N_{\mathcal{S}}$ is the number of points in $\mathcal{S}$; when no subscript is shown, the average is taken over the interior mini-batch. Here, the averages are taken over the collocation mini-batch and the normalization factors are defined in Section~\ref{sec:residual_norm}. The pressure-reference loss is
\begin{equation}
\mathcal{L}_{p,\mathrm{BC}}
=
\left\langle
\left(\frac{p}{p_{\mathrm{scale}}}\right)^2
\right\rangle_{\mathrm{outlet}},
\end{equation}
which removes the additive pressure null space by imposing the outlet reference condition $p=0$. The thermal boundary contribution is written as
\begin{equation}
\mathcal{L}_{\theta,\mathrm{BC}}
=
\mathcal{L}_{\theta,\mathrm{wall}}
+
\mathcal{L}_{\theta,\mathrm{in}}
+
w_{\mathrm{out}}\,
\mathcal{L}_{\theta,\mathrm{out}},
\label{eq:thermal_bc_loss}
\end{equation}
where the wall Dirichlet term and inlet zero-flux term are satisfied by the hard-constrained ansatz and are retained primarily as diagnostic measures. Their values remain close to numerical round-off throughout training. The outlet contribution is
\[
\mathcal{L}_{\theta,\mathrm{out}}
=
\left\langle
\left(\partial_z\theta\right)^2
\right\rangle_{\mathrm{outlet}},
\]
with the P\'eclet-dependent weight
\begin{equation}
w_{\mathrm{out}}
=
\frac{1}{1+(Pe/Pe_{\mathrm{ref}})^2},
\qquad
Pe_{\mathrm{ref}}=50.
\label{eq:outlet_weight}
\end{equation}
Thus, the outlet zero-flux condition is enforced strongly at lower $Pe$ and progressively down-weighted as the problem becomes increasingly advection dominated. To discourage excursions outside the physically admissible temperature interval, we additionally use
\begin{equation}
\mathcal{L}_{\mathrm{range}}
=
\left\langle
\mathrm{ReLU}(\theta-1)^2
+
\mathrm{ReLU}(-\theta)^2
\right\rangle .
\label{eq:range_penalty}
\end{equation}
This term is inactive whenever $0\leq\theta\leq1$ and contributes only when the predicted temperature violates the prescribed physical range. The velocity boundary residuals at the wall, inlet, and axis are also monitored but are not included in the optimization objective, since these conditions are satisfied analytically by construction. The active loss terms depend on the training stage. The momentum stage uses
\begin{equation}
\mathcal{L}_{\mathrm{mom}}
=
\mathcal{L}_{g,r}
+
\mathcal{L}_{g,z}
+
\beta\mathcal{L}_{p,\mathrm{BC}},
\end{equation}
whereas the thermal stage uses
\begin{equation}
\mathcal{L}_{\mathrm{th}}
=
\lambda_{\mathrm{energy}}\mathcal{L}_{g,\theta}
+
\beta\lambda_{\mathrm{th,BC}}\mathcal{L}_{\theta,\mathrm{BC}}
+
\lambda_{\mathrm{range}}\mathcal{L}_{\mathrm{range}}.
\end{equation}
The joint-polish stage and the subsequent L-BFGS refinement use the full objective defined in Eq.~\eqref{eq:total_loss}. Residual terms that are inactive in a given stage are not evaluated, which also reduces the associated computational and memory cost. The weights
\[
\beta=5,
\qquad
\lambda_{\mathrm{energy}}
=
\lambda_{\mathrm{th,BC}}
=
\lambda_{\mathrm{range}}
=
1
\]
are kept fixed for all fluids and operating conditions. Because the governing residuals are already scaled according to Section~\ref{sec:residual_norm}, no case-specific or adaptive loss reweighting is required. Although gradient-based and self-adaptive weighting strategies are available \cite{xiang2021selfadaptive}, they are not used here in order to retain a single, fixed optimization formulation across the complete parameter sweep. Figure~\ref{fig:loss_phases} summarizes the loss components active in each training stage.

\subsection{Training Strategy}
\label{sec:training_strategy}
\begin{figure}[htbp]
\centering
\includegraphics[width=\textwidth]{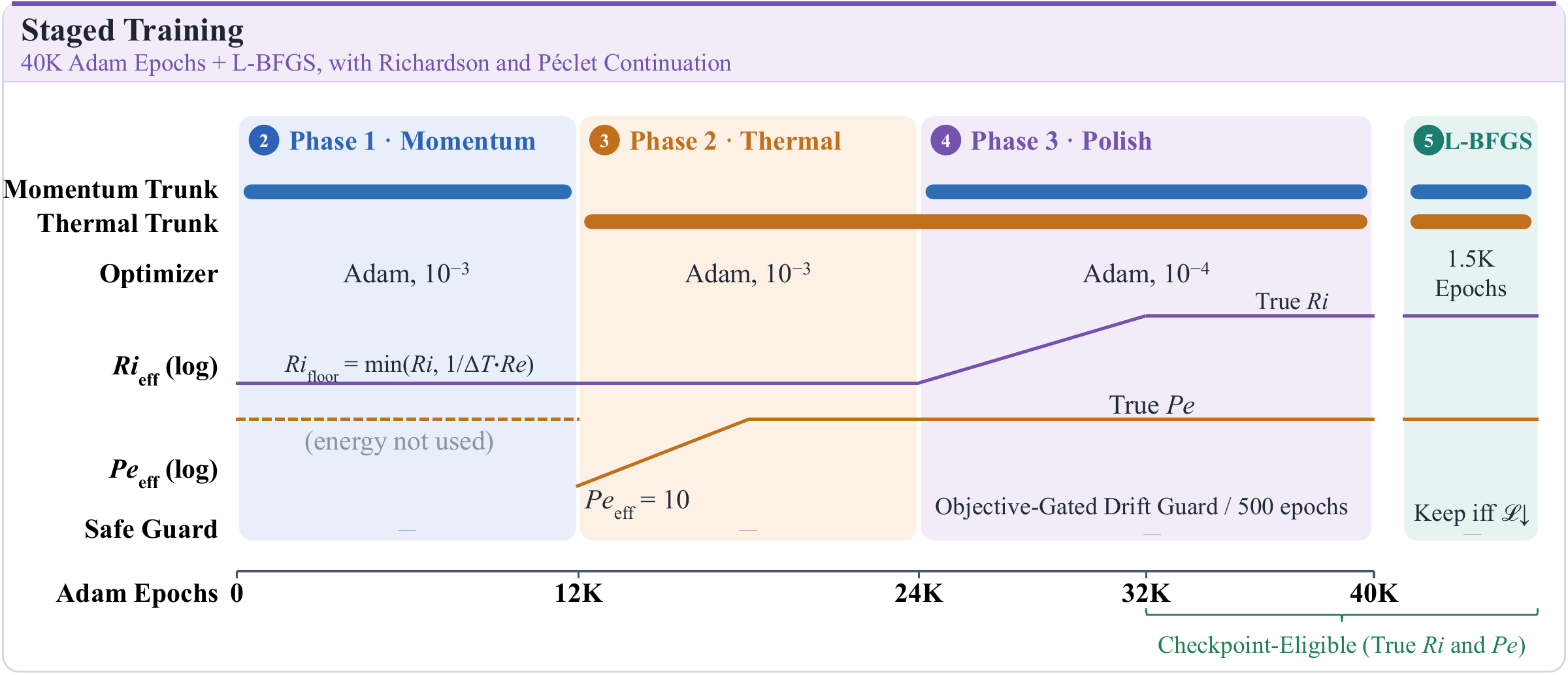}
\caption{Training schedule: Momentum Stage (epochs $1$-$12K$), Thermal Stage with P\'eclet Continuation ($12K$-$24K$), Joint Polish Stage with Richardson Continuation ($24K$--$40K$), and L-BFGS Refinement ($1{,}500$ Epochs) at the true $Ri$ and $Pe$.}
\label{fig:train_schedule}
\end{figure}

The network parameters are optimized using Adam with an initial learning rate of $\eta_0=10^{-3}$ in the momentum and thermal stages and $\eta_{\mathrm{polish}}=10^{-4}$ in the joint-polish stage. Gradient-norm clipping with a maximum norm of $1$ is applied throughout Adam training. A fresh optimizer and learning-rate scheduler are initialized at the beginning of each stage. Within a stage, the learning rate is reduced by a factor $\gamma=0.5$ whenever the monitored loss fails to improve by more than a relative tolerance $\epsilon_{\mathrm{tol}}=10^{-4}$ for $1500$ consecutive optimization steps, following a loss-plateau strategy similar to Ref.~\cite{roy2025wavy}. The learning rate is not reduced below $10^{-6}$. The three Adam stages use $30\%$, $30\%$, and $40\%$ of the total $40{,}000$ optimization steps, corresponding to steps $1$--$12{,}000$ for the momentum stage, $12{,}001$--$24{,}000$ for the thermal stage, and $24{,}001$--$40{,}000$ for the joint-polish stage. The joint stage is assigned the largest share because both network trunks are updated there and the Richardson-number continuation is completed. At each Adam step, a mini-batch of $N_{\mathrm{batch}}=10{,}000$ interior collocation points is sampled without replacement from the fixed set of $47{,}000$ points. Thus, the reported $40{,}000$ training steps correspond to $40{,}000$ mini-batch updates rather than full passes through the complete collocation set. The same step-based convention is used for the learning-rate patience and the drift-guard interval. After completion of the Adam stages, both trunks are refined using limited-memory BFGS (L-BFGS) with a strong-Wolfe line search and a history size of $50$. The refinement is performed for at most $1500$ L-BFGS iterations at the target values of $Ri$ and $Pe$ using a fixed subset of $4{,}000$ collocation points. A fixed point set is used because L-BFGS requires a deterministic objective for consistent curvature estimation. The model parameters are saved before refinement, and the L-BFGS solution is retained only if it decreases the total objective; otherwise, the pre-refinement model is restored. Intermediate models are stored after the momentum, thermal, and joint-polish stages, together with the final Adam model, the accepted L-BFGS-refined model, and the best checkpoint selected after both continuation parameters have reached their target values. The complete optimization schedule is summarized in Fig.~\ref{fig:train_schedule}.

\subsection{Proposed Algorithm}
\label{sec:algorithm}
The complete SAGE-PINN training procedure is summarized in Algorithm~\ref{alg:training}. All hyperparameters are fixed across the $40$ cases considered in Section~\ref{sec:numerical_experiment}. The algorithm combines data-point generation, network initialization, construction of the normalized stage-wise objectives, staged Adam optimization with P\'eclet and Richardson continuation and drift control, L-BFGS refinement, and final model evaluation. At each forward pass, the normalized input $x$ defined in Eq.~\eqref{eq:input} is supplied to the two network trunks, the physical fields are reconstructed through the constrained ans\"atze of Sections~\ref{sec:mom_ansatz}-\ref{sec:theta_ansatz}, and the derivatives required for the governing residuals are evaluated by automatic differentiation as described in Section~\ref{sec:autodiff}. The optimizer settings, stage lengths, continuation parameters, and safeguard parameters are identical for all cases.

\par\medskip
\refstepcounter{algorithm}\label{alg:training}

\begin{tcolorbox}[
  breakable,
  enhanced,
  colback=white,
  colframe=black!65,
  boxrule=0.6pt,
  arc=1.2mm,
  left=3mm,
  right=3mm,
  top=4mm,
  bottom=4mm,
  before skip=0pt,
  after skip=6pt,
  title={\bfseries Algorithm \thealgorithm: SAGE-PINN Training},
  coltitle=white,
  colbacktitle=black!65,
  fonttitle=\bfseries
]
\small
\begin{algorithmic}[1]

\Statex
\Statex \textbf{Step 1: Data and Collocation Points}
\State Generate $N_c = 47{,}000$ Interior Collocation Points by Latin Hypercube Sampling (for more details read Section ~\ref{sec:dataset_generation}).
\State Generate $N_b = 7{,}000$ boundary points: wall $4{,}000$, axis $1{,}500$, inlet $750$, outlet $750$ (not mini-batched) (Section ~\ref{sec:dataset_generation}).
\State Fix geometry, fluid properties, $Re$, $Ri$, $Pe$, and the random seed $42$.

\Statex
\Statex \textbf{Step 2: Architecture and Initialization}
\State Build the Momentum-Pressure Trunk ($6\times128$; outputs $\hat \varphi$, $\hat{p}$) and the Thermal Trunk ($4\times64$; output $\hat \theta$) as $\tanh$ MLPs (Eq.~\ref{eq:mlp}).
\State Initialize all Hidden Weights with Xavier normal and Zero biases and set the output heads to zero.
\State Define the field Reconstruction: Input $x=(2\xi-1,\ z/z_{\max})$, Momentum Ansatz Eq.~\eqref{eq:shi_ansatz}, Thermal Ansatz Eq.~\eqref{eq:theta_ansatz}, $p = p_{\mathrm{scale}}\hat p$.

\Statex
\Statex \textbf{Step 3: Composite Loss}
\State Compute Residuals $\mathcal{R}_r$, $\mathcal{R}_z$, $\mathcal{R}_\theta$ (Eqs.~\eqref{eq:res_r}--\eqref{eq:res_theta}) by automatic differentiation.
\State Normalize residuals by $s_{\mathrm{mom}}$, $s_{\mathrm{energy}}$ (Eq.~\ref{eq:res_scales}) and define $\mathcal{L}_{g,r}$, $\mathcal{L}_{g,z}$, $\mathcal{L}_{g,\theta}$ over the mini-batch,
\Statex \hspace{\algorithmicindent}$\mathcal{L}_{\mathrm{BC}} = \mathcal{L}_{p,\mathrm{BC}} +
\lambda_{\mathrm{th,BC}}\mathcal{L}_{\theta,\mathrm{BC}}$
with outlet weight $w_{\mathrm{out}}$ (Eq.~\ref{eq:outlet_weight}), and
\Statex \hspace{\algorithmicindent}$\mathcal{L}_{\mathrm{range}}$ in original form (Eq.~\ref{eq:range_penalty}).
\State Define Stage Objectives:
\begin{equation*}
\begin{aligned}
\mathcal{L}_{\mathrm{mom}} 
  &= \mathcal{L}_{g,r}+\mathcal{L}_{g,z}+\beta\mathcal{L}_{p,\mathrm{BC}},\\
\mathcal{L}_{\mathrm{th}} 
  &= \lambda_{\mathrm{energy}}\mathcal{L}_{g,\theta}
     +\beta\lambda_{\mathrm{th,BC}}\mathcal{L}_{\theta,\mathrm{BC}}
     +\lambda_{\mathrm{range}}\mathcal{L}_{\mathrm{range}},\\
\mathcal{L}_{\mathrm{total}}
  &= \mathcal{L}_{\mathrm{mom}}+\mathcal{L}_{\mathrm{th}}.
\end{aligned}
\label{eq:total_loss_2ndtime}
\end{equation*}

\Statex
\Statex \textbf{Step 4: Staged Adam Optimization}
\For{stage $\in \{$ Momentum, Thermal, Joint Polish $\}$}
  \State Activate Trunks: Momentum-only OR Thermal-only OR Both (Polish).
  \State Set Adam Learning Rate: $\eta_0$ (Momentum, Thermal) or $\eta_{\mathrm{polish}}$ (Polish) and Attach Loss-Scheduler.
  \State Set $Ri_{\mathrm{eff}}=Ri_{\mathrm{floor}}$ (Momentum, Thermal) or ramp it geometrically to $Ri$ over the first $50\%$ of the Stage (Polish) (Eq.~\ref{eq:ri_ramp}).
  \If{Stage $=$ Thermal}
    \State Ramp $Pe_{\mathrm{eff}}$ geometrically from $Pe_0=10$ to $Pe$ over the first $50\%$ of the Stage (Eq.~\ref{eq:pe_ramp}).
  \EndIf
  \For{Epoch $= 1$ to Stage-Epochs}
    \State Sample a Mini-Batch of $N_{\mathrm{batch}}$ Collocation Points.
    \State Forward pass: Reconstruct the Fields and compute the Active Objective with $Ri_{\mathrm{eff}}$, $Pe_{\mathrm{eff}}$ and the Scales computed from them.
    \State Back-Propagate, Clip the Grad-Norm to $1.0$, and take one Adam step.
    \State Update the Scheduler
    \If{Epoch mod 500 = 0 and Stage $\ne$ Momentum}
      \State Compute the Velocity Drift vs.\ the Momentum snapshot on $8{,}000$ Points (relative $L_2$ of $u_r$, $u_z$).
      \If{Drift $> \tau$ (Eq.~\ref{eq:drift_budget}) \textbf{and} $\mathcal{L} \ge \mathcal{L}_{\text{stage start}}$}
        \State Rollback the Whole Model.
      \EndIf
    \EndIf
    \If{both Continuations at their true values \textbf{and} $\mathcal{L}$ is the lowest so far}
      \State Save the Best Checkpoint.
    \EndIf
    \If{Epoch mod 10 = 0 or Epoch $=1$}
      \State Log Losses and Diagnostics.
    \EndIf
  \EndFor
  \If{Stage $=$ Momentum}
    \State Store the Momentum velocity snapshot.
  \EndIf
\EndFor

\Statex
\Statex \textbf{Step 5: L-BFGS Refinement}
\State Train both Trunks with up to $1{,}500$ L-BFGS Epochs (strong-Wolfe) on $4{,}000$ points at the true $Ri$ and $Pe$.
\State Accept the Refinement only if $\mathcal{L}_{\mathrm{total}}$ decreases otherwise roll back.

\Statex
\Statex \textbf{Step 6: Evaluation}
\State Store the Stage, Final, Refined, and Best Models.
\State Report Continuity Residual, Flux Drift, Thermal-Range diagnostics, and per-Radial-Bin Residuals.
\end{algorithmic}
\end{tcolorbox}

In summary, Step~4 constitutes the core of the SAGE-PINN training procedure. The momentum stage first establishes the velocity and pressure fields with the thermal trunk inactive, after which the thermal stage determines the temperature field using the frozen velocity under P\'eclet continuation. The joint-polish stage then reactivates both trunks at a reduced learning rate and progressively restores the full buoyancy coupling through Richardson continuation, while the drift guard protects against unstable departures from the previously converged flow field. Step~5 subsequently applies an L-BFGS quasi-Newton refinement at the target $Ri$ and $Pe$, and checkpoint gating ensures that only models corresponding to the target physical parameters are eligible for final selection. Together, these stages provide a unified training procedure that is applied without case-specific hyperparameter tuning across all five fluids and eight inlet velocities. The fixed hyperparameters and solver settings used for all $40$ cases are summarized in Table~\ref{tab:hyperparams}.

\begin{table}[htbp]
\centering
\setlength{\tabcolsep}{18pt}
\renewcommand{\arraystretch}{1.4}
\begin{tabular}{lclc}
\toprule
Parameter & Value & Parameter & Value \\
\midrule
$N_{\mathrm{epochs}}$ & $40{,}000$ &  $f_{\mathrm{mom}}$ & $0.3$\\
$N_{\mathrm{batch}}$ & $10{,}000$ & $f_{\mathrm{th}}$ & $0.3$ \\
$\lambda_{\mathrm{energy}}$ & $1.0$ & $f_{\mathrm{polish}}$ & $0.4$ \\
$\lambda_{\mathrm{range}}$ & $1.0$ & LR Patience & $1{,}500$ \\
$\lambda_{\mathrm{th,BC}}$& $1.0$ & LR Floor & $10^{-6}$ \\
$\beta$ & $5.0$ & LR Factor & $0.5$ \\
$\epsilon_{\mathrm{tol}}$ & $10^{-4}$ & LR ($\eta_{\mathrm{polish}}$) & $10^{-4}$ \\
$Pe_0$ & $10$ & LR ($\eta_0$) & $10^{-3}$  \\
$f_{\mathrm{ramp}}$ ($Pe$, $Ri$) & $0.5$  & Grad Clip & $1.0$ \\
$Pe_{\mathrm{ref}}$ (Outlet Weight) & $50$ & Thermal Ansatz$(q,\,k,\,A)$ & $(2,\,4,\,6)$ \\
Drift Budget $\tau$ & $\max(0.05,\,0.006 \,|Ri\, \Delta T\, Re|)$ &  L-BFGS Points & $4{,}000$\\
Drift Points & $8{,}000$ & L-BFGS Epochs & $1{,}500$ \\
Drift Interval & $500$ Epochs & Seed & $42$ \\
\bottomrule
\end{tabular}
\caption{Training Hyperparameters and Solver Settings used in Algorithm~\ref{alg:training}. Identical for all $40$ runs.}
\label{tab:hyperparams}
\end{table}

\subsection{Automatic Differentiation}
\label{sec:autodiff}

All derivatives appearing in the residuals of Eqs.~(\ref{eq:cont_nd})-(\ref{eq:energy_nd}) and Eqs.~(\ref{eq:res_r})-(\ref{eq:res_theta}) are obtained by automatic differentiation (AD): derivatives are computed by recursive application of the chain rule on the computational graph of the network, rather than by numerical differentiation. For a scalar output $f(r,z)$, the derivative with respect to the input coordinate $r$ is propagated through the hidden layers $\hbar^{(\ell)}$, where $\ell=1,\ldots,N$ denotes the hidden-layer index.  Specifically,
\begin{equation}
\frac{\partial \hbar^{(\ell)}}{\partial r}
=
\operatorname{diag}\!\left[
f'\!\left(W^{(\ell)}\hbar^{(\ell-1)}+b^{(\ell)}\right)
\right]
W^{(\ell)}
\frac{\partial \hbar^{(\ell-1)}}{\partial r},
\qquad \ell=1,\ldots,N,
\end{equation}
and, for the scalar network output,
\begin{equation}
\frac{\partial f}{\partial r}
=
W^{(N+1)}
\frac{\partial \hbar^{(N)}}{\partial r},
\qquad
\frac{\partial^2 f}{\partial r^2}
=
\frac{\partial}{\partial r}
\left(
\frac{\partial f}{\partial r}
\right),
\label{eq:autodiff}
\end{equation}
with analogous expressions for $z$. Higher-order derivatives are obtained by differentiating the computational graph again, which requires the intermediate derivative graph to be retained (``differentiate-through-differentiation''); thus, each derivative evaluation is exact up to floating-point round-off, with no truncation error, stencil, or mesh involved. In the transformed residuals of Eqs.~(\ref{eq:res_r})--(\ref{eq:res_theta}), the reconstructed stream-function field $\varphi$ is differentiated up to third order through terms such as $A_{ss}$, $B_{sz}$, and $A_{zz}$, while $T$ is differentiated up to second order and $p$ up to first order; retained-graph AD is therefore essential. As the momentum--pressure and thermal trunks are separate networks, their derivatives are computed independently, which also allows gradients from one trunk to be detached from the other during the staged training of Section~\ref{sec:training_strategy}.

\section{Numerical Experiment}
\label{sec:numerical_experiment}

\subsection{Dataset Generation}
\label{sec:dataset_generation}
\begin{figure}[htbp]
\centering
\includegraphics[width=0.8\textwidth]{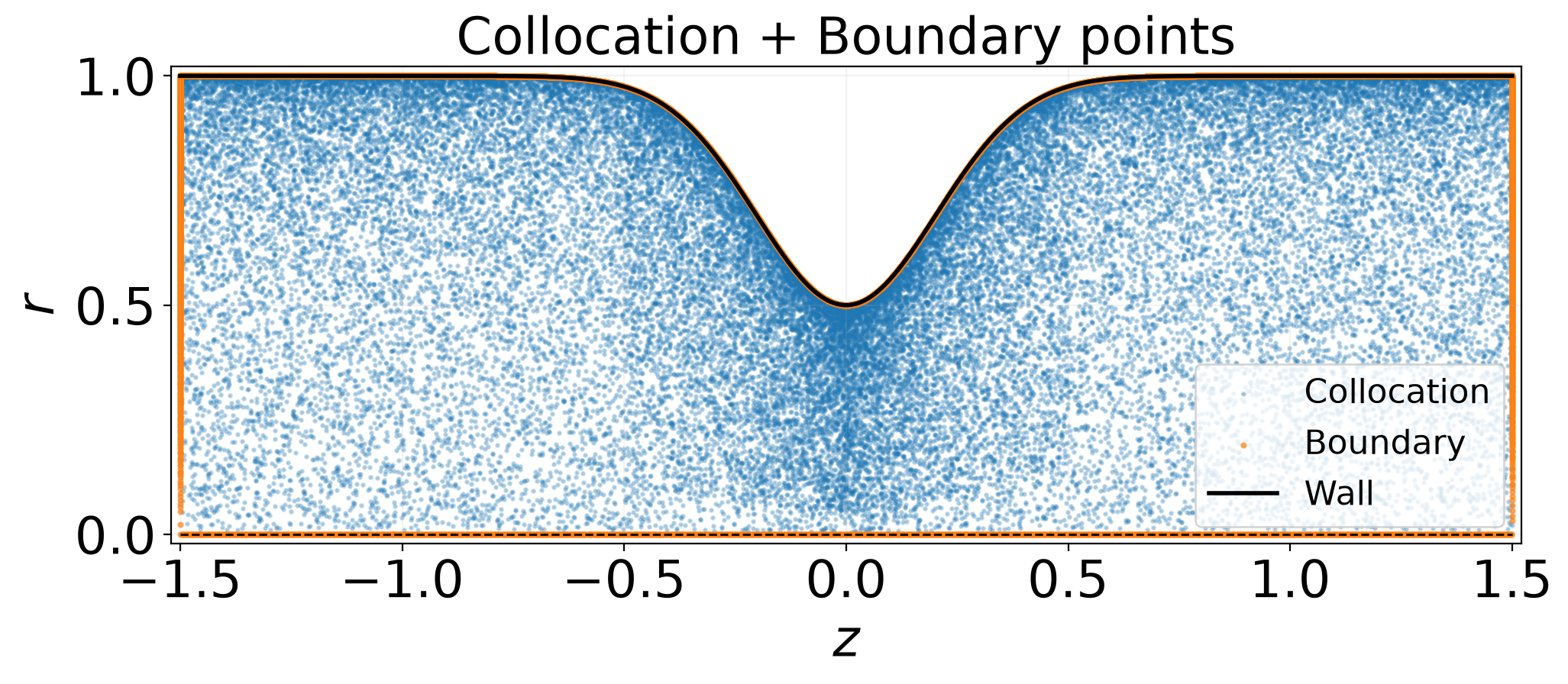}
\caption{Collocation and boundary points used for training a representative case.}
\label{fig:points}
\end{figure}
No labelled solution data are used for training; the network is trained entirely from the governing-equation residuals and boundary constraints. The required data therefore consist only of collocation points at which the residuals are evaluated. A fixed set of $N_c=47{,}000$ interior points is generated using Latin hypercube sampling (LHS). Of these, $N_{c1}=40{,}000$ bulk points provide coverage of the full computational domain, while an additional $N_{c2}=7{,}000$ sensitive-zone points are concentrated near the stenosed wall and throat region, $z\in[-0.5,0.5]$, where stronger spatial gradients are expected. The radial coordinates of these enriched points are generated as
$r=\eta(\xi)\,h(z)$, where $\eta(\xi)\in[0,1]$ denotes a wall-biased mapping that increases the sampling density towards $\xi=1$. The boundary set contains an additional $N_b=7{,}000$ points, distributed as $4{,}000$ on the wall, $1{,}500$ on the axis, $750$ at the inlet, and $750$ at the outlet, and generated independently using LHS. These boundary points are evaluated at every training step and are not mini-batched. The resulting collocation and boundary-point distributions are shown in Fig.~\ref{fig:points}.

\subsection{Test Problem}
\label{sec:test_problem}
We solve the coupled flow and energy equations for five working fluids: pure blood and $3\%$ Cu, Ag, Au, and Al$_2$O$_3$ nanofluids at eight inlet velocities, $u_0^*=
0.0001,\,
0.0002,\,
0.0005,\,
0.001,\,
0.002,\,
0.005,\,
0.01,\,
0.02~\mathrm{m/s},$ giving $40$ cases in total. Because the effective thermophysical properties differ among the fluids (Supplementary Tables~S3 and S6), the same inlet velocity corresponds to different values of the dimensionless groups $(Re,Ri,Pe,Gr)$. Across the complete parameter sweep, the cases span approximately
$0.46\leq Re\leq128$,
$10.5\leq Pe\leq2330$, and
$0.2\lesssim Ri\lesssim2.8\times10^{4}$, ranging from strongly buoyancy-dominated conditions at low inlet velocity to increasingly inertia- and advection-dominated conditions at higher velocity. Field-wise comparisons are presented for four representative inlet velocities,
$u_0^*=10^{-4}$,
$5\times10^{-4}$,
$5\times10^{-3}$, and
$2\times10^{-2}$~m/s,
which sample the principal regimes across the full parameter range, while the derived wall and throat quantities are evaluated for all $40$ cases. The corresponding dimensionless groups are listed in Supplementary Table~S5. Reference solutions are computed using COMSOL Multiphysics 6.4 with the same axisymmetric governing equations and stress-divergence momentum formulation as Eqs.~\eqref{eq:cont_nd}-\eqref{eq:energy_nd} \cite{nadeem2016drug,cai2021heattransfer}.

\subsection{Metrics}
\label{sec:metrics}
SAGE-PINN predictions are compared with the COMSOL reference fields at the same evaluation nodes. These nodes are generated independently by COMSOL and are therefore distinct from the collocation points used for training. For a field $\mathcal{F}\in\{u_r,u_z,\theta,p\}$, with SAGE-PINN prediction $\mathcal{F}_{\mathrm{PINN}}$ and reference solution $\mathcal{F}_{\mathrm{ref}}$, the mean absolute error (MAE), root-mean-square error (RMSE), relative $L_2$ error, and maximum absolute error are defined as
\begin{equation}
\mathrm{MAE}
=
\frac{1}{N}
\sum_{i=1}^{N}
\left|
\mathcal{F}_{\mathrm{PINN},i}
-
\mathcal{F}_{\mathrm{ref},i}
\right|,
\qquad
\mathrm{RMSE}
=
\sqrt{
\frac{1}{N}
\sum_{i=1}^{N}
\left(
\mathcal{F}_{\mathrm{PINN},i}
-
\mathcal{F}_{\mathrm{ref},i}
\right)^2
},
\label{eq:mae_rmse}
\end{equation}
and
\begin{equation}
\epsilon_{\mathrm{rel}}
=
\frac{
\left\|
\mathcal{F}_{\mathrm{PINN}}
-
\mathcal{F}_{\mathrm{ref}}
\right\|_2
}{
\left\|
\mathcal{F}_{\mathrm{ref}}
\right\|_2
},
\qquad
\epsilon_{\max}
=
\max_i
\left|
\mathcal{F}_{\mathrm{PINN},i}
-
\mathcal{F}_{\mathrm{ref},i}
\right|.
\label{eq:relative_metrics}
\end{equation}
Since the stream-function representation satisfies continuity by construction and the continuity equation is not included in the training loss, mass conservation is monitored independently through the continuity residual
\begin{equation}
C(r,z)
=
\frac{1}{r}
\frac{\partial (r u_r)}{\partial r}
+
\frac{\partial u_z}{\partial z},
\qquad r>0,
\label{eq:continuity_metric}
\end{equation}
with its value at $r=0$ understood through the regular axis limit. In addition, conservation of volumetric flow rate is assessed using
\begin{equation}
\epsilon_Q(z)
=
\left|
\frac{
Q_{\mathrm{PINN}}(z)-Q_{\mathrm{in}}
}{
Q_{\mathrm{in}}
}
\right|,
\label{eq:flux_metric}
\end{equation}
where $Q_{\mathrm{in}}$ is the imposed inlet flow rate. These two diagnostics provide independent measures of mass conservation and are reported in Section~\ref{sec:mass_conservation}.

\subsection{Reference Solution and Validation}
\label{sec:comparison_methods}
All SAGE-PINN reconstructions are compared with the COMSOL Multiphysics 6.4 \cite{comsol62} finite-element solution described above, which serves as the reference solution for the field-error metrics and the derived quantities reported in Section~\ref{sec:derived_quantities}. The reference fields are evaluated at the nodes of the COMSOL mesh, and the SAGE-PINN solution is evaluated at the same locations. No labelled data from the COMSOL solution are used during training. Since no existing PINN solution is available for the present coupled axisymmetric problem, the validation is carried out in stages. First, the COMSOL reference solution is verified through a mesh-independence study and comparison with an independent analytical solution for the same axisymmetric stenosis configuration. The SAGE-PINN results are then interpreted in the context of previously established PINN solutions for related Cartesian flow and heat-transfer problems. The COMSOL solution was verified on five successively refined structured meshes, for which the throat centreline velocity changes by less than $0.1\%$, and the finest mesh is used for all reference solutions (Supplementary Fig.~S14a). It also agrees closely with the analytical solution of Nadeem and Ijaz \cite{nadeem2016drug} at $Re=25.797$ and $257.97$, deviating only at $Re=2579.7$, where the simplifying assumptions of the analytical model become less accurate (Supplementary Fig.~S14b). Finally, the present PINN implementation was verified by reproducing the Cartesian wavy-channel problem of Roy et al.~\cite{roy2025wavy} (Supplementary Section~S1).

\begin{figure}[htbp]
\centering
\newlength{\vanh}\setlength{\vanh}{0.322\linewidth}%
\newsavebox{\vanbox}%
\newcommand{\vf}[2]{\sbox{\vanbox}{\fieldimg[\vanh]{Figures/vanilla/#1.pdf}}\usebox{\vanbox}\llap{\makebox[\wd\vanbox][c]{\raisebox{-1.1em}[0pt][0pt]{\footnotesize #2}}}}%
\newcommand{\vl}[1]{\alignedleg{Figures/vanilla/#1.pdf}{\vanh}}%
\makebox[\linewidth][c]{\vf{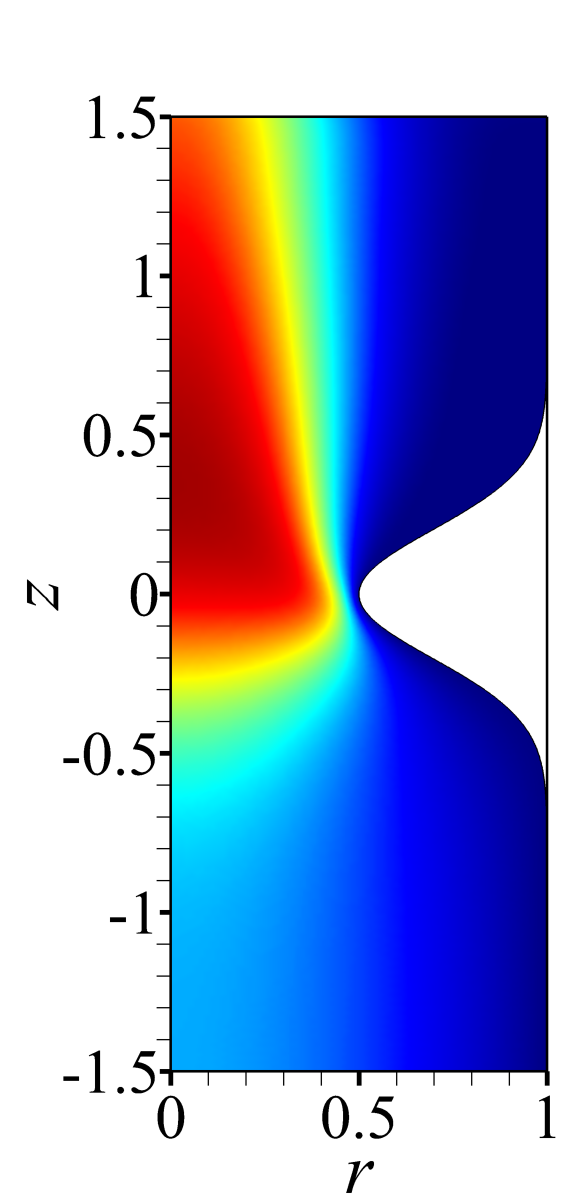}{True COMSOL}\vf{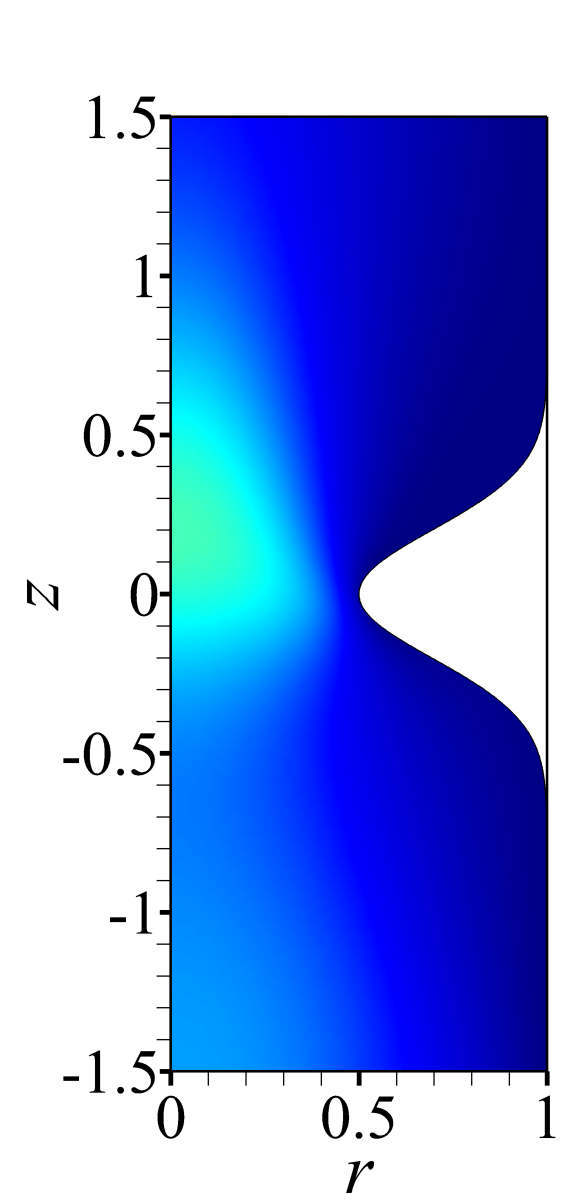}{Vanilla-PINN}\vf{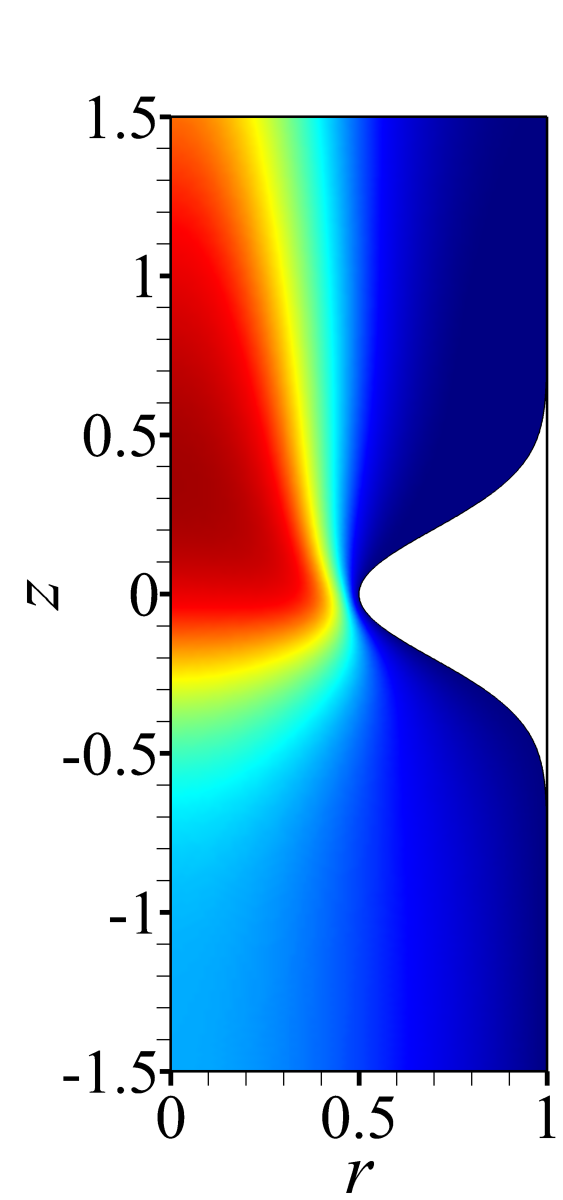}{SAGE-PINN}\vl{legend_true_w}\hspace{2pt}\vf{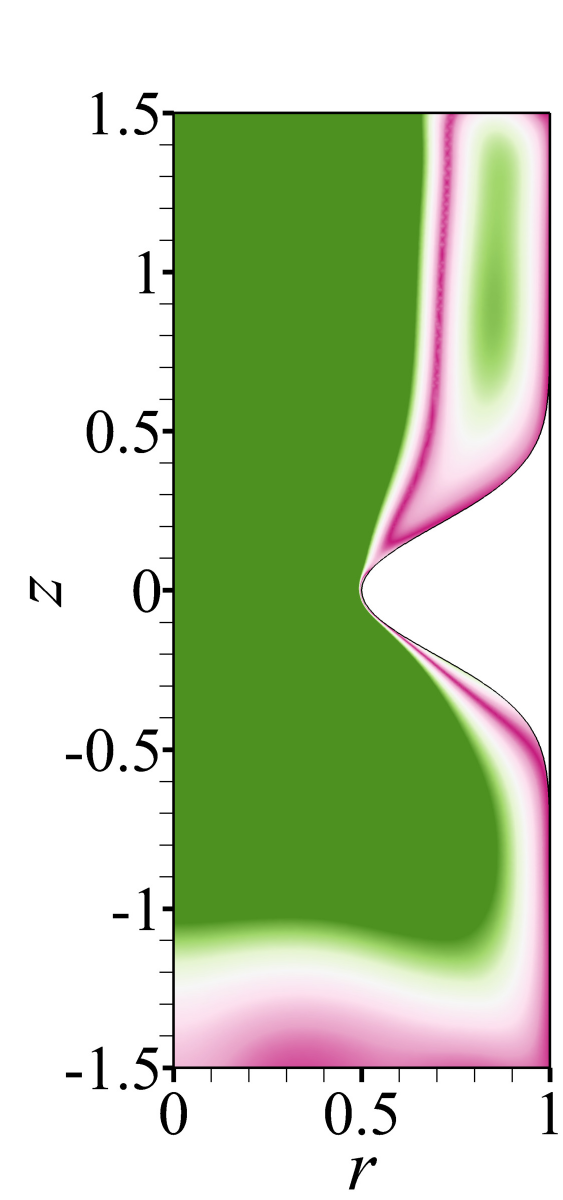}{Vanilla Error}\vl{vanilla_w_err_leg}\hspace{2pt}\vf{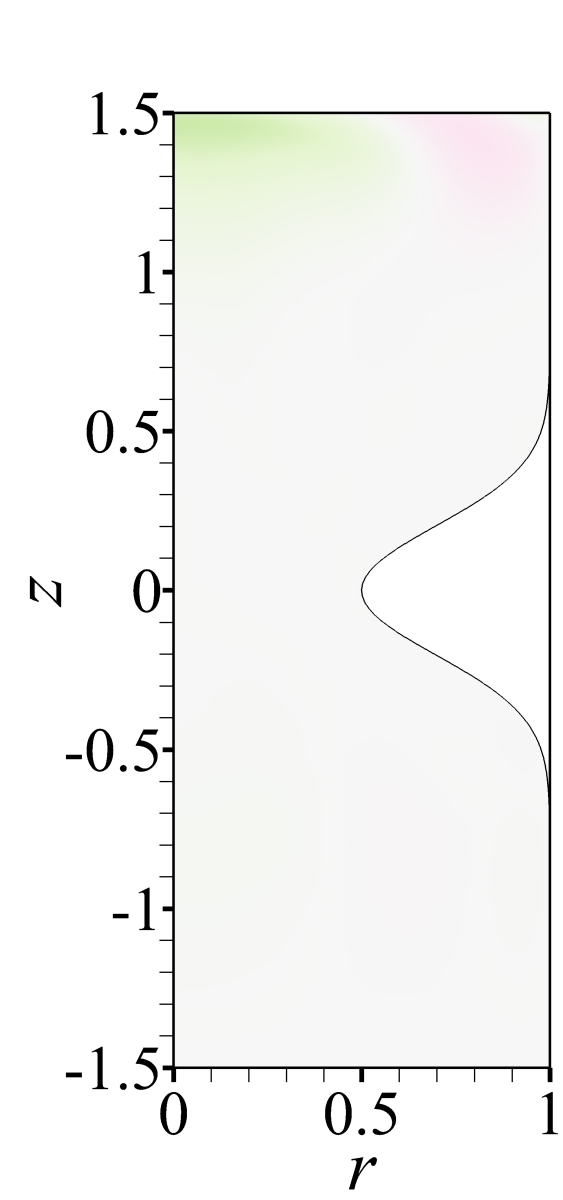}{SAGE Error}\vl{sage_legend_w_Err}}
\par\vspace{-1.5em}
\caption{Comparison between the COMSOL reference solution, the vanilla PINN, and SAGE-PINN for pure blood at $u_0^*=5\times10^{-3}$~m/s ($Re=22.8$, $Pe=582$). From left to right: COMSOL reference field, vanilla-PINN prediction, SAGE-PINN prediction, absolute error of the vanilla PINN, and absolute error of SAGE-PINN. The comparison highlights the failure of the vanilla formulation to reproduce the throat jet and its substantially larger field-wise error.}
\label{fig:vanilla_pinn}
\end{figure}

\textbf{Vanilla PINN in cylindrical coordinates:}
To isolate the effect of the cylindrical formulation, a standard PINN was trained for pure blood at $u_0^*=5\times10^{-3}$~m/s ($Re=22.8$, $Pe=582$), using the same data budget, network capacity, optimizer, number of Adam steps, and L-BFGS refinement as SAGE-PINN. The vanilla PINN predicts $(u_r,u_z,p,\theta)$ directly from $(r,z)$, minimizes the original cylindrical residuals including continuity, and imposes all boundary conditions through equally weighted soft penalties. To avoid direct evaluation of the coordinate singularity, residual points are restricted to $r\geq10^{-3}$. Although the training loss decreases by nearly three orders of magnitude, the resulting solution remains substantially inaccurate, as shown in Fig.~\ref{fig:vanilla_pinn}. In particular, the throat jet is not reproduced correctly, and the relative $L_2$ errors are $54.5\%$ in $u_z$, $83.5\%$ in $u_r$, and $22.7\%$ in $\theta$, compared with $0.67\%$, $4.47\%$, and $1.74\%$, respectively, for SAGE-PINN. The wall no-slip penalty also stagnates at approximately $6\times10^{-3}$, leading to measurable wall leakage: the integrated flux varies by as much as $55\%$, while the mean continuity residual remains at $1.8\times10^{-2}$, approximately five orders of magnitude larger than the round-off-level residual obtained with SAGE-PINN. These results show that a low aggregate training loss alone does not ensure a physically consistent solution in the present cylindrical problem. The hard boundary constraints, exact mass conservation, and singularity-free residual formulation are therefore essential components of the improved performance of SAGE-PINN.

\subsection{Hyperparameters}
\label{sec:hyperparams}

The hyperparameters used for all $40$ cases are listed in Table~\ref{tab:hyperparams}. Apart from the dimensionless groups $(Re,Ri,Pe)$, which vary with fluid composition and inlet velocity, no hyperparameter is re-tuned from one case to another. The corresponding configurations therefore differ only in the working fluid, $u_0^*$, and the resulting values of $Re$, $Ri$, and $Pe$. Case-dependent quantities such as the residual-normalization scales, the Richardson-number floor, the drift threshold, and the outlet thermal weight are computed directly from $(Re,Ri,Pe)$ using the relations introduced in Section~\ref{sec:proposed_method}, rather than selected manually for individual cases. The same network architecture, optimizer settings, training schedule, continuation parameters, and loss weights are consequently retained throughout the complete set of $40$ simulations.

\subsection{Software, Hardware, and Implementation}
\label{sec:software}
The SAGE-PINN framework is implemented in PyTorch \cite{paszke2019pytorch}, with all residual and boundary derivatives evaluated using automatic differentiation. Training is performed on a single NVIDIA GeForce RTX 5080 GPU with 16~GB of VRAM (CUDA 13.2), on a workstation equipped with an Intel Core Ultra 7 155U processor and 32~GB of RAM. A typical run requires approximately $90$~minutes for the $40{,}000$ Adam epochs and a further $18$~minutes, on average, for the $1{,}500$ L-BFGS epochs, giving a total wall-clock time of about $108$~minutes per case. For the final full-field evaluation, the collocation points are processed in chunks of $4{,}000$ to limit the peak memory required for constructing and retaining the automatic-differentiation graph. The COMSOL Multiphysics 6.4 reference solutions are computed separately on a workstation equipped with an Intel Core Ultra 9 processor and 64~GB of RAM. The implementation uses fixed scripts, configuration files, and random seeds for all cases; these materials are available from the corresponding author upon request.

\section{Results and Discussion}
\label{sec:results_discussion}
The five working fluids: pure blood and $3\%$ Ag, Au, Al$_2$O$_3$, and Cu nanofluids---have different effective densities, viscosities, heat capacities, and thermal conductivities (Supplementary Table~S3), and therefore correspond to different values of $Re$, $Ri$, and $Pe$ at the same inlet velocity. The results are consequently organized by fluid, with the inlet velocity varied within each fluid, rather than by comparing different fluids at a common nominal Reynolds number. For the field-wise comparisons, four representative inlet velocities are selected to span the parameter range: $u_0^*=10^{-4}$~m/s, corresponding to the strongest buoyancy influence and lowest $Pe$; $5\times10^{-4}$~m/s, representing an intermediate low-velocity regime; $5\times10^{-3}$~m/s, where inertial and advective effects become increasingly important; and $2\times10^{-2}$~m/s, corresponding to the highest $Re$ and strongly advection-dominated thermal regime. Derived wall and throat quantities are subsequently reported for all $40$ cases in Section~\ref{sec:derived_quantities}. For conciseness, the complete two-dimensional field comparisons are presented here only for the $3\%$ Cu nanofluid (Section~\ref{sec:copper_results}); the corresponding velocity, pressure, temperature, and stream-function fields for pure blood and the $3\%$ Ag, Al$_2$O$_3$, and Au nanofluids are provided in the Supplementary Reading (Section~S2).

\subsection{Global Accuracy}
\label{sec:test_matrix}
\begin{figure}[htbp]
\centering
\includegraphics[width=0.9\textwidth]{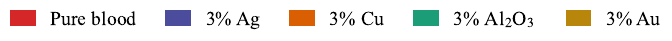}\\[2pt]
\begin{subfigure}[t]{0.49\textwidth}
\centering
\includegraphics[width=\linewidth]{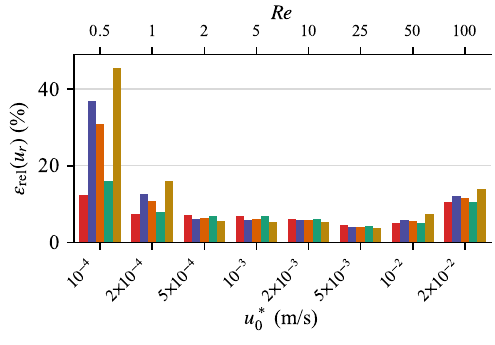}
\caption{Radial velocity $u_r$}
\label{fig:errors_bar_ur}
\end{subfigure}\hfill
\begin{subfigure}[t]{0.49\textwidth}
\centering
\includegraphics[width=\linewidth]{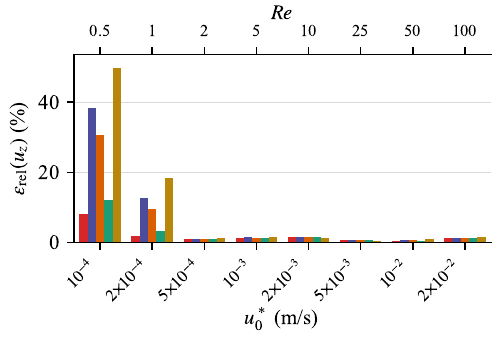}
\caption{Axial velocity $u_z$}
\label{fig:errors_bar_uz}
\end{subfigure}\\[4pt]
\begin{subfigure}[t]{0.49\textwidth}
\centering
\includegraphics[width=\linewidth]{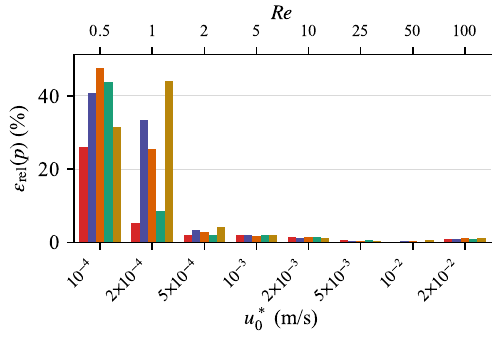}
\caption{Pressure $p$}
\label{fig:errors_bar_p}
\end{subfigure}\hfill
\begin{subfigure}[t]{0.49\textwidth}
\centering
\includegraphics[width=\linewidth]{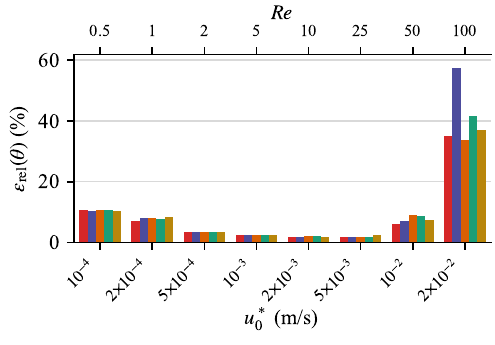}
\caption{Normalized temperature $\theta$}
\label{fig:errors_bar_theta}
\end{subfigure}
\caption{Relative $L_2$ errors (\%) of SAGE-PINN with respect to the COMSOL reference for all $40$ cases: (a) radial velocity $u_r$, (b) axial velocity $u_z$, (c) offset-free pressure $p$, and (d) normalized temperature $\theta$. The bars are grouped by inlet velocity (bottom axis; the top axis gives the corresponding approximate Reynolds number), with one bar per fluid, and each panel uses its own linear scale. The corresponding numerical values are listed in Supplementary Table S7.}
\label{fig:errors_bar}
\end{figure}
Figure~\ref{fig:errors_bar} reports the relative $L_2$ errors of the predicted fields for all $40$ cases; the corresponding dimensionless groups are listed in Supplementary Table~S5. The errors are evaluated at the $44{,}817$ nodes of the COMSOL reference mesh. Since the reference mesh is refined towards the wall, we also compared the standard nodal norm with an area-weighted norm; the two differ by only $0.2$ percentage points in the median and by at most about $4$ percentage points across the complete dataset. Three accuracy regimes are evident. In the intermediate range, $2\times10^{-4}\leq u_0^*\leq10^{-2}$~m/s ($30$ cases, approximately $0.9\lesssim Re\lesssim64$ and $53\lesssim Pe\lesssim1165$), the axial-velocity error averages $1.05\%$ and does not exceed $1.6\%$, while the corresponding radial-velocity errors are $5.7\%$ on average and at most $7.4\%$. Over the same range, the temperature error averages $3.5\%$ and remains below $9.1\%$. At the lowest inlet velocity, $u_0^*=10^{-4}$~m/s, the temperature error is comparatively uniform across fluids, ranging from $10.3\%$ to $10.8\%$, whereas the axial-velocity error varies much more strongly, from $8.0\%$ for pure blood to $49.7\%$ for the Au nanofluid. At the highest inlet velocity, $u_0^*=2\times10^{-2}$~m/s, the axial velocity remains well reproduced, with $\epsilon_{\mathrm{rel}}(u_z)\leq1.7\%$, while the temperature error increases substantially to $33.6$--$57.3\%$. The overall error trend is therefore lowest over the intermediate velocity range and increases towards both ends of the parameter sweep. The mechanisms responsible for the low-velocity and high-velocity deterioration are different and are examined in Section~\ref{sec:discussion}.

The field-wise comparisons for the remaining four fluids follow the same overall trends and are summarized here. Pure blood has the weakest buoyancy coupling, with $Ri\,\Delta T\,Re\approx60$ at $u_0^*=10^{-4}$~m/s, and its axial velocity is reproduced with relative $L_2$ errors of $8.0\%$ at the lowest velocity and below $1.8\%$ at all higher velocities (Section~S2.1). The $3\%$ Al$_2$O$_3$ nanofluid remains the closest to pure blood in its effective properties ($Ri\,\Delta T\,Re=84$), with a moderate axial-velocity error of $12.1\%$ at the lowest velocity (Section~S2.3). The $3\%$ Ag and Au nanofluids exhibit the strongest buoyancy coupling ($Ri\,\Delta T\,Re\approx226$ and $294$, respectively); at $u_0^*=10^{-4}$~m/s, buoyancy restructures the throat flow into an M-shaped axial-velocity profile, which SAGE-PINN captures qualitatively, but with axial-velocity errors of $38.3\%$ and $49.7\%$, respectively. For both fluids, these errors decrease rapidly with increasing inlet velocity and fall to approximately $1\%$ from $u_0^*=5\times10^{-4}$~m/s onwards (Sections~S2.2 and S2.4). In all four fluids, the temperature error is largest at the highest inlet velocity, ranging from $34.9\%$ for pure blood to $57.3\%$ for the Ag nanofluid, the largest temperature error among the $40$ cases, while the stream-function field remains closely reproduced across all cases, consistent with the hard flux constraint.

\subsection{3\% Copper (Cu)}
\label{sec:copper_results}
\begin{figure}[htbp]
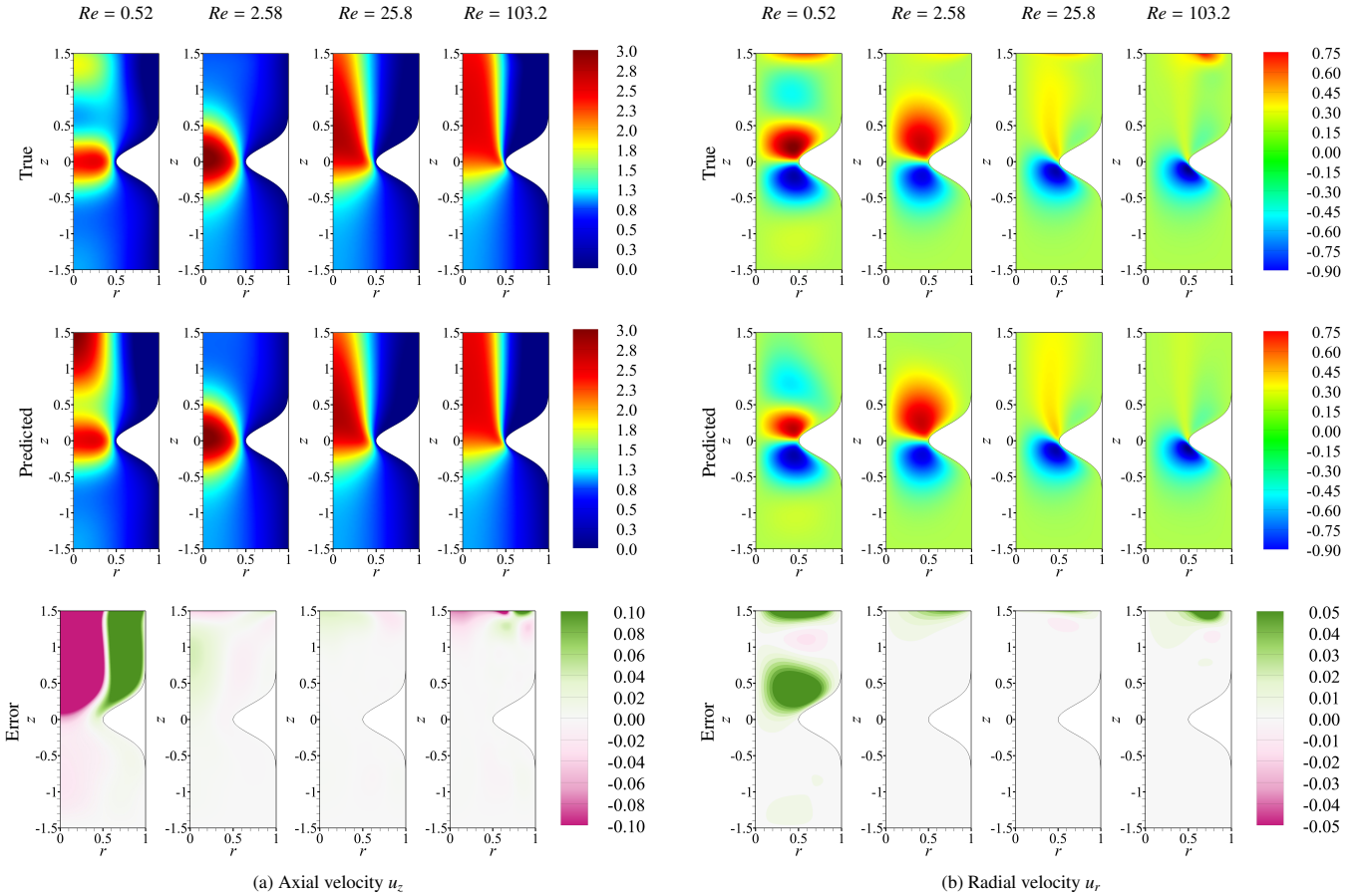

\centering
\begin{subfigure}[t]{0.495\textwidth}
\centering
\fieldblock{All_copper}{w}{1}
\caption{Axial velocity $u_z$}
\label{fig:copper_uz}
\end{subfigure}\hfill
\begin{subfigure}[t]{0.495\textwidth}
\centering
\fieldblock{All_copper}{u}{1}
\caption{Radial velocity $u_r$}
\label{fig:copper_ur}
\end{subfigure}
\caption{Axial and radial velocity fields for the $3\%$ Cu nanofluid. The first, second, and third rows show the COMSOL reference solution, the SAGE-PINN prediction, and the corresponding error (COMSOL$-$PINN), respectively. Columns correspond to $u_0^* = 10^{-4}$, $5\times10^{-4}$, $5\times10^{-3}$, and $2\times10^{-2}$~m/s, with corresponding $Re = 0.52$, $2.58$, $25.8$, and $103.2$, respectively.}
\label{fig:copper_vel}
\end{figure}

The $3\%$ Cu nanofluid follows the same overall trend as the Ag case, but with slightly weaker buoyancy coupling ($Ri\,\Delta T\,Re = 185$ at $u_0^* = 10^{-4}$~m/s). Accordingly, the most difficult case is again the lowest-velocity one: at $u_0^* = 10^{-4}$~m/s, the relative $L_2$ errors in the axial and radial velocities are $30.7\%$ and $30.8\%$, respectively, with the discrepancy concentrated mainly in the core region. As the inlet velocity increases, the velocity field is recovered much more accurately; the axial-velocity error drops to $9.7\%$ at $u_0^* = 2\times10^{-4}$~m/s and remains within $0.6$--$1.5\%$ for all higher velocities. The representative velocity fields are shown in Fig.~\ref{fig:copper_vel}.

\begin{figure}[htbp]
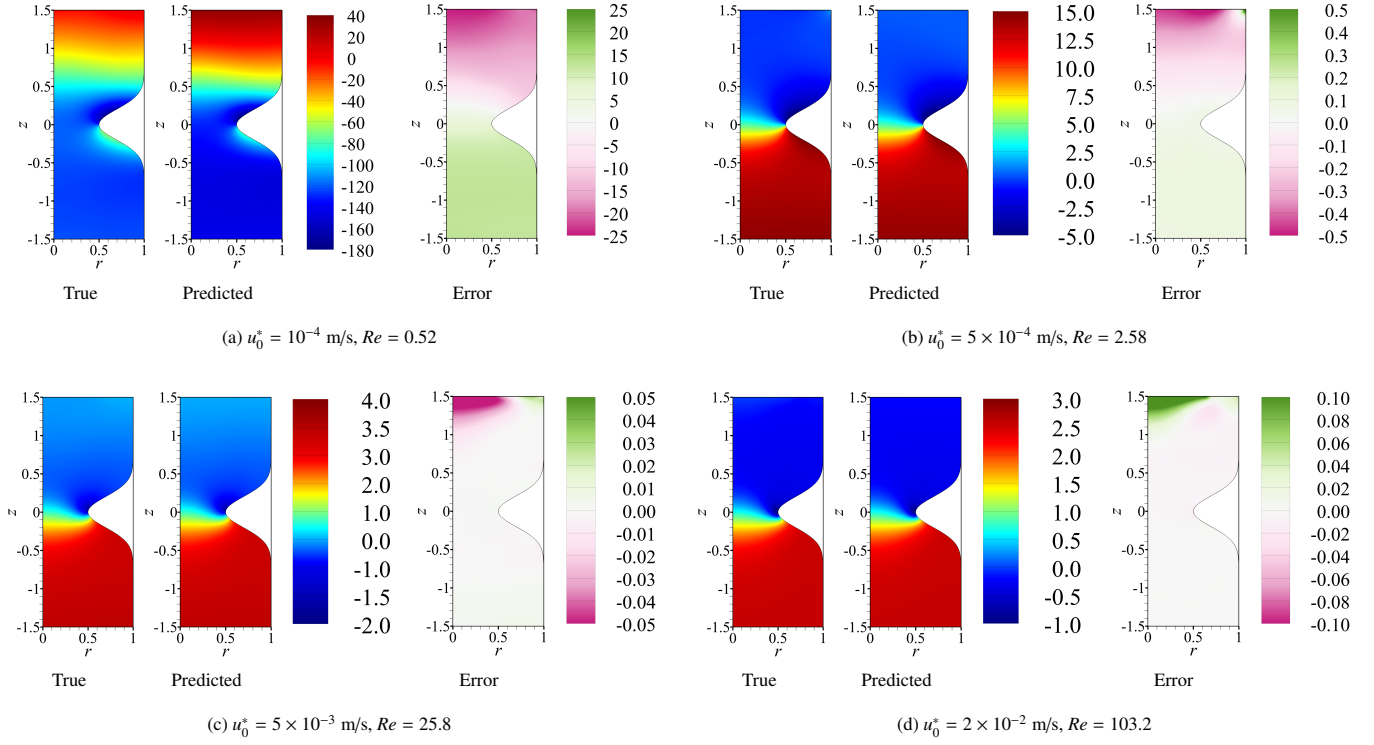

\centering
\begin{subfigure}[t]{0.495\textwidth}
\centering
\presrow{All_copper}{u0001}
\caption{$u_0^*=10^{-4}$~m/s, $Re=0.52$}
\label{fig:copper_p_u0001}
\end{subfigure}\hfill
\begin{subfigure}[t]{0.495\textwidth}
\centering
\presrow{All_copper}{u0005}
\caption{$u_0^*=5\times10^{-4}$~m/s, $Re=2.58$}
\label{fig:copper_p_u0005}
\end{subfigure}\\[6pt]
\begin{subfigure}[t]{0.495\textwidth}
\centering
\presrow{All_copper}{u005}
\caption{$u_0^*=5\times10^{-3}$~m/s, $Re=25.8$}
\label{fig:copper_p_u005}
\end{subfigure}\hfill
\begin{subfigure}[t]{0.495\textwidth}
\centering
\presrow{All_copper}{u02}
\caption{$u_0^*=2\times10^{-2}$~m/s, $Re=103.2$}
\label{fig:copper_p_u02}
\end{subfigure}
\caption{Pressure field $p$ for the $3\%$ Cu nanofluid at four representative inlet velocities. In each panel, the COMSOL reference solution, SAGE-PINN prediction, and corresponding error (COMSOL$-$PINN) are shown. Separate colour scales are used for each velocity because the pressure magnitude varies substantially across the parameter range.}
\label{fig:copper_p}
\end{figure}

The pressure field is also reproduced well once the solution moves away from the most strongly buoyancy-influenced regime. For $5\times10^{-4} \le u_0^* \le 10^{-2}$~m/s, the offset-free relative error remains below $2.8\%$, and at the fastest case it is $1.09\%$ (cf. Fig.~\ref{fig:copper_p}). As in the previous fluids, the low-velocity pressure discrepancy follows the deterioration of the momentum solution and is discussed further in Section~\ref{sec:discussion}.
\begin{figure}[htbp]
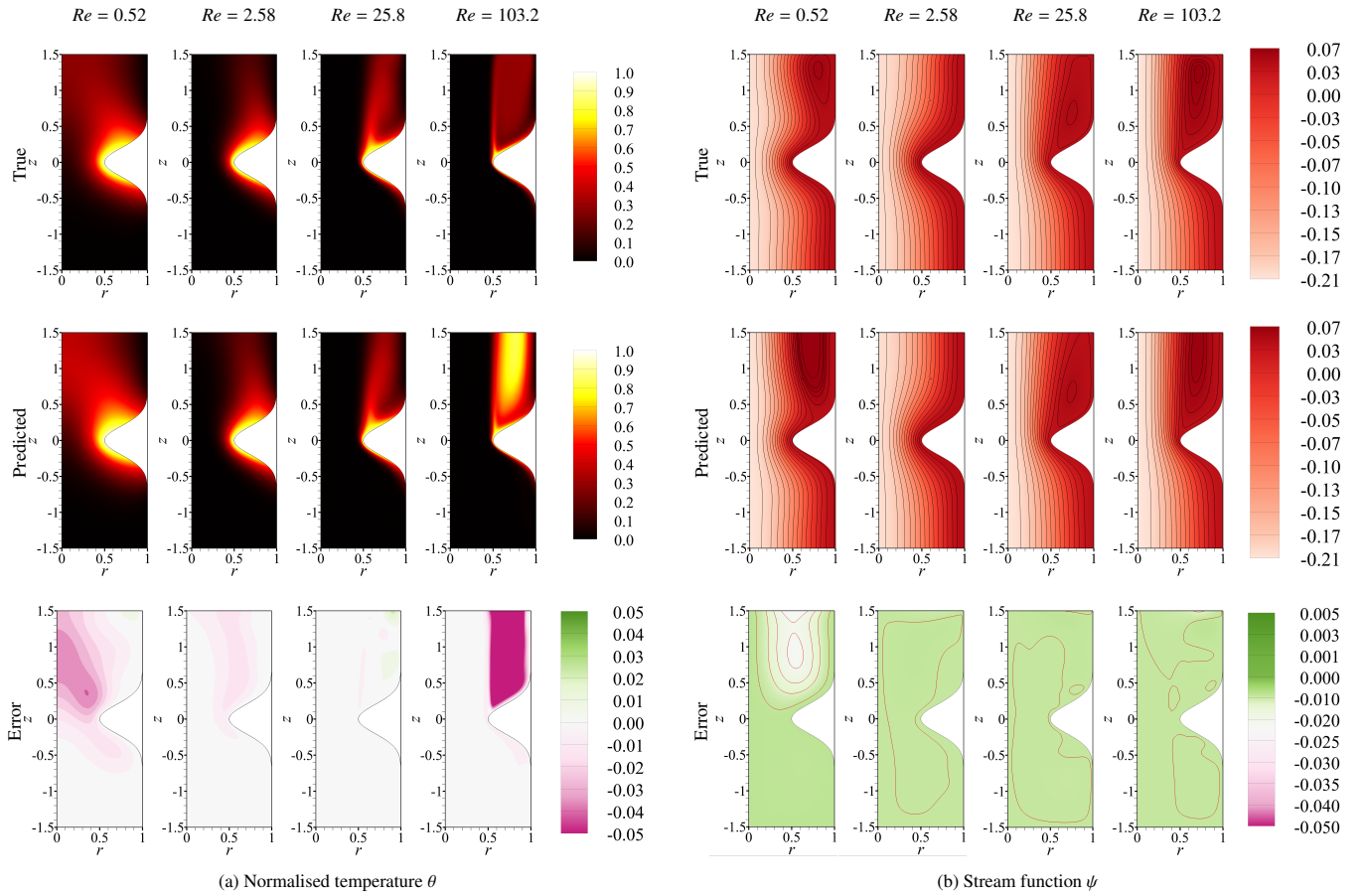

\centering
\begin{subfigure}[t]{0.495\textwidth}
\centering
\fieldblock{All_copper}{theta}{1}
\caption{Normalised temperature $\theta$}
\label{fig:copper_theta}
\end{subfigure}\hfill
\begin{subfigure}[t]{0.495\textwidth}
\centering
\fieldblock{All_copper}{stream}{1}
\caption{Stream function $\psi$}
\label{fig:copper_psi}
\end{subfigure}
\caption{Normalized Temperature and Stream function for the $3\%$ Cu nanofluid. Rows show the COMSOL reference, the SAGE-PINN prediction, and the Error and Columns correspond to $u_0^* = 10^{-4},\ 5\times10^{-4},\ 5\times10^{-3},\ 2\times10^{-2}$~m/s ($Re = 0.52,\ 2.58,\ 25.8,\ 103.2$).}
\label{fig:copper_thermal}
\end{figure}
\FloatBarrier

The normalized temperature field shows the same qualitative evolution seen in the previous cases: at low $Pe$, thermal diffusion spreads the heating over much of the cross-section, whereas at high $Pe$ the heated region is compressed into a thin downstream layer adjacent to the wall. SAGE-PINN captures this transition well at the first three representative velocities, with relative $L_2$ errors of $10.7\%$, $3.6\%$, and $1.8\%$. At $u_0^* = 2\times10^{-2}$~m/s, however, the temperature error rises to $33.6\%$, although this is still the smallest high-$Pe$ thermal error among the five fluids. The stream-function field, which carries the hard flux constraint, is reproduced closely in all four cases as shown in Fig.~\ref{fig:copper_thermal}).

\subsection{Derived Wall and Throat Quantities}
\label{sec:derived_quantities}
Field-wise error norms provide a direct measure of the agreement between the SAGE-PINN and COMSOL solutions, but many quantities of practical interest in stenosed-flow problems are derived from local extrema, cross-sectional averages, pressure differences, or wall-normal gradients. These quantities provide a complementary assessment of the solution quality and, in the case of gradient-based measures, constitute a more stringent test of the reconstructed fields. Accordingly, we examine the throat axial velocity, stenosis-averaged friction factor and pressure-drop coefficient, and wall heat-transfer characteristics for all $40$ cases. Each quantity is computed from the SAGE-PINN solution and compared with the corresponding COMSOL reference value. The discussion focuses on the extent to which the derived quantities reproduce the reference trends across the complete parameter sweep and on the regimes in which the endpoint inaccuracies identified in the field-wise comparisons reappear.

\subsubsection{Throat Velocity}
\label{sec:throat_velocity}

\begin{figure}[htbp]
\centering
\begin{subfigure}[t]{0.32\textwidth}
\centering
\includegraphics[width=\linewidth]{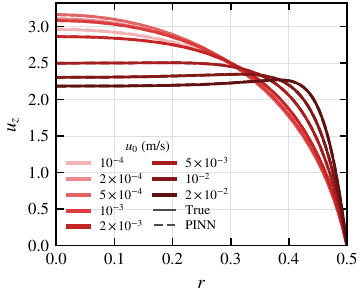}
\caption{Pure Blood}
\label{fig:throat_profile_blood}
\end{subfigure}\hfill
\begin{subfigure}[t]{0.32\textwidth}
\centering
\includegraphics[width=\linewidth]{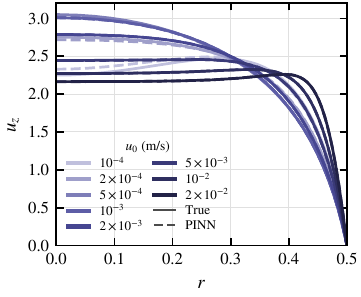}
\caption{3\% Silver}
\label{fig:throat_profile_silver}
\end{subfigure}\hfill
\begin{subfigure}[t]{0.32\textwidth}
\centering
\includegraphics[width=\linewidth]{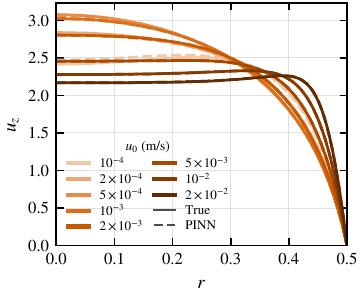}
\caption{3\% Copper}
\label{fig:throat_profile_copper}
\end{subfigure}\\[4pt]
\begin{subfigure}[t]{0.32\textwidth}
\centering
\includegraphics[width=\linewidth]{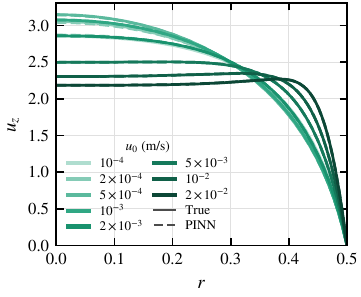}
\caption{3\% Aluminum Oxide}
\label{fig:throat_profile_aluminium}
\end{subfigure}\hspace{0.02\textwidth}
\begin{subfigure}[t]{0.32\textwidth}
\centering
\includegraphics[width=\linewidth]{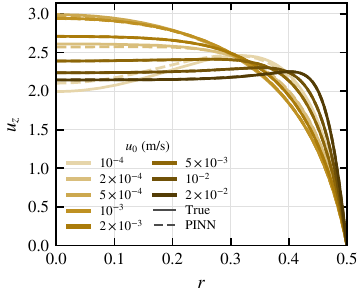}
\caption{3\% Gold}
\label{fig:throat_profile_gold}
\end{subfigure}
\caption{Axial-velocity profiles across the stenosis throat, $u_z(r,z=0)$, for the five fluids at the eight inlet velocities. Each panel corresponds to one fluid, with progressively darker shades indicating increasing inlet velocity. Solid lines denote the COMSOL reference solutions and dashed lines denote the SAGE-PINN predictions.}
\label{fig:throat_profile}
\end{figure}

Figure~\ref{fig:throat_profile} compares the axial-velocity profiles across the stenosis throat for all five fluids and eight inlet velocities. For pure blood, the throat profile evolves progressively with increasing $Re$, with the centreline velocity first increasing and then decreasing as the profile becomes flatter at higher inlet velocities. The nanofluids show a similar progression over the intermediate and high-velocity range, while the largest departures from the blood profiles occur at the lowest inlet velocities, where buoyancy effects are strongest. SAGE-PINN reproduces the reference profiles closely over most of the parameter range, with the most visible discrepancies confined to the lowest-velocity nanofluid cases. Taken together, these trends show that the throat velocity is recovered very accurately once the flow leaves the strongest buoyancy-dominated regime, while the flux constraint keeps the cross-sectional mean essentially invariant across all cases.

\begin{figure}[htbp]
\centering
\includegraphics[width=0.9\textwidth]{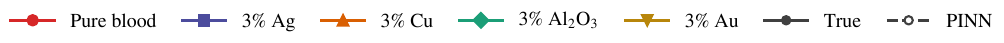}\\[2pt]
\begin{subfigure}[t]{0.49\textwidth}
\centering
\includegraphics[width=\linewidth]{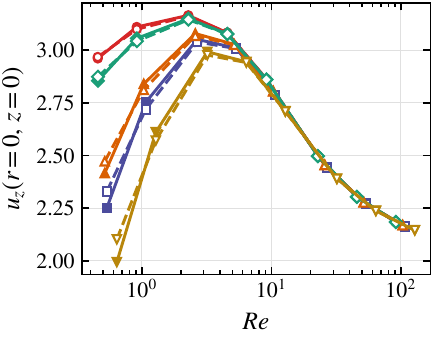}
\caption{Centreline axial velocity}
\label{fig:throat_uz_centre}
\end{subfigure}\hfill
\begin{subfigure}[t]{0.49\textwidth}
\centering
\includegraphics[width=\linewidth]{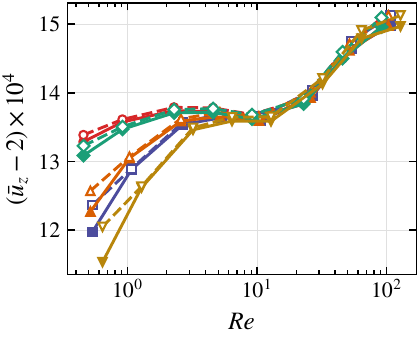}
\caption{Cross-sectional mean axial velocity}
\label{fig:throat_uz_mean}
\end{subfigure}
\caption{(a) Centreline axial velocity at $r=0,\,z=0$ and (b) deviation of the cross-sectional mean axial velocity $\bar{u}_z$ for the five fluids as a function of Reynold number. Filled markers with solid lines denote the COMSOL reference results, while open markers with dashed lines denote the SAGE-PINN predictions.}
\label{fig:throat_velocity}
\end{figure}

The corresponding centreline and cross-sectional mean velocities are summarized in Fig.~\ref{fig:throat_velocity}a,b, respectively. For pure blood, the centreline velocity increases from approximately $2.96$ at $Re=0.46$ to a maximum of about $3.17$ at $Re\approx2.3$, and subsequently decreases as $Re$ increases further. At the lowest velocity, the corresponding centreline values for the Au, Ag, and Cu nanofluids are approximately $1.99$, $2.25$, and $2.42$, respectively, reflecting the stronger buoyancy influence in these cases. SAGE-PINN predicts the centreline throat velocity with an error below $0.52\%$ for $u_0^*\geq5\times10^{-4}$~m/s, while the largest discrepancy, approximately $5.4\%$, occurs for the Au nanofluid at the slowest velocity. The cross-sectional mean axial velocity at the throat remains close to the flux-constrained value $\bar{u}_z=2$ for all cases. The small deviations visible in Fig.~\ref{fig:throat_uz_mean} arise from the numerical radial quadrature used to evaluate the mean velocity rather than from a violation of the continuity or flux constraints. Overall, these results indicate that the SAGE-PINN preserves the principal throat-flow characteristics across the parameter range, with noticeable deterioration only in the most strongly buoyancy-influenced low-velocity cases.

\subsubsection{Friction and Pressure Drop}
\label{sec:friction}
To quantify the hydraulic response of the stenosed region, we consider the stenosis-averaged Darcy friction factor and a pressure-drop coefficient based on the cross-section-averaged pressure. The local Darcy friction factor is evaluated from the wall-tangential velocity gradient as
\begin{equation}
f_D(z)
=
\frac{8\mu_{nf}}
{\rho_{bf} e_0^* u_0^*}\, \tau_w(z),
\label{eq:fd_local}
\end{equation}
where $\tau_w(z)=-\left.\partial_{\mathbf n}\left(\mathbf{u}\cdot\mathbf{t}\right)\right|_{r=h(z)}$ is the dimensionless wall-shear measure. Here, $\mathbf{t}$ and $\mathbf{n}$ denote the local unit tangent and outward unit normal to the wall, respectively, $\mu_{nf}$ is the effective nanofluid viscosity, $\rho_{bf}$ is the base-fluid density, and $e_0^*$ is the healthy-artery radius. The minus sign here is associated with the adopted outward-normal convention. The quantity used for comparison is the average over the stenosed wall,
\begin{equation}
\bar{f}_D
=
\frac{1}{2}
\int_{-1}^{1} f_D(z)\,dz .
\label{eq:fd_average}
\end{equation}
The pressure-based quantity is obtained from the cross-sectional mean pressure,
\begin{equation}
\bar{p}(z)
=
\frac{
\displaystyle\int_0^{h(z)} p(r,z)\,r\,dr
}{
\displaystyle\int_0^{h(z)} r\,dr
},
\label{eq:p_average}
\end{equation}
where the $r$-weighting accounts for the axisymmetric cross-sectional area. The pressure-drop coefficient is then defined as
\begin{equation}
f_p
=
\frac{\bar{p}(z_2)-\bar{p}(z_1)}
{z_2-z_1},
\qquad
z_1=-0.75,\qquad z_2=0.75 ,
\label{eq:fp}
\end{equation}
so that the sign of $f_p$ follows directly from the axial variation of the dimensionless cross-sectional pressure.

\begin{figure}[htbp]
\centering
\includegraphics[width=0.8\textwidth]{Figures/One_D_Plots/legend_fluids.pdf}\\[2pt]
\begin{subfigure}[t]{0.4\textwidth}
\centering
\includegraphics[width=\linewidth]{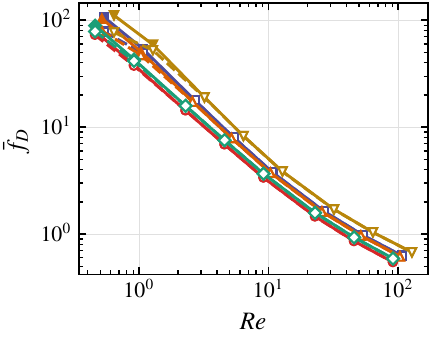}
\caption{Friction factor, stenosis mean}
\label{fig:friction_fD}
\end{subfigure}\hspace{0.5cm}
\begin{subfigure}[t]{0.4\textwidth}
\centering
\includegraphics[width=\linewidth]{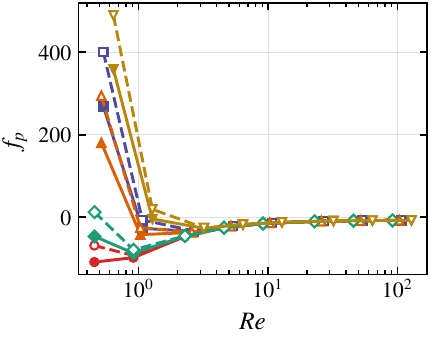}
\caption{Pressure-drop coefficient}
\label{fig:friction_fp}
\end{subfigure}
\caption{(a) Stenosis-averaged friction factor $\bar{f}_D$ over $|z|\leq1$ and (b) pressure-drop coefficient $f_p$ based on the cross-section-averaged pressure over $|z|\leq 0.75$ as a function of Reynolds number. Filled markers with solid lines denote the COMSOL reference results, while open markers with dashed lines denote the SAGE-PINN predictions.}
\label{fig:friction}
\end{figure}

Figure~\ref{fig:friction}a,b, respectively compares $\bar{f}_D$ and $f_p$ for all five fluids over the complete range of Reynolds number. The stenosis-averaged friction factor decreases monotonically with increasing Reynolds number and exhibits an approximately inverse dependence over much of the parameter range. The nanofluid curves lie systematically above that of pure blood, reflecting the combined influence of the effective mixture viscosity and the base-fluid-density normalization adopted in Eq.~\eqref{eq:fd_local}. SAGE-PINN reproduces this trend closely over most of the parameter sweep. For $u_0^*\geq5\times10^{-4}$~m/s, the deviation from the COMSOL reference remains below $0.6\%$, whereas the largest discrepancy occurs for the Au nanofluid at the lowest inlet velocity.

The pressure-drop coefficient exhibits a stronger sensitivity to the low-velocity regime. At $u_0^*=10^{-4}$~m/s, substantial differences occur among the fluids, including a change in sign between the weaker- and stronger-buoyancy cases. Since $f_p$ is defined from the axial variation of the cross-section-averaged pressure Eq.~\eqref{eq:fp}, this change reflects a substantial modification of the pressure balance when buoyancy becomes important relative to the imposed flow. The SAGE-PINN predictions show their largest deviations from the COMSOL reference in this same regime, consistent with the increased momentum-field errors discussed in Section~\ref{sec:copper_results} and Supplementary Section~S2. With increasing inlet velocity, the relative influence of buoyancy decreases and the agreement improves rapidly; for $u_0^*\geq10^{-3}$~m/s, $f_p$ is reproduced to within $1.1\%$. Overall, $\bar{f}_D$ remains accurately recovered over most of the parameter range, whereas $f_p$ provides a more sensitive measure of the deterioration of the momentum solution in the strongly buoyancy-influenced low-velocity regime.

\subsubsection{Wall Heat Transfer}
\label{sec:wall_heat_transfer}
\begin{figure}[htbp]
\centering
\includegraphics[width=0.25\textwidth]{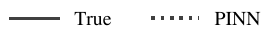}\\[2pt]
\setlength{\tabcolsep}{1pt}
\renewcommand{\arraystretch}{0.6}
\begin{tabular}{cccccc}
 & \footnotesize $u_0^*=$$10^{-4}$ & \footnotesize $u_0^*=$$5\times10^{-4}$ & \footnotesize $u_0^*=$$10^{-3}$ & \footnotesize $u_0^*=$$5\times10^{-3}$ & \footnotesize $u_0^*=$$2\times10^{-2}$ \\
\rotatebox{90}{\footnotesize Pure blood} & \includegraphics[width=0.188\textwidth,valign=c]{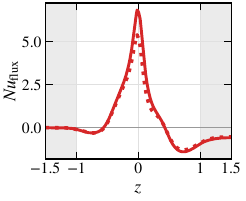} & \includegraphics[width=0.188\textwidth,valign=c]{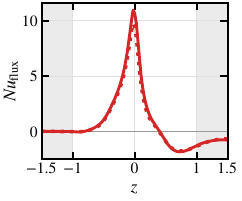} & \includegraphics[width=0.188\textwidth,valign=c]{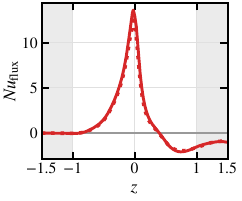} & \includegraphics[width=0.188\textwidth,valign=c]{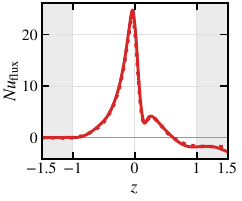} & \includegraphics[width=0.188\textwidth,valign=c]{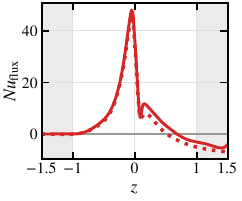} \\
\rotatebox{90}{\footnotesize 3\% Ag} & \includegraphics[width=0.188\textwidth,valign=c]{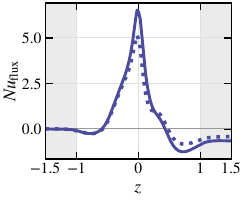} & \includegraphics[width=0.188\textwidth,valign=c]{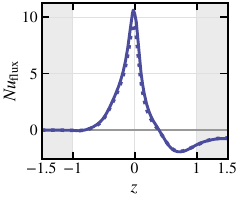} & \includegraphics[width=0.188\textwidth,valign=c]{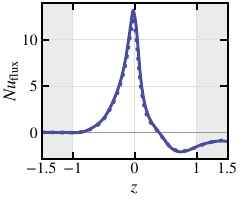} & \includegraphics[width=0.188\textwidth,valign=c]{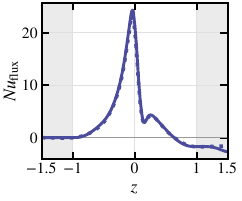} & \includegraphics[width=0.188\textwidth,valign=c]{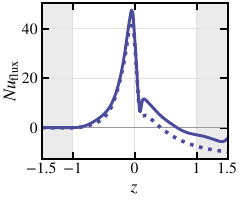} \\
\rotatebox{90}{\footnotesize 3\% Cu} & \includegraphics[width=0.188\textwidth,valign=c]{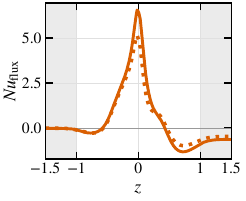} & \includegraphics[width=0.188\textwidth,valign=c]{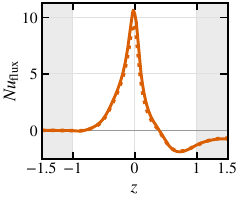} & \includegraphics[width=0.188\textwidth,valign=c]{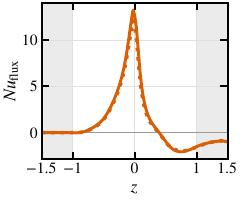} & \includegraphics[width=0.188\textwidth,valign=c]{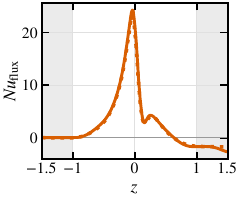} & \includegraphics[width=0.188\textwidth,valign=c]{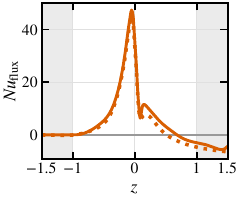} \\
\rotatebox{90}{\footnotesize 3\% Al$_2$O$_3$} & \includegraphics[width=0.188\textwidth,valign=c]{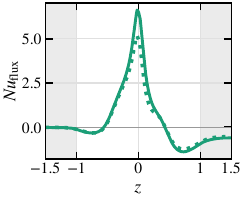} & \includegraphics[width=0.188\textwidth,valign=c]{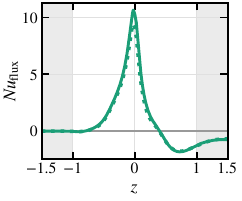} & \includegraphics[width=0.188\textwidth,valign=c]{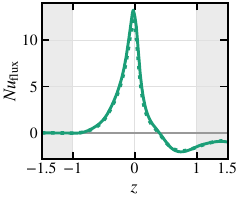} & \includegraphics[width=0.188\textwidth,valign=c]{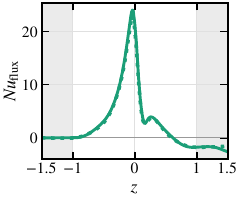} & \includegraphics[width=0.188\textwidth,valign=c]{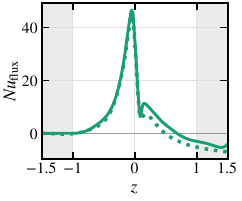} \\
\rotatebox{90}{\footnotesize 3\% Au} & \includegraphics[width=0.188\textwidth,valign=c]{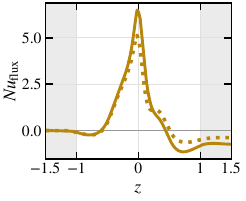} & \includegraphics[width=0.188\textwidth,valign=c]{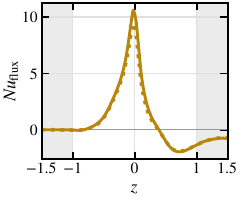} & \includegraphics[width=0.188\textwidth,valign=c]{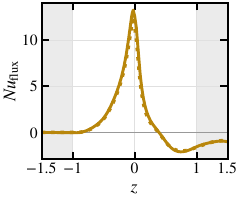} & \includegraphics[width=0.188\textwidth,valign=c]{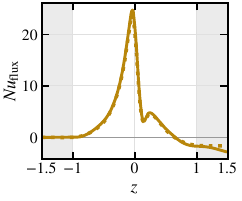} & \includegraphics[width=0.188\textwidth,valign=c]{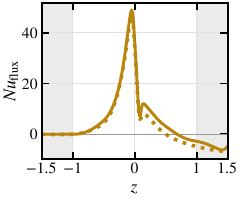} \\
\end{tabular}
\caption{Local wall Nusselt number $Nu_{\mathrm{flux}}(z)$ for the five fluids at five representative inlet velocities. Rows correspond to the working fluids and columns to the inlet velocities. The shaded regions indicate the unheated portions of the wall, $|z|>1$. Solid lines denote the COMSOL reference solutions and dotted lines denote the SAGE-PINN predictions.}
\label{fig:nu_flux_z}
\end{figure}

The wall heat-transfer characteristics are quantified through the local Nusselt number based on the wall-normal temperature gradient. For the wall $r=h(z)$, the outward unit normal is $\mathbf{n}=(1,-h')/\sqrt{1+h'^{2}}$, with $ h'=d_z\,h$. Using the normalized temperature $\theta=T/\Delta T$, the local wall Nusselt number is defined as
\begin{equation}
Nu_{\mathrm{flux}}(z)
=
\left.
\mathbf{n}\cdot\nabla\theta
\right|_{r=h(z)}
=
\left.
\frac{\mathbf{n}\cdot\nabla T}{\Delta T}
\right|_{r=h(z)}.
\label{eq:nu_flux}
\end{equation}
Here, the spatial coordinates are scaled by the healthy-artery radius $e_0^*$ and $\Delta T=(T_S-T_0)/T_0$ is the dimensionless peak wall-temperature difference. With the adopted outward-normal convention, $Nu_{\mathrm{flux}}>0$ corresponds to heat transfer from the heated wall into the fluid, whereas $Nu_{\mathrm{flux}}<0$ indicates heat transfer from the fluid back towards the wall. For comparison across the complete parameter sweep, a stenosis-averaged value is obtained by weighting the local Nusselt number with the axisymmetric wall-area element. Since $ds=\sqrt{1+h'^2}\,dz,$ the averaged quantity is defined as
\begin{equation}
\overline{Nu}_{\mathrm{flux}}
=
\frac{
\displaystyle\int_{-1}^{1}
Nu_{\mathrm{flux}}(z)\,
h(z)\,ds
}{
\displaystyle\int_{-1}^{1}
h(z)\,ds
}.
\label{eq:nu_flux_avg}
\end{equation}
The same quantity is used subsequently in the parity comparison of Section~\ref{sec:parity}.

Figure~\ref{fig:nu_flux_z} shows the spatial variation of the local wall Nusselt number. At the lower inlet velocities, the distribution is comparatively broad, reflecting the stronger role of thermal diffusion in transporting heat away from the heated wall. As the inlet velocity increases, the thermal field becomes increasingly advection dominated and the wall-normal temperature gradients become more localized. Correspondingly, the maximum $Nu_{\mathrm{flux}}$ increases from values of order $7$ at the lowest velocity to approximately $48$ at the highest velocity. A notable feature of the distributions is the change in sign downstream of the heated region. After the fluid is heated while passing through the stenosis, its local temperature can exceed that of the downstream unheated wall, causing the wall-normal temperature gradient to reverse and hence producing $Nu_{\mathrm{flux}}<0$. At the larger inlet velocities, an additional local maximum develops downstream of the throat as the thermal field is convected through the constricted region.

SAGE-PINN reproduces the principal spatial features of the COMSOL wall-heat-transfer distributions across all five fluids, including the increase in peak heat transfer with inlet velocity, the downstream sign reversal, and the emergence of the secondary maximum. The largest visible discrepancies occur near regions of strong wall-normal temperature gradients, particularly at the lowest and highest velocities. At the high-$Pe$ end of the sweep, these deviations are consistent with the thin downstream thermal layer identified in the field-wise comparisons of Section~\ref{sec:copper_results} and Supplementary Section~S2. Because $Nu_{\mathrm{flux}}$ depends directly on a spatial derivative of the temperature field, it is more sensitive to local approximation errors than the temperature itself. The larger discrepancies in the wall Nusselt number relative to the field-wise temperature errors of Fig.~\ref{fig:errors_bar} are therefore consistent with the greater sensitivity of this gradient-based quantity, particularly near the two ends of the parameter sweep.

\subsubsection{Parity of Derived Quantities}
\label{sec:parity}
\begin{figure}[htbp]
\centering
\includegraphics[width=0.9\textwidth]{Figures/One_D_Plots/legend_fluids.pdf}\\[2pt]
\begin{subfigure}[t]{0.47\textwidth}
\centering
\includegraphics[width=\linewidth]{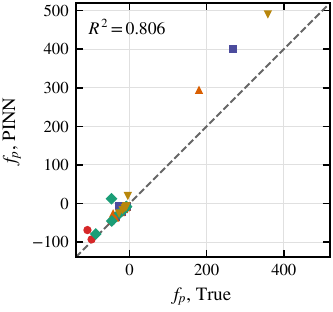}
\caption{Pressure-drop coefficient $f_p$}
\label{fig:parity_fp}
\end{subfigure}
\begin{subfigure}[t]{0.45\textwidth}
\centering
\includegraphics[width=\linewidth]{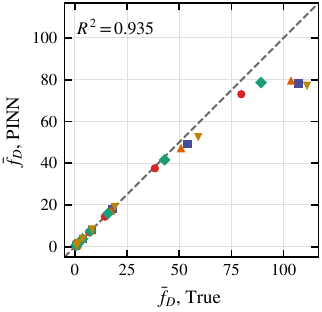}
\caption{Friction factor $\bar{f}_D$}
\label{fig:parity_fD}
\end{subfigure}\\[4pt]
\begin{subfigure}[t]{0.45\textwidth}
\centering
\includegraphics[width=\linewidth]{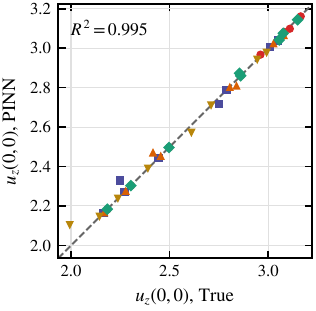}
\caption{Throat Centreline velocity $u_z(0, 0)$}
\label{fig:parity_uz_centre}
\end{subfigure}
\begin{subfigure}[t]{0.45\textwidth}
\centering
\includegraphics[width=\linewidth]{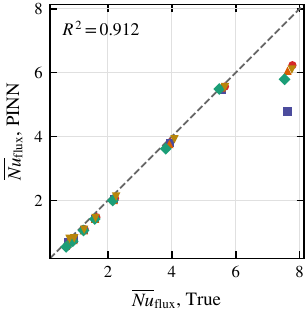}
\caption{Nusselt number $\overline{Nu}_{\mathrm{flux}}$}
\label{fig:parity_NuD}
\end{subfigure}
\caption{Parity of derived scalar quantities for all $40$ cases: SAGE-PINN prediction against the COMSOL value. The dashed line is $y=x$, and $R^2$ is the coefficient of determination of Eq.~\eqref{eq:r2}.}
\label{fig:parity}
\end{figure}

To provide an overall assessment of the derived quantities across the complete parameter sweep, Fig.~\ref{fig:parity} presents parity plots for the throat centreline velocity $u_z(0,0)$, stenosis-averaged friction factor $\bar{f}_D$, pressure-drop coefficient $f_p$, and stenosis-averaged Nusselt number $\overline{Nu}_{\mathrm{flux}}$. For each of the $40$ cases, the SAGE-PINN prediction $\hat{y}_i$ is plotted against the corresponding COMSOL reference value $y_i$, so that exact agreement lies on the line $y=x$.
\FloatBarrier
The overall agreement is quantified using
\begin{equation}
R^2
=
1-
\frac{
\displaystyle\sum_{i=1}^{40}
\left(\hat{y}_i-y_i\right)^2
}{
\displaystyle\sum_{i=1}^{40}
\left(y_i-\bar{y}\right)^2
},
\qquad
\bar{y}
=
\frac{1}{40}
\sum_{i=1}^{40} y_i ,
\label{eq:r2}
\end{equation}
where $R^2=1$ corresponds to exact agreement and $R^2\leq0$ indicates that the predictions perform no better, in this global measure, than using the mean reference value. The throat centreline velocity shows the strongest overall agreement, with $R^2=0.995$, followed by the friction factor ($R^2=0.935$) and the stenosis-averaged Nusselt number ($R^2=0.912$). The pressure-drop coefficient exhibits the largest scatter, with $R^2=0.806$, consistent with its greater sensitivity to the low-velocity momentum solution. The most pronounced departures from the diagonal in $u_z(0,0)$, $\bar{f}_D$, and $f_p$ occur primarily for the strongly buoyancy-influenced low-velocity nanofluid cases, whereas the largest deviation in $\overline{Nu}_{\mathrm{flux}}$ occurs at the high-$Pe$ end of the sweep. Taken together, the parity plots confirm that the derived quantities are reproduced accurately over the central parameter range, while the deviations remain concentrated in the same limiting regimes identified independently from the field-wise comparisons.

\subsection{Training Behaviour}
\label{sec:training_behaviour}
\subsubsection{Training Loss}
\label{sec:training_loss}
\begin{figure}[htbp]
\centering
\includegraphics[width=0.85\textwidth]{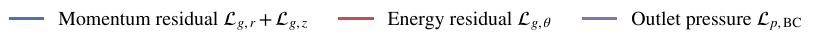}\\[2pt]
\begin{subfigure}[t]{0.49\textwidth}
\centering
\includegraphics[width=\linewidth]{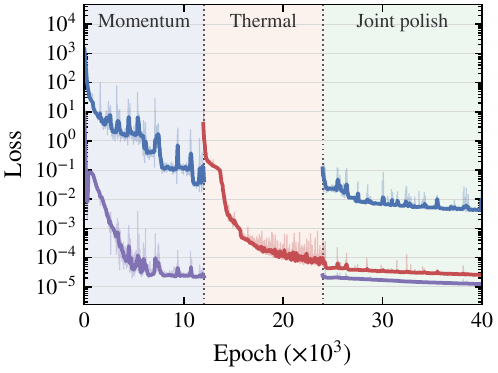}
\caption{3\% Cu, $u_0^*=2\times10^{-3}$~m/s ($Re\approx10.3$)}
\label{fig:loss_cu}
\end{subfigure}\hfill
\begin{subfigure}[t]{0.49\textwidth}
\centering
\includegraphics[width=\linewidth]{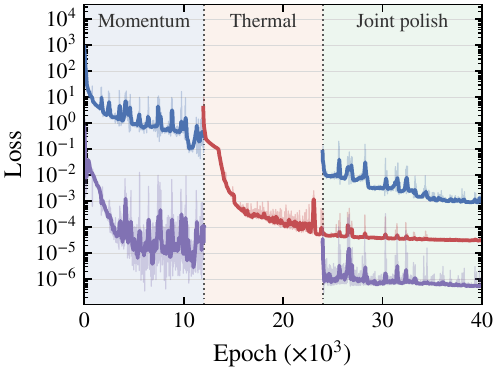}
\caption{3\% Au, $u_0^*=5\times10^{-3}$~m/s ($Re\approx32$)}
\label{fig:loss_au}
\end{subfigure}
\caption{Training histories of SAGE-PINN for two representative runs. Faint lines denote the raw loss values and bold lines the corresponding rolling medians. Each loss component is shown only during the stages in which it is active in the optimization objective. The temporary increase in the momentum residual at the beginning of the joint-polish stage coincides with the Richardson-number continuation from $Ri_{\mathrm{floor}}$ towards the target $Ri$.}
\label{fig:loss_curves}
\end{figure}

Figure~\ref{fig:loss_curves} shows the Adam-stage training histories for two representative cases. The same stage-wise structure is observed across the complete set of $40$ simulations: the momentum--pressure trunk is first optimized independently, the thermal trunk is then trained with the velocity field frozen, and both trunks are subsequently updated together during the joint-polish stage. For the $3\%$ Cu case at $u_0^*=2\times10^{-3}$~m/s (Fig.~\ref{fig:loss_cu}), the combined normalized momentum residual, $\mathcal{L}_{g,r}+\mathcal{L}_{g,z}$, decreases from approximately $1.7\times10^{3}$ to $6.5\times10^{-2}$ during the momentum stage, corresponding to a reduction of about $4.4$ orders of magnitude. Over the same stage, the outlet-pressure loss decreases from approximately $4.0$ to $2.3\times10^{-5}$. During the thermal stage, with the momentum trunk frozen, the normalized energy residual decreases from about $2.6$ to $7.3\times10^{-5}$ while $Pe_{\mathrm{eff}}$ is progressively increased to its target value.

At the beginning of the joint-polish stage, the momentum residual increases temporarily as the effective Richardson number is ramped from $Ri_{\mathrm{floor}}$ towards the target $Ri$ and both network trunks are allowed to adjust simultaneously. The residual then decreases again, reaching approximately $4.5\times10^{-3}$ by the end of the Adam training, while the energy residual decreases further to about $2.6\times10^{-5}$. A similar evolution is observed for the $3\%$ Au case at $u_0^*=5\times10^{-3}$~m/s (Fig.~\ref{fig:loss_au}), representative of the high-accuracy intermediate regime. The momentum residual decreases from approximately $1.3\times10^{3}$ to below $10^{-3}$ by the end of the joint-polish stage, while the energy residual decreases from about $2.7$ to $3.1\times10^{-5}$ and the outlet-pressure loss reaches $5.5\times10^{-7}$. The short-scale fluctuations in the raw loss traces arise primarily from mini-batch resampling, whereas the rolling median reveals the underlying convergence trend. The learning-rate scheduler acts separately by reducing the learning rate when the monitored objective remains on a sustained plateau. These histories show that the staged optimization produces systematic multi-order reductions in both momentum and thermal residuals without case-specific changes to the training hyperparameters.

\subsubsection{Effect of the Continuations}
\label{sec:pe_effect}
The continuation strategies are intended to guide the optimization from better-conditioned auxiliary problems towards the target thermal and momentum operators. Their behaviour is illustrated in Figs.~\ref{fig:pe_ramp} and~\ref{fig:ri_ramp}. Figure~\ref{fig:pe_ramp} shows the P\'eclet continuation for pure blood at $u_0^*=2\times10^{-2}$~m/s, for which $Pe=2329.6$. During the first half of the thermal stage, $Pe_{\mathrm{eff}}$ is increased geometrically from $Pe_0=10$ to the target value. Over the same interval, the normalized energy residual decreases by several orders of magnitude, allowing the thermal trunk to approach the strongly advection-dominated target problem from a smoother, more diffusion-influenced auxiliary solution. Once $Pe_{\mathrm{eff}}$ reaches the prescribed value, the remainder of the thermal stage proceeds at the true $Pe$. The continuation is inactive when $Pe\leq1.5Pe_0$, for which the auxiliary and target operators are already sufficiently close.

The Richardson continuation plays the analogous role for the momentum equations at the low-velocity, strongly buoyancy-influenced end of the parameter range. Figure~\ref{fig:ri_ramp} shows the $3\%$ Au case at $u_0^*=10^{-4}$~m/s, where the target Richardson number is $Ri=2.84\times10^4$. During the momentum and thermal stages, the effective Richardson number is held at $Ri_{\mathrm{floor}}=\min\left(Ri,\,1/\Delta T\,Re\right),$ so that, whenever the reduction is active, $Ri_{\mathrm{floor}}\Delta T Re=1$. This limits the characteristic buoyancy-to-viscous coefficient ratio in the auxiliary momentum problem while the thermal field is not yet fully established. During the first half of the joint-polish stage, $Ri_{\mathrm{eff}}$ is then increased geometrically from $Ri_{\mathrm{floor}}$ to the target $Ri$. The temporary rise in the momentum residual observed during this interval reflects the progressive restoration of the full buoyancy coupling and the simultaneous readjustment of both network trunks. After the target value is reached, the second half of the joint stage, as well as the subsequent L-BFGS refinement, is performed at the physical value of $Ri$. The two continuations are therefore activated only when required by the corresponding transport regime, while the same training schedule is retained across all $40$ cases. In the present parameter sweep, the P\'eclet and Richardson continuations are active in $35$ and $32$ cases, respectively.

\subsubsection{Mass conservation}
\label{sec:mass_conservation}
\begin{table}[htbp]
\centering
\setlength{\tabcolsep}{15pt}
\renewcommand{\arraystretch}{1.4}
\begin{tabular}{lcc}
\toprule
Fluid & Mean Continuity Residual & Max Relative Flux Drift \\
\midrule
Pure blood & $2.7$--$4.4\times10^{-7}$ & $2.7$--$5.2\times10^{-5}$ \\
$3\%$ Ag & $2.6$--$4.4\times10^{-7}$ & $2.8$--$5.5\times10^{-5}$ \\
$3\%$ Cu & $2.8$--$4.7\times10^{-7}$ & $2.8$--$5.4\times10^{-5}$ \\
$3\%$ Al$_2$O$_3$ & $2.6$--$4.4\times10^{-7}$ & $2.7$--$5.2\times10^{-5}$ \\
$3\%$ Au & $2.7$--$4.9\times10^{-7}$ & $2.9$--$6.0\times10^{-5}$ \\
\bottomrule
\end{tabular}
\caption{Mass-conservation diagnostics per fluid over the eight inlet velocities: Range of the Mean Absolute Continuity Residual and of the Maximum Relative Flux Drift obtained by Radial Quadrature.}
\label{tab:mass}
\end{table}

Because the velocity field is reconstructed from the axisymmetric stream function, continuity is satisfied identically by construction and is not included in the training loss. Across all $40$ cases, the mean absolute continuity residual remains between $2.6\times10^{-7}$ and $4.9\times10^{-7}$, with a maximum pointwise value below $1.1\times10^{-5}$, while the wall stream-function drift is approximately $6.0\times10^{-8}$. Independent radial quadrature gives a maximum relative flux drift of $2.7\times10^{-5}$--$6.0\times10^{-5}$ (cf. Table~\ref{tab:mass}), reflecting numerical integration and finite-precision evaluation rather than a violation of the underlying flux constraint. These results confirm that mass conservation is enforced structurally across all fluids and inlet velocities without requiring an additional continuity-penalty term.

\subsection{Discussion}
\label{sec:discussion}
\begin{figure}[htbp]
\centering
\includegraphics[width=0.7\textwidth]{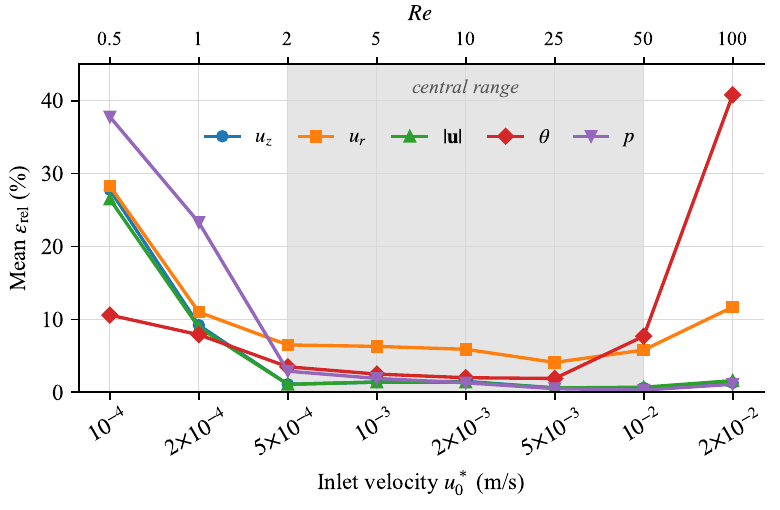}
\caption{Accuracy envelope of SAGE-PINN: relative $L_2$ error (\%) of each field, averaged over the five fluids, as a function of the inlet velocity (bottom axis) and the corresponding approximate Reynolds number (top axis). Pressure errors are offset-free. The shaded region marks the central range $5\times10^{-4}\leq u_0^*\leq10^{-2}$~m/s. The corresponding values, together with the mean and maximum errors over the central range, are listed in Supplementary Table~S2.}
\label{fig:envelope}
\end{figure}

Figure~\ref{fig:envelope} summarizes the accuracy of SAGE-PINN across the complete parameter sweep. The most consistent performance is obtained over the central range,
$5\times10^{-4}\leq u_0^*\leq10^{-2}$~m/s, comprising $25$ cases. Within this range, the mean relative $L_2$ errors are $1.05\%$ for $u_z$, $1.05\%$ for $|\mathbf{u}|$, $5.71\%$ for $u_r$, $3.53\%$ for $\theta$, and $1.41\%$ for the offset-free pressure; the corresponding maximum errors remain below $1.6\%$, $1.6\%$, $7.4\%$, $9.1\%$, and $4.2\%$, respectively. The larger relative error in $u_r$ should be interpreted in view of its substantially smaller magnitude compared with $u_z$, whereas the velocity-magnitude error remains close to that of the axial component.

The deterioration at the two ends of the sweep occurs in different fields. At $u_0^*=10^{-4}$~m/s, the axial-velocity error increases systematically from blood to Al$_2$O$_3$, Cu, Ag, and Au as the characteristic buoyancy-to-viscous coefficient ratio $Ri\,\Delta T\,Re$ increases from approximately $60$ to $294$, while the temperature error remains nearly unchanged at $10.3$--$10.8\%$. The corresponding discrepancies are concentrated predominantly in the core region and are accompanied by larger pressure errors. The ablation study reported in Supplementary Section~S4 (Supplementary Table~S1) indicates that neither the detached buoyancy-gradient path nor the P\'eclet-weighted outlet treatment is responsible for this low-velocity deterioration. At the opposite limit, $u_0^*=2\times10^{-2}$~m/s ($Pe\approx2110$--$2330$), the velocity and offset-free pressure remain comparatively accurate, whereas the temperature error rises to $33.6$--$57.3\%$. These errors are concentrated mainly in the thin downstream thermal layer, indicating that the high-$Pe$ thermal field is substantially more difficult for the present smooth network representation and collocation strategy to resolve.

Despite these endpoint limitations, all $40$ cases were obtained using the same network architecture, collocation strategy, loss weights, optimizer settings, and training schedule, with the operating-point-dependent quantities determined directly from $(Re,Ri,Pe)$. Mass conservation remains structurally enforced throughout the sweep, and the derived throat, friction, pressure-drop, and heat-transfer quantities reproduce the COMSOL trends over the regime in which the underlying fields are accurately resolved. The results therefore define a clear accuracy envelope for the present formulation: robust performance over the intermediate regime, with further improvement required for strongly buoyancy-influenced slow flow and highly advection-dominated thermal transport.

\section{Limitations and Future Work}
\label{sec:limitations_future}
The present results identify two principal accuracy limitations. At the lowest inlet velocities, strong buoyancy coupling leads to substantial errors in the momentum fields for the Cu, Ag, and Au nanofluids, whereas at $Pe\gtrsim2000$ the temperature and wall-heat-transfer predictions deteriorate because of the increasingly thin downstream thermal layer. Gradient-based quantities such as the Nusselt number are consequently more sensitive than the underlying temperature field. The radial velocity also exhibits larger relative errors than $u_z$ because of its considerably smaller magnitude, while pressure remains subject to the softly imposed reference level at the outlet. Beyond these numerical limitations, the present study is restricted to steady, laminar, axisymmetric flow through a single bell-shaped stenosis with one heating configuration. The nanofluid is represented as a homogeneous single-phase mixture with constant properties and fixed volume fraction, neglecting particle migration, Brownian motion, thermophoresis, and temperature-dependent material properties.

Future developments should therefore focus on the two limiting transport regimes and on extending the generality of the framework. High-$Pe$ thermal transport may benefit from boundary-layer-adapted collocation, wall-normal coordinate stretching, and improved continuation strategies, while the strongly buoyancy-influenced regime may require more effective coupled optimization of the momentum and thermal fields. Treating $(Re,Ri,Pe)$ as additional network inputs would provide a natural route towards a parametric solver spanning multiple operating conditions \cite{naderibeni2024parametric}. The geometry-encoded formulation can also be extended to other stenosis shapes, severities, and related axisymmetric configurations, although each extension will require corresponding modification of the geometric representation and validation. Further directions include unsteady and pulsatile flow, additional nanofluid transport physics, temperature-dependent properties, and inverse problems based on sparse measurements \cite{arzani2021nearwall,zhang2025ces,yang2025reconstruction,wang2025gan}. Finally, the connection between the $s=r^2$ transformation and weighted axisymmetric formulations \cite{nowakowski2023weighted}, together with lower-derivative-order residual approaches \cite{su_fvpin}, provides a useful basis for future mathematical error analysis and computational acceleration.

\section{Conclusion}
\label{sec:conclusion}
This work introduced SAGE-PINN, a singularity-free, axisymmetric, geometry-encoded physics-informed neural network for coupled flow and heat transfer in cylindrical coordinates. To the best of our knowledge, this is the first PINN framework developed specifically for this class of axisymmetric multiphysics problems. The method is built around three central ideas: a parity-preserving coordinate transformation, $s=r^2$, combined with a modified stream-function representation that removes the axis singularity from the trained residuals and enforces mass conservation identically; boundary-fitted analytical ans\"atze that satisfy the axis, wall, inlet, flux, and wall-heating constraints by construction; and a staged training strategy with residual normalisation and P\'eclet- and Richardson-number continuation to stabilize the strongly coupled momentum-thermal optimization across widely varying transport regimes.

The framework was assessed over a broad range of stenosed-flow conditions and an extensive selection of Reynolds number, Richardson number, and P\'eclet number, using a single network and training configuration without case-specific tuning. Over the central operating range, the method reproduces the principal flow and thermal fields with good accuracy, while the derived throat, friction, pressure-drop, and wall-heat-transfer quantities follow the COMSOL reference trends. More importantly, the cylindrical validation against a conventional PINN highlights the benefit of the proposed formulation: for pure blood at a specific inlet velocity $5\times10^{-3}$~m/s, the vanilla PINN gives relative $L_2$ errors of $54.5\%$, $83.5\%$, and $22.7\%$ in axial velocity, radial velocity, and temperature, respectively, whereas SAGE-PINN reduces these to $0.67\%$, $4.47\%$, and $1.74\%$. The largest remaining discrepancies occur only in the limiting regimes of strongly buoyancy-influenced slow flow and highly advection-dominated thermal transport.

More broadly, the study demonstrates that many of the difficulties encountered by PINNs in cylindrical multiphysics problems can be alleviated by embedding coordinate regularity, conservation structure, and boundary physics directly into the solution representation rather than relying exclusively on penalty-based learning. SAGE-PINN therefore provides a practical basis for further developments in axisymmetric flow and heat-transfer problems, including more complex geometries, parametric formulations, and inverse problems.

\section*{Declaration of Generative AI and AI-Assisted Technologies in the Manuscript preparation process}
During the preparation of this work, the authors used large language models solely to improve the readability and language of the manuscript and to assist with routine code debugging. The research ideas, methodology, mathematical formulation, implementation, simulations, analysis, interpretation of the results, and conclusions were developed by the authors. The authors take full responsibility for the content of the article.

\section*{Data Availability}
The trained models, source code for training and evaluation, and the COMSOL reference data supporting the findings of this study are available from the corresponding author upon reasonable request.

\section*{CRediT authorship contribution statement}
\textbf{Raj Maurya:} Methodology, Software, Formal analysis, Investigation, Validation, Visualization, Writing -- original draft. \textbf{Joyprakash Akhuli:} Investigation, Validation, Visualization, Writing -- original draft. \textbf{Doyel Pandey:} Conceptualization, Formal analysis, Visualization, Supervision, Writing -- original draft.

\section*{Acknowledgements}
D.P. acknowledges financial support from the Anusandhan National Research Foundation (ANRF), Government of India, under Grant No.~ANRF/ECRG/2025/002630/PMS.

\bibliographystyle{elsarticle-num} 
\bibliography{Bib_file}

\appendix

\section{Stream-Function Transformation}
\label{sec:appendix_stream_transform}
Let $A = \varphi_{z}$ and $B = u_z = 2\varphi + 2s\,\varphi_{s}$ be as in Section~\ref{sec:modified_equation}. This appendix proves that the ansatz $\psi = r^{2}\varphi(s,z)$ satisfies continuity identically, working directly in the transformed variables $(s,z)$ with the chain rule $\partial/\partial r = 2r\,\partial/\partial s$.

First,
\begin{equation}
r u_r = -\big(r\big)^{2} A = -sA.
\end{equation}
Then,
\medskip
\begin{equation}
\begin{aligned}
\frac{\partial (ru_r)}{\partial r} = -\frac{\partial (sA)}{\partial r} = -\left(\frac{\partial s}{\partial r}A + s\frac{\partial A}{\partial r}\right) = -\big(2rA + s(2r)A_s\big) = -2r\big(A + sA_s\big)
\end{aligned}
\label{eq:continuity_derivation}
\end{equation}
so that,
\medskip
\begin{equation}
\frac{1}{r}\frac{\partial (ru_r)}{\partial r} = -2\big(A + sA_s\big) = -2A - 2sA_s.
\label{eq:appA_term1}
\end{equation}
Second, differentiating $u_z=B=2\,\varphi+2s\,\varphi_s$ with respect to $z$ at fixed $s$,
\begin{equation}
\frac{\partial u_z}{\partial z} = B_z = 2\,\varphi_z + 2s\,(\varphi_s)_z = 2A + 2sA_s,
\label{eq:appA_term2}
\end{equation}
using equality of mixed partials, $(\varphi_s)_z = (\varphi_z)_s = A_s$. Adding Eqs.~\eqref{eq:appA_term1} and \eqref{eq:appA_term2}:
\begin{equation}
\frac{1}{r}\frac{\partial (ru_r)}{\partial r} + \frac{\partial u_z}{\partial z}
= \big(-2A - 2sA_s\big) + \big(2A + 2sA_s\big) = 0,
\end{equation}
confirming continuity is satisfied identically for ansatz used, for any $\varphi(s,z)$. The identity, $B_z = 2A + 2sA_s$, is used repeatedly in further derivations. \\
\textit{Remark.} Equations~\eqref{eq:appA_term1} and~\eqref{eq:appA_term2} are equal in magnitude and opposite in sign for any sufficiently smooth $\varphi(s,z)$. Their exact cancellation confirms that the transformed stream-function representation preserves the axisymmetric continuity identity.

\section{Transformation to Modified Equations}
\label{sec:modified_eqns_derivations}
This appendix derives the residual expressions of Eqs.~\eqref{eq:res_r}--\eqref{eq:res_theta} from the dimensionless governing equations, Eqs.~\eqref{eq:zmom_nd}, \eqref{eq:rmom_nd}, and \eqref{eq:energy_nd}, using the substitutions $s = r^{2}$, $u_r = -rA$, and $u_z = B$ introduced in \ref{sec:appendix_stream_transform}. Throughout, the chain rule $\partial/\partial r = 2r\,\partial/\partial s$ is applied, together with the exact identity
\begin{equation}
B_z = 2A + 2sA_s,
\label{eq:appB_identity1}
\end{equation}
which follows from Eq.~\eqref{eq:appA_term2}, and its further derivatives
\begin{equation}
B_{sz} = 4A_s + 2sA_{ss}, \qquad B_{zz} = 2A_z + 2sA_{sz},
\label{eq:appB_identity2}
\end{equation}
obtained by differentiating Eq.~\eqref{eq:appB_identity1} with respect to $s$ and $z$, respectively. \\

\textbf{Reduction of the viscous terms.} The momentum equations~\eqref{eq:rmom_nd} and \eqref{eq:zmom_nd} are written in stress-divergence form. For an incompressible field with constant viscosity they reduce to the vector-Laplacian form used below. Writing $D = \partial u_r/\partial r + u_r/r = (1/r)\,\partial(ru_r)/\partial r$, continuity gives $D = -\partial u_z/\partial z$. In the radial equation, expanding the bracket gives
\begin{equation*}
\frac{1}{r}\frac{\partial}{\partial r}\!\left(2r\frac{\partial u_r}{\partial r}\right) + \frac{\partial^{2} u_r}{\partial z^{2}} + \frac{\partial}{\partial r}\!\left(\frac{\partial u_z}{\partial z}\right) - \frac{2u_r}{r^{2}}
= \frac{\partial}{\partial r}\!\left(\frac{1}{r}\frac{\partial (ru_r)}{\partial r}\right) + \frac{\partial^{2} u_r}{\partial z^{2}} + \frac{\partial}{\partial r}\big(D + \partial_z u_z\big),
\end{equation*}
and in the axial equation
\begin{equation*}
\frac{1}{r}\frac{\partial}{\partial r}\!\left(r\frac{\partial u_r}{\partial z} + r\frac{\partial u_z}{\partial r}\right) + 2\frac{\partial^{2} u_z}{\partial z^{2}}
= \frac{1}{r}\frac{\partial}{\partial r}\!\left(r\frac{\partial u_z}{\partial r}\right) + \frac{\partial^{2} u_z}{\partial z^{2}} + \frac{\partial}{\partial z}\big(D + \partial_z u_z\big).
\end{equation*}
The last term vanishes in both cases by continuity, which the stream-function ansatz satisfies identically (\ref{sec:appendix_stream_transform}). The viscous operators are therefore the vector Laplacians used in Eqs.~\eqref{eq:appB_z_visc} and \eqref{eq:appB_r_visc}. In each case the residual is assembled by substituting the inertial, pressure, and viscous contributions below into the corresponding governing equation and moving all terms to one side, so that $\mathcal{R}$ vanishes at the exact solution.

\subsection{Axial-Momentum Residual}
\label{sec:axial_appendix}

\textbf{Inertial terms.} Using $u_r = -rA$ and $\partial u_z/\partial r = 2rB_s$ (chain rule),
\begin{equation}
\begin{aligned}
u_r\frac{\partial u_z}{\partial r} + u_z\frac{\partial u_z}{\partial z} = (-rA)(2rB_s) + BB_z = -2r^2 A B_s + BB_z.
\end{aligned}
\label{eq:appB_z_inertial}
\end{equation}

\medskip
\textbf{Viscous (Laplacian) term.} With $r\,\partial u_z/\partial r = 2sB_s$,
\begin{equation}
\begin{aligned}
\frac{\partial}{\partial r}\!\left(r\frac{\partial u_z}{\partial r}\right) = \frac{\partial (2sB_s)}{\partial r} = 2r(2B_s) + 2r(2sB_{ss}) = 4rB_s + 4rsB_{ss},
\end{aligned}
\label{eq:diff_uz}
\end{equation}
so that
\begin{equation}
\frac{1}{r}\frac{\partial}{\partial r}\!\left(r\frac{\partial u_z}{\partial r}\right) = 4B_s + 4sB_{ss}, \qquad
\frac{\partial^{2}u_z}{\partial z^{2}} = B_{zz},
\end{equation}
and the full axial viscous operator becomes
\begin{equation}
\frac{1}{r}\frac{\partial}{\partial r}\!\left(r\frac{\partial u_z}{\partial r}\right)
+ \frac{\partial^{2}u_z}{\partial z^{2}}
= 4B_s + 4sB_{ss} + B_{zz}.
\label{eq:appB_z_visc}
\end{equation}

\medskip
\textbf{Assembled residual.} Substituting Eqs.~\eqref{eq:appB_z_inertial} and \eqref{eq:appB_z_visc} into the axial-momentum equation, Eq.~\eqref{eq:zmom_nd}, and moving all terms to one side gives
\begin{equation}
\mathcal{R}_z
= BB_z - 2sAB_s + p_z
- \frac{1}{Re}\bigl(4B_s + 4sB_{ss} + B_{zz}\bigr)
- Ri\,T = 0,
\label{eq:appB_z_final}
\end{equation}
where $p_z = \partial p/\partial z$. This residual is algebraically identical to Eq.~\eqref{eq:res_z}: writing $E = 2B_s - A_z$, one obtains $2(E + sE_s) + 2B_{zz} = 4B_s + 4sB_{ss} - 2A_z - 2sA_{sz} + 2B_{zz}$; since $B_{zz} = 2A_z + 2sA_{sz}$ (Eq.~\ref{eq:appB_identity1}, differentiated), the terms $-2A_z - 2sA_{sz}$ cancel against one copy of $B_{zz}$, leaving exactly $4B_s + 4sB_{ss} + B_{zz}$, as in Eq.~\eqref{eq:appB_z_visc}.

\subsection{Radial-Momentum Residual}
\label{sec:radial_appendix}

The radial-momentum equation, Eq.~\eqref{eq:rmom_nd}, contains explicit $1/r$ and $1/r^2$ factors that are singular at the axis $r=0$. To regularise the residual, the entire equation is divided by $r$ before assembly; because $u_r = -rA$ vanishes linearly at the axis, this operation is well defined and introduces no new singularity.

\medskip
\textbf{Inertial terms.} Using $\partial u_r/\partial r = -A - 2sA_s$ (obtained from $u_r = -rA$ by the chain rule) and $\partial u_r/\partial z = -rA_z$,
\begin{equation}
\begin{aligned}
u_r\frac{\partial u_r}{\partial r} + u_z\frac{\partial u_r}{\partial z}
&= (-rA)(-A-2sA_s) + B(-rA_z) \\
&= r\bigl(A^{2} + 2sAA_s - BA_z\bigr),
\end{aligned}
\label{eq:radial_inertial}
\end{equation}
and after division by $r$,
\begin{equation}
\frac{1}{r}\left[u_r\frac{\partial u_r}{\partial r} + u_z\frac{\partial u_r}{\partial z}\right]
= A^{2} + 2sAA_s - BA_z.
\label{eq:appB_r_inertial}
\end{equation}

\medskip
\textbf{Pressure term.} Using $\partial p/\partial r = 2rp_s$,
\begin{equation}
\frac{1}{r}\frac{\partial p}{\partial r} = 2p_s.
\label{eq:appB_r_pressure}
\end{equation}

\medskip
\textbf{Viscous term.} With $ru_r = -sA$,
\begin{equation}
\begin{aligned}
\frac{\partial (ru_r)}{\partial r} &= -2r\bigl(A + sA_s\bigr) \\
&\Longrightarrow \quad
\frac{1}{r}\frac{\partial (ru_r)}{\partial r} = -2\bigl(A + sA_s\bigr),
\end{aligned}
\label{eq:continuity_divide}
\end{equation}
and differentiating once more with respect to $r$,
\begin{equation}
\begin{aligned}
\frac{\partial}{\partial r}\bigl[-2(A + sA_s)\bigr]
&= -2\frac{\partial A}{\partial r} - 2\frac{\partial (sA_s)}{\partial r} \\
&= -2(2rA_s) - 2\bigl(2rA_s + s(2r)A_{ss}\bigr) \\
&= -8rA_s - 4rsA_{ss},
\end{aligned}
\label{eq:div_continuity}
\end{equation}
so that
\begin{equation}
\frac{\partial}{\partial r}\!\left[\frac{1}{r}\frac{\partial (ru_r)}{\partial r}\right]
= -8A_s - 4sA_{ss}, \qquad
\frac{\partial^{2}u_r}{\partial z^{2}} = -rA_{zz}.
\end{equation}
Dividing the complete viscous operator by $r$ gives
\begin{equation}
\frac{1}{r}\left[\frac{\partial}{\partial r}\!\left(\frac{1}{r}\frac{\partial (ru_r)}{\partial r}\right)
+ \frac{\partial^{2}u_r}{\partial z^{2}}\right]
= -8A_s - 4sA_{ss} - A_{zz}.
\label{eq:appB_r_visc}
\end{equation}

\medskip
\textbf{Assembled residual.} Combining Eqs.~\eqref{eq:appB_r_inertial}--\eqref{eq:appB_r_visc}, the residual of Eq.~\eqref{eq:rmom_nd}, divided by $r$, is
\begin{equation}
\mathcal{R}_r
= A^{2} + 2sAA_s - BA_z + 2p_s
- \frac{1}{Re}\bigl(-8A_s - 4sA_{ss} - A_{zz}\bigr) = 0.
\label{eq:appB_r_final}
\end{equation}
This residual is algebraically identical to Eq.~\eqref{eq:res_r}: writing $C_s = -3A_s - 2sA_{ss}$, one obtains $4(C_s - A_s) = -16A_s - 8sA_{ss}$; adding $2B_{sz} = 8A_s + 4sA_{ss}$ (using $B_{sz} = 4A_s + 2sA_{ss}$) gives $-8A_s - 4sA_{ss}$, and subtracting $A_{zz}$ recovers exactly $-8A_s - 4sA_{ss} - A_{zz}$, as in Eq.~\eqref{eq:appB_r_visc}.

\subsection{Energy Residual}
\label{sec:energy_appendix}
\textbf{Advective terms.} Using $\partial T/\partial r = 2rT_s$,
\begin{equation}
u_r\frac{\partial T}{\partial r} + u_z\frac{\partial T}{\partial z} = (-rA)(2rT_s) + BT_z = -2sAT_s + BT_z.
\label{eq:appB_e_adv}
\end{equation}
Unlike the momentum equations, no division by $r$ is required: the two factors of $r$ arising from $u_r = -rA$ and $\partial T/\partial r = 2rT_s$ combine directly into the coefficient $2s$.

\medskip
\textbf{Diffusive (Laplacian) term.} By the same steps as in Eq.~\eqref{eq:appB_z_visc}, with $T$ in place of $u_z$,
\begin{equation}
\frac{1}{r}\frac{\partial}{\partial r}\!\left(r\frac{\partial T}{\partial r}\right) + \frac{\partial^{2}T}{\partial z^{2}} = 4T_s + 4sT_{ss} + T_{zz}.
\label{eq:appB_e_diff}
\end{equation}

\medskip
\textbf{Assembled residual.} Substituting Eqs.~\eqref{eq:appB_e_adv} and \eqref{eq:appB_e_diff} into the energy equation, Eq.~\eqref{eq:energy_nd}, gives
\begin{equation}
\mathcal{R}_{\theta} = -2sAT_s + BT_z - \frac{1}{Pe}\bigl(4T_s + 4sT_{ss} + T_{zz}\bigr) = 0,
\end{equation}
which is identical to Eq.~\eqref{eq:res_theta}.

\section{Proof of Hard Boundary-Condition Satisfaction}
\label{sec:appendix_hard_bc}
This appendix shows that the flow ansatz, Eq.~\eqref{eq:shi_ansatz}, and the thermal ansatz, Eq.~\eqref{eq:theta_ansatz}, satisfy the axis, wall, and inlet conditions exactly, or to an exponentially small residual. With $\xi = s/S(z)$, the derivatives used below are $\partial\xi/\partial s = 1/S$ and $\partial\xi/\partial z = -\xi S'/S$, where $S' = dS/dz$. The velocities follow from Eq.~\eqref{eq:vel_modified} as $u_z = 2\varphi + 2s\varphi_s$ and $u_r = -r\varphi_z$. \\

\textbf{1. Axis ($r=0$, exact):} Since $u_r = -r\varphi_z$ and $\varphi_z$ is finite, $u_r|_{r=0} = 0$ for any network output. Every other field ($u_z$, $p$, $\theta$) is a smooth function of $s=r^2$, so $\partial(\cdot)/\partial r = 2r\,\partial(\cdot)/\partial s$ vanishes at $r=0$, and the symmetry conditions hold identically. \\

\textbf{2. Wall ($\xi=1$, exact):} Write $\varphi = \varphi_{\mathrm{base}} + \varphi_{\mathrm{NN}}$ as in Eq.~\eqref{eq:shi_ansatz}. For the base term, $\varphi_{\mathrm{base}} = Q(2-\xi)/S$, so
\begin{equation}
\varphi_{\mathrm{base},s} = -\frac{Q}{S^{2}}, \qquad
\varphi_{\mathrm{base},z} = \frac{2QS'}{S^{2}}\,(\xi-1).
\label{eq:appC_base}
\end{equation}
At $\xi = 1$ ($s = S$) this gives $u_z = 2Q/S - 2S\,Q/S^{2} = 0$ and $\varphi_{\mathrm{base},z} = 0$. For the network term, $\varphi_{\mathrm{NN}} = G_z(z)(1-\xi)^{2}\hat\varphi$, and
\begin{equation}
\varphi_{\mathrm{NN},s} = G_z\Big[-\frac{2(1-\xi)}{S}\,\hat\varphi + (1-\xi)^{2}\hat\varphi_s\Big], \qquad
\varphi_{\mathrm{NN},z} = G_z'(1-\xi)^{2}\hat\varphi + G_z\Big[\frac{2(1-\xi)\xi S'}{S}\,\hat\varphi + (1-\xi)^{2}\hat\varphi_z\Big].
\label{eq:appC_nn}
\end{equation}
Every term contains a factor $(1-\xi)$, so $\varphi_{\mathrm{NN}}$, $\varphi_{\mathrm{NN},s}$, and $\varphi_{\mathrm{NN},z}$ all vanish at $\xi=1$. Hence $u_z|_{\mathrm{wall}} = 0$ and $u_r|_{\mathrm{wall}} = -r\varphi_z = 0$. \\

\textbf{3. Inlet ($z=-z_{\max}$, exact):} The axial gate $G_z = [(z+z_{\max})/(2z_{\max})]^{2}$ and its derivative $G_z' = (z+z_{\max})/(2z_{\max}^{2})$ both vanish at $z=-z_{\max}$. By Eq.~\eqref{eq:appC_nn}, every network contribution to $\varphi$, $\varphi_s$, and $\varphi_z$ is therefore zero at the inlet. The inlet velocity follows from the base term alone:
\begin{equation}
u_z\big|_{\mathrm{inlet}} = \frac{2Q(2-\xi)}{S} - \frac{2sQ}{S^{2}} = \frac{4Q}{S}\left(1-\frac{s}{S}\right).
\label{eq:appC_inlet}
\end{equation}
With $Q = u_{z,\mathrm{inlet}}/4$, this equals the target profile $u_{z,\mathrm{inlet}}(1-s)$ exactly when $S(-z_{\max}) = 1$. Since $S = h^{2}$ with $h = 1-\delta e^{-(\omega z)^{2}}$, $S(-z_{\max})$ differs from unity by $O\big(e^{-(\omega z_{\max})^{2}}\big)$; for $\delta = 0.5$, $\omega = 3.5$, and $z_{\max} = 1.5$ this is about $10^{-12}$, below single-precision round-off. Likewise $u_r|_{\mathrm{inlet}} = -r\varphi_{\mathrm{base},z} = -2rQS'(\xi-1)/S^{2}$, and $S'(-z_{\max})$ is of the same exponentially small order. \\

\textbf{4. Thermal wall and inlet:} In Eq.~\eqref{eq:theta_ansatz}, at the wall ($\xi=1$) the wall factor $1-\xi^{k}$ vanishes and $\xi^{q} = 1$, so $\theta|_{\mathrm{wall}} = \Omega_w(z)$, equivalently $T|_{\mathrm{wall}} = \Delta T\,\Omega_w(z)$. At the inlet, the gate $G(z) = 1-\exp\{-[(z+z_{\max})/(2z_{\max})]^{2}\}$ satisfies $G(-z_{\max}) = 0$ and $G'(-z_{\max}) = 0$, so its axial derivative vanish there. Because $\Omega_w$ is zero outside the heated patch $|z|\le z_{\mathrm{heat}}$, $\Omega_w(-z_{\max}) = \Omega_w'(-z_{\max}) = 0$, and therefore  $\partial_z\theta|_{\mathrm{inlet}} = 0$ hold exactly. \\

\textbf{Remark.} The axis, wall, and inlet conditions on velocity and temperature are thus built into the ans\"atze: exactly for every network-dependent term, and to an exponentially small residual for the analytic inlet profile. The corresponding boundary losses are therefore not trained; they are evaluated only as diagnostic checks of this construction. Only the outlet pressure condition is imposed as a soft loss. \\

\section{Nomenclature}
{\setlength{\tabcolsep}{4pt}
\renewcommand{\arraystretch}{1.12}
\setlength{\LTpre}{2pt}\setlength{\LTpost}{0pt}
\begin{longtable}{@{}l p{0.37\textwidth} @{\hspace{0.03\textwidth}} l p{0.37\textwidth}@{}}
\multicolumn{2}{@{}l}{\textit{Latin symbols}} & \multicolumn{2}{@{}l}{\textit{Greek symbols}} \\
$c_p$ & specific heat (J\,kg$^{-1}$\,K$^{-1}$) & $\alpha$ & thermal-expansion coefficient (K$^{-1}$) \\
$e_0^* $ & healthy artery radius (m) & $\Delta T$ & peak temperature ratio, $(T_S-T_0)/T_0$ \\
$G_z$, $G$ & flow and thermal inlet gates & $\delta^*$, $\omega $ & stenosis height and shape parameters \\
$g$ & gravitational acceleration (m\,s$^{-2}$) & $\theta$ & normalised temperature, $T/\Delta T$ \\
$h(z)$ & dimensionless wall radius & $\hat\theta$, $\hat\varphi$ & raw thermal and momentum network outputs \\
$K$ & thermal conductivity (W\,m$^{-1}$\,K$^{-1}$) & $\mu$ & dynamic viscosity (Pa\,s) \\
$L_{0}^{*}$ & axial length of the stenosis (m) & $\xi$ & boundary-fitted coordinate, $s/S(z)$ \\
$p$ & dimensionless pressure & $\rho$ & density (kg\,m$^{-3}$) \\
$Pe$, $Pe_{\mathrm{eff}}$ & P\'eclet number; ramped value & $\sigma$, $\sigma^*$ & heating width; dimensionless value \\
$Q$ & dimensionless volumetric flux & $\tau$ & drift-guard budget \\
$r$, $z$ & dimensionless radial, axial coordinates & $\Phi$ & nanoparticle volume fraction \\
$Re$, $Re_{bf}$ & Reynolds number; base-fluid value & $\varphi$ & modified stream function, $\psi=s\varphi$ \\
$Ri$, $Ri_{\mathrm{floor}}$ & Richardson number; reduced value & $\psi$ & Stokes stream function \\
$s$ & parity-preserving coordinate, $s=r^2$ & $\Omega_w(z)$ & wall-heating profile \\
$S(z)$ & squared wall radius, $h(z)^2$ & \multicolumn{2}{@{}l}{\textit{Subscripts and abbreviations}} \\
$s_{\mathrm{mom}}$, $s_{\mathrm{energy}}$ & residual scales & $f$, $s$, $nf$ & base fluid, particle, nanofluid \\
$T^*$, $T_0$, $T_S$ & temperature; reference; peak wall value (K) & $\mathcal{L}$, $\mathcal{R}$ & loss term; residual \\
$T$ & dimensionless temperature, $(T^*-T_0)/T_0$ & $Nu_{\mathrm{flux}}$ & wall Nusselt number \\
$u_r$, $u_z$ & dimensionless radial, axial velocity & $f_D$, $f_p$ & friction factor; pressure-drop coefficient \\
$u_0^*$ & inlet centreline velocity (m\,s$^{-1}$) &  &  \\
$z_{\max}$, $z_{\mathrm{heat}}$ & domain half-length; heated half-length &  &  \\
\end{longtable}
}

\end{document}